\documentclass[11pt]{article}
\pdfoutput=1
\usepackage{gensymb}
\usepackage{jheppub} \usepackage{braket,slashed,bm}
\usepackage{booktabs}
\usepackage{array,multirow}
\usepackage{amsmath}
\usepackage[normalem]{ulem}
\usepackage[T1]{fontenc}
\usepackage{mathrsfs}
\usepackage{tikz}
\usepackage{stmaryrd}
\usepackage{subcaption}
\usetikzlibrary{arrows,decorations.pathmorphing,backgrounds,positioning,fit,petri,automata,shadows,calendar,mindmap,
decorations.markings,calc}
\usepackage{lscape}
\usepackage{hyperref}
\usepackage{graphicx,wrapfig}
\usepackage{floatrow}
\newfloatcommand{capbtabbox}{table}[][\FBwidth]

\def\nn{\nonumber\\ }

\def\lsix{ \mathcal{L}^{(6)}}

\def\lsix{ \mathcal{L}^{(6)}}

\def\that{{\hat \theta}}

\def\sc{{\overline s}}

\newcommand{\M}{\ensuremath{\mathcal{M}}}

\def\bea{\begin{eqnarray}}
\def\eea{\end{eqnarray}}

\title{Incorporating doubly resonant $W^\pm$ data in a global fit of SMEFT parameters to lift flat directions.}
\author{
Laure Berthier, Mikkel Bj\o rn and Michael Trott\\
Niels Bohr International Academy,
University of Copenhagen, Blegdamsvej 17, DK-2100 Copenhagen, Denmark}

\abstract{We calculate the double pole contribution to two to four fermion scattering through $W^{\pm}$ currents at tree level in the Standard Model Effective Field Theory (SMEFT).
We assume all fermions to be massless, $\rm U(3)^5$ flavour and $\rm CP$ symmetry. Using this result, we update the global constraint picture on SMEFT parameters including LEPII data on these charged current processes, and also modifications to our fit procedure motivated by a companion paper focused on $W^{\pm}$ mass extractions. The fit reported is now to 177 observables and emphasises the need for a consistent inclusion of theoretical errors, and a consistent treatment of observables. Including charged current data lifts the two-fold degeneracy previously encountered in LEP (and lower energy) data, and allows us to set simultaneous constraints on 20 of 53 Wilson coefficients in the SMEFT, consistent with our assumptions.  This allows the model independent inclusion of LEP data in SMEFT studies at LHC, which are projected into the SMEFT in a consistent fashion. We show how stronger constraints can be obtained by using some combinations of Wilson coefficients, when making assumptions on the UV completion of the Standard Model, or in an inconsistent analysis.
We explain why strong bounds at the per-mille or sub-per-mille level on some combinations of Wilson coefficients in the Effective Lagrangian can be artificially enhanced in fits of this form in detail.
This explains some of the different claims present in the literature.}
\begin{document}
\maketitle

\section{Introduction} \label{sec:intro}

What is the shape of possible physics beyond the Standard Model? This question has been returned to with renewed vigor in recent years,
after the discovery of a Higgs like $J^P = 0^+$ boson at LHC. In this paper we investigate this question using the Standard Model Effective Field Theory (SMEFT)
formalism. We assume that  $\rm SU(2)_L \times U(1)_Y$ is spontaneously broken to $\rm U(1)_{em}$ by the vacuum expectation value ($\langle H^\dagger H \rangle  \equiv \bar{v}_T^2/2$) of the Higgs field, and the observed $0^+$ scalar is embedded in a doublet of $\rm SU(2)_L$. We also assume a mass gap to the scale of new physics $\sim \Lambda$. The SMEFT Lagrangian that follows from this assumption, is the sum of the Standard Model (SM) Lagrangian and a series of $\rm SU(3)_C \times SU(2)_L \times U(1)_Y$ invariant higher dimensional operators built out of SM fields
\bea
\mathcal{L}_{SMEFT} = \mathcal{L}_{SM} + \mathcal{L}^{(5)} + \mathcal{L}^{(6)} + \mathcal{L}^{(7)} + ...
\eea
where $\mathcal{L}^{(k)}$ contains the dimension $k$ operators $O_i^{(k)}$. The number of non redundant operators in $\mathcal{L}^{(5)}$, $\mathcal{L}^{(6)}$, $\mathcal{L}^{(7)}$ and $\mathcal{L}^{(8)}$ is known \cite{Buchmuller:1985jz,Grzadkowski:2010es,Weinberg:1979sa,Abbott:1980zj,Lehman:2014jma,Lehman:2015coa,Henning:2015alf} and a general algorithm to determine operator bases at higher orders has been established in Ref.\cite{Lehman:2015coa,Henning:2015alf}.
We adopt a naive power counting in mass dimension so that the operators  $O_i^{(k)}$ will be suppressed by $k-4$ powers of the cutoff scale $\Lambda$;
\bea
\mathcal{L}^{(k)}= \sum_{i = 1}^{n_k} \frac{C_i^{(k)}}{\Lambda^{k-4}} O_i^{(k)} \hspace{0.25cm} \text{ for $k > 4$, }
\eea
where the $C_i^{(k)}$ are the Wilson coefficients\footnote{Note that in this paper we generally absorb the
cut off scale into the Wilson coefficients associated to the dimension 6 operators, which then have mass dimension $-2$ unless otherwise noted.} associated to the operators $O_i^{(k)}$.
This approach conforms with the standard and well validated understanding of model independent EFT. It is unnecessary to adopt more restricted UV assumptions to globally constrain the SMEFT from data, however, we will also illustrate that once general model independent results are obtained, how these results
project into a variety of more restricted scenarios.

Indirectly constraining physics beyond the SM is of great value. This is clearly the case when there is no
direct collider evidence of new physics to guide model building. Even when partial hints of physics beyond the SM exist, such an approach is
still critical to globally understand the data set.  Broadly speaking, global constraint works can be grouped into the following categories:
\begin{itemize}
\item{The $\rm STU$ core. In advance of LEP data, the utility of parameterizations of vacuum polarization effects to indirectly constrain
the source of Electroweak Symmetry Breaking was appreciated in a series of papers \cite{Kennedy:1988sn,Altarelli:1990zd,Altarelli:1991fk,Golden:1990ig,Holdom:1990tc,Peskin:1990zt,Peskin:1991sw}. The capstone of these developments was the work of Peskin and Takeuchi establishing the modern $\rm STU$ formalism in
Ref.\cite{Peskin:1991sw}. The STU approach has had manifest utility over
the years. On the other hand, the $\rm STU$ approach is defined with conditions that are not field redefinition invariant, considering an operator
level EFT interpretation of EWPD.\footnote{Attempts to deal with this situation by restricting
ones attention to classes of UV theories that are consistent with the STU defining assumptions, do allow  model dependent interpretations of
EWPD in the $\rm STU$ framework of course.} Despite this limitation, the $\rm STU$ approach was efficient at constraining indirectly the possible mass of the Higgs Boson
when the SM is assumed, by construction. The validation of the inferred Higgs mass with the discovered $0^+$ state's mass is further support for the historical importance of the $\rm STU$ approach.}
\item{The LEP and post-LEP interpretation and $\rm STU$ extension phase.
Immediately after the establishment of the $\rm STU$  approach, extensions to this parameterization were advanced in the literature. These extensions
allow the mass scale of new physics to be lower \cite{Maksymyk:1993zm,Burgess:1993vc} than implicitly assumed in the $\rm STU$ formalism,
or for a set of data off the $Z$ pole to be accommodated  in some limited cases of physics beyond the SM. Several of these works focused on the
potential of LEPII data  \cite{Barbieri:2004qk,Barbieri:2000gf}, and measurements sensitive to Triple Gauge Couplings (TGCs) in an EFT framework \cite{Gaemers:1978hg,DeRujula:1991ufe,Hagiwara:1993ck},
with Ref. \cite{Hagiwara:1986vm} being a core reference.}
\item{The development of the SMEFT analysis. This approach is advanced further in this work, and was developed in parallel to some of the developments above.
Immediately following the initial  $\rm STU$ analysis works, Ref.\cite{Grinstein:1991cd} performed an operator EFT analysis of electroweak precision data.
The next major advance in this effort was achieved in Ref.\cite{Han:2004az} where a global analysis similar to the work presented here was performed.
These efforts were hampered by the lack of any non-redundant minimal operator basis for $\mathcal{L}_6$.
With the establishment of this basis in Ref.\cite{Grzadkowski:2010es}, progress in SMEFT global analyses
was reinvigorated. Refs.\cite{Ciuchini:2013pca,
Ciuchini:2014dea,Buchalla:2013wpa,Durieux:2014xla,Petrov:2015jea,Cirigliano:2009wk,Gonzalez-Alonso:2016etj,Grinstein:2013vsa,Wells:2014pga,Trott:2014dma,Henning:2014wua,deBlas:2015aea,Corbett:2015ksa,Buckley:2015lku,Cirigliano:2016nyn} made contributions to this effort and
Ref.\cite{Berthier:2015oma,Berthier:2015gja,David:2015waa} has recently formed a line of developments that are distinct from past analyses, in their consideration and treatment of theoretical
errors in the SMEFT. The conclusions reached in these works, are that model independent constraints on parameters in $\mathcal{L}_6$ require a careful consideration
of theoretical errors in the SMEFT, and that such a consideration can weaken model independent bounds to the percent level on the combinations of parameters $C_i \bar{v}_T^2/\Lambda^2$.
However we stress, as was also stressed in Ref.\cite{Berthier:2015oma,Berthier:2015gja,David:2015waa} that if this relaxation occurs, or not, strongly depends on the unknown UV physics underlying the SMEFT.
Nevertheless, as general model independent bounds are intended to cover all UV cases consistent with analysis assumptions, this can still dictate a model independent statement.}
\end{itemize}

The past results of two of us, were limited by the presence of two flat directions in the Wilson coefficient space in the global fit
\cite{Berthier:2015oma,Berthier:2015gja}. In this work we address this issue in a consistent and reproducible manner in the SMEFT. Doing so, it is important to calculate the full cross sections for charged current LEPII data that we report, and not use a TGC parameterization as effectively an observable.
An off-shell TGC vertex is not an observable in the sense that such a vertex is gauge dependent and is not trivially mapped to the $S$ matrix due to its off-shellness.
The problems introduced when not using an observable to constrain the SMEFT parameter space
model independently were emphasized in Ref.\cite{Trott:2014dma}.
To overcome these issues, it is required to calculate two to four fermion scattering through $W^{\pm}$ currents in order to fit LEPII data at leading order in the SMEFT power counting.  In this paper we perform this calculation in the SMEFT using the Warsaw basis \cite{Grzadkowski:2010es} for $\mathcal{L}_6$ and perform this fit.

Our results include the consistent redefinition of the set of parameters used in making the two to four fermion scattering observables and assume massless initial and final state fermions, $\rm U(3)^5$ flavour and $\rm CP$ symmetry, but are not limited to formally on-shell intermediate $W^{\pm}$ bosons, or a TGC parameterization.
With these assumptions in mind, we calculate the CC03 production cross section\footnote{The CC03 cross section is a subclass of the full set of diagrams appearing at tree level, motivated by the different scaling and pole structure of the various diagrams contributing to the processes. See Refs.\cite{Achard:2004zw,Abbiendi:2007rs,Heister:2004wr} for more discussion.} {\it utilizing the double pole approximation in the SMEFT} to define the off-shell two to four fermion scattering through $W^\pm$ charged currents in the SMEFT.
We present the calculation and results in Section \ref{sec:calculation}. LEPII results on the CC03 cross sections are extracted from measured $e^+ e^- \rightarrow 4f$-events \cite{Achard:2004zw,Abbiendi:2007rs,Heister:2004wr}. Using our results and the measurements in Tables \ref{L3Data}, \ref{OPALData}, \ref{ALEPHData} we update the global fit initiated in \cite{Berthier:2015oma,Berthier:2015gja} and give model independent constraints on 20 Wilson coefficients in Section \ref{sec:global_fit}. In a companion paper, we explain how
$m_W$ measurements were included in past fit efforts in a manner that was not optimal for the SMEFT. We also adopt the recommendations of
\cite{todaypaper} and incorporate extractions of the $m_W$ mass from the Tevatron, and related LEPII data in a more consistent manner in this work.

The general model independent results we report must be interpreted with care. We explain and illustrate how significantly different conclusions have been reached in the literature
for effective combinations of Wilson coefficients present in the SMEFT Lagrangian when it is rotated to mass eigenstate for the $W^\pm,Z$ bosons, or Wilson coefficients in the Warsaw basis. Essentially, these different conclusions
are related to different (usually implicit) assumptions about the UV physics underlying the SMEFT allowing significant cancelations between parameters (in the case of the Warsaw basis) or not (in the case of mass eigenstate parameters). The limited theoretical development\footnote{The lack of a consistent set of one loop results for LEP data interpreted in the SMEFT in particular.} of the SMEFT to date also limits the strength of the model independent bounds that can be drawn. The differing conclusions are most difficult to reconcile when the results are presented as general model independent bounds, that are intended to
span all possible UV cases. This difficulty is relaxed when theoretical errors for the SMEFT itself are considered and included in a fit of this form \cite{Berthier:2015oma,Berthier:2015gja}. We demonstrate how UV assumptions strongly enhance the strength of bounds on the Wilson coefficients of the individual operators in the SMEFT, and how fits to some combinations of Wilson coefficients in the Effective Lagrangian\footnote{We generally refer to these parameters as $\delta X$ parameters, mass eigenstate parameters, or "core shifts" in this paper. This is consistent with our previous usage for these parameters in Ref.~\cite{Berthier:2015oma,Berthier:2015gja}. These parameters correspond to combinations of Wilson coefficients that appear in a number of Feynman diagrams. However,  it is important to note that these combinations of parameters do not constitute an operator basis for the SMEFT \cite{Passarino:2138031}.} can be subject to significant theoretical errors. Our results clearly explain the discrepancies present in the literature and support
the conclusion (already argued in Refs.\cite{Berthier:2015oma,Berthier:2015gja}) that the SMEFT analysis of LEP data should be further developed theoretically, in order to robustly develop model independent results with sub-dominant theory errors. The global $\chi^2$ constructed is fully reproducible from the results reported in this paper and the Fisher information matrices, that are available from the authors upon request.

\section{Four fermion production in the SMEFT} \label{sec:calculation}

This paper further develops the results reported in Refs.~\cite{Alonso:2013hga,Berthier:2015oma,Berthier:2015gja}. Our notation and conventions are consistent with these works. We use bar superscripts for parameters in the canonically normalized SMEFT, and hat superscripts for measured input parameters, or parameters
directly related to input parameters at tree level using SM relations. We use the input parameter set $\{\hat{G}_F,\hat{m}_Z,\hat{\alpha} \}$. %\subsection{Redefinitions of Electro-Weak Parameters} % (fold)
%\label{sub:redefinitions_of_electro_weak_parameters}
% section redefinitions_of_electro_weak_parameters (end)

%\begin{align}
   % \delta_{\hat s_\theta}^2 =- \frac{s_{\hat \theta}c_{\hat \theta}}{2\sqrt{s}\hat G_F (1-2 s_{\hat \theta}^2)}\left[ s_{\hat \theta}c_{\hat \theta}\left(C_{HD}+4C_{H\ell}^{(3)}-2C_{\ell \ell}\right) + 2 C_{HWB} \right]
%\end{align}

\begin{figure}[t]
	\centering
	\begin{subfigure}{0.40 \textwidth}
	\includegraphics[width=0.95\textwidth]{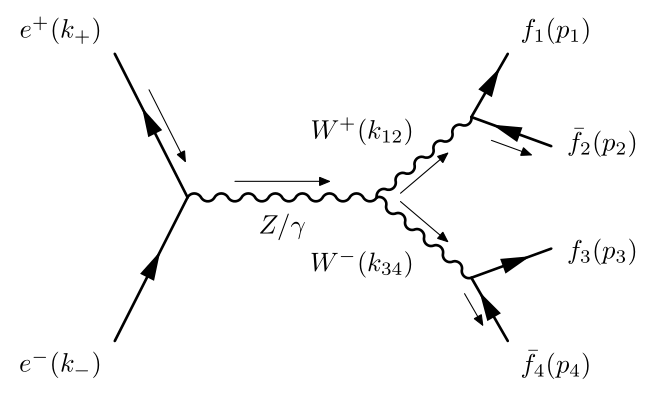}
	\caption{}
	\label{fig:CC03_s_channel}
	\end{subfigure}
	\begin{subfigure}{0.40 \textwidth}
	\includegraphics[width=0.95\textwidth]{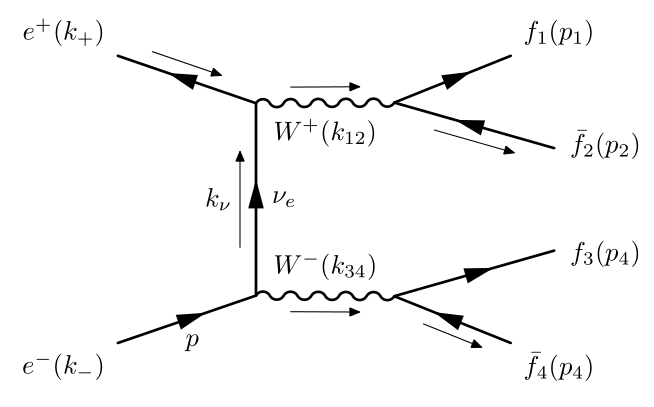}
	\caption{}
	\label{fig:CC03_t_channel}
	\end{subfigure}

	\caption{ The $s$-channel (a) and $t$-channel (b) CC03 Feynman diagrams contributing to $e^+ e^- \rightarrow W^+ W^- \rightarrow f_1 \bar{f}_2 f_3 \bar{f}_4$. The diagrams can be understood either in the SM, or in the SMEFT by taking couplings and gauge boson vector masses to be redefined as described in Section \ref{sec:calculation} and Appendix \ref{CoreShifts}. }
	\label{fig:CC03}
\end{figure}

\subsection{The CC03 Matrix Elements}
\label{sub:matrx_element_results}
The theoretical cross section of the double pole contribution to the process
\bea
e^+ \left(k_{+}, \lambda_{+}\right) e^-\left(k_{-},\lambda_{-}\right) \rightarrow  W^+ \left(k_{12},\lambda_{12}\right) W^- \left(k_{34},\lambda_{34}\right) \rightarrow f_1^{\lambda_1} \left(p_1\right) \bar{f}_2^{\lambda_2} \left(p_2\right) f_3^{\lambda_3} \left(p_3\right) \bar{f}_4^{\lambda_4}\left(p_4\right), \nonumber
\eea
is computed from the CC03 set of three (Charged Current) Feynman diagrams shown in Figure \ref{fig:CC03}.  Final states can be either fully hadronic $\left(q,q,q,q\right)$, semi-leptonic $\left(\ell, \nu, q, q \right)$ or fully leponic $\left(\ell,\nu,\ell,\nu\right)$. We use the  the spinor helicity formalism of  Ref.\cite{Hagiwara1986Helicity,Hagiwara:1986vm}, where the helicities are labeled $\{\lambda_{\pm},\lambda_{12},\lambda_{34},\lambda_{1,2,3,4}\}$.  In Appendix \ref{CoreShifts} we list some of the results of Refs.\cite{Berthier:2015oma,Berthier:2015gja} that are used directly in this work to make the paper self contained.  We also summarize some simple parameter redefinitions in the SMEFT that are used in the cross section results in the Appendix. Further, Appendix \ref{Parametrisation} gives details on the phase space definitions, which also defines some of our notation.
The matrix elements corresponding to each diagram are decomposed into separate factors for $W^+W^-$ production and decay
\begin{align}
	\M_{\nu} &=  \bar D^W(s_{12}) \bar D^W(s_{34}) \M_\nu^{\lambda_i}  \M_{W^+}^{\lambda_{12}}  \M_{W^-}^{\lambda_{34}} \\
	\M_{V} &=  \bar D^W(s_{12}) \bar D^W(s_{34})  \M_V^{\lambda_i}  \M_{W^+}^{\lambda_{12}}  \M_{W^-}^{\lambda_{34}}
\end{align}
with $V=\{A,Z\}$ and the sub-amplitudes
\bea
\M_\nu^{\lambda_i} =  \M_{ee \rightarrow WW,\nu}^{\lambda_{12}\lambda_{34}\lambda_+\lambda_-}, \hspace{0.20cm}  \M_V^{\lambda_i} = \M_{ee \rightarrow WW,V}^{\lambda_{12}\lambda_{34}\lambda_+\lambda_-}, \hspace{0.20cm}  \M_{W^+}^{\lambda_{12}} = \M_{W^+\rightarrow f_1\bar f_2}^{\lambda_{12}}, \hspace{0.20cm}
 \M_{W^-}^{\lambda_{34}} =\M_{W^-\rightarrow f_3 \bar f_4}^{\lambda_{34}}
\eea
are given in the Tables  \ref{Wf1f2}, \ref{Wf3f4}, \ref{nu-exchange} and \ref{V-exchange-+} and $\bar D^W({s_{ij}})$. The $W^\pm$-propagators are denoted
\begin{align}
     \bar{D}^W\left(s_{ij}\right) &=\frac{1}{s_{ij} - \bar{m}_W^2 + i \bar{\Gamma}_W \bar{m}_W + i \epsilon}.
\end{align}
and we have chosen to define the width in an $s$ independent manner. The challenge of defining gauge invariant expressions for this process,
due to the requirement of defining the propagator of the unstable $W^\pm$ bosons, is well known \cite{Veltman:1963th,Stuart:1991xk,Grunewald:2000ju}. We return to this point below.

The $W^\pm$-decay matrix elements $\mathcal{M}_{W^{+} \rightarrow f_1 \bar{f}_2}^{\lambda_{12}}$ and $\mathcal{M}_{W^{-} \rightarrow f_3 \bar{f}_4}^{\lambda_{34}}$ are shown in Table \ref{tab:WDecay} for the helicity values $\lambda_{ij}= \{0,+,-,L\}$. We denote the longitudinal polarization of the virtual $W^\pm$ bosons with an $L$,
which vanishes in the case of massless fermions, so we subsequently neglect it. In obtaining the expressions for the helicity amplitudes, we have checked against Ref.\cite{Hagiwara:1986vm}, finding agreement with the SM expressions. We give details on the calculation in Appendix \ref{SpinorHelicity} including the extension to the SMEFT case. In Table.\ref{tab:WDecay} we show the results for the decomposition of the $W^\pm$ decay amplitudes into Helicity eigenstates to briefly familiarise the uninitiated reader with this formalism.
\begin{table}
\begin{subtable}{0.45 \textwidth}

\begin{tabular}{|c|c|}
\hline
$\lambda_{12}$ & $\mathcal{M}_{W^{+} \rightarrow f_1 \bar{f}_2}^{\lambda_{12}}/C \sqrt{2\pi \hat{\alpha}}$   \\ 
\hline \hline
$0$&$ \frac{- 2 \bar{g}_V^{W,f_1}}{s_{\that}}\sqrt{s_{12}}\sin \tilde{\theta}_{12}$ \\
$+$&$ \frac{\bar{g}_V^{W,f_1}}{s_{\that}}\sqrt{s_{12}}\sqrt{2}\left(1-\cos \tilde{\theta}_{12}\right) \, e^{i \tilde{\phi}_{12}}$\\
$-$&$ \frac{\bar{g}_V^{W,f_1}}{s_{\that}}\sqrt{s_{12}}\sqrt{2}\left(1+\cos \tilde{\theta}_{12}\right) \, e^{-i \tilde{\phi}_{12}}$\\
$L$&$0$ \\
\hline
\end{tabular}

\caption{}
\label{Wf1f2}
\end{subtable}
\begin{subtable}{0.45 \textwidth}
%%%%%%%%%%%%%%%%%%%%%%%%%%%%%%%

%%%%%%%%%%%%%%%%%%%%%%%%%%%%%%%%%%%
\begin{tabular}{|c|c|}
\hline
$\lambda_{34}$ & $\mathcal{M}_{W^{-} \rightarrow f_3 \bar{f}_4}^{\lambda_{34}}/C'\sqrt{2\pi \hat{\alpha}}$   \\ 
\hline \hline
$0$&$ \frac{2 \, \bar{g}_V^{W,f_3}}{s_{\that}}\sqrt{s_{34}}\sin \tilde{\theta}_{34}$ \\
$+$&$ \frac{- \bar{g}_V^{W,f_3}}{s_{\that}}\sqrt{s_{34}}\sqrt{2}\left(1-\cos \tilde{\theta}_{34}\right) \, e^{-i \tilde{\phi}_{34}}$\\
$-$&$ \frac{-\bar{g}_V^{W,f_3}}{s_{\that}}\sqrt{s_{34}}\sqrt{2}\left(1+\cos \tilde{\theta}_{34}\right) \, e^{+i \tilde{\phi}_{34}} $\\
$L$&$0$ \\
\hline
\end{tabular}

%%%%%%%%%%%%%%%%%%%%%%%%%%%%%%%%%%%%%%
\caption{}
\label{Wf3f4}
\end{subtable}
\caption{The $W^\pm$-decay amplitudes decomposed in helicity Eigenstates. $C' = \{1, \sqrt{3}\}$ for leptons and quarks respectively.} \label{tab:WDecay}
\end{table}

\subsection{The double pole approximation in the SM}

In the SM, the definition of gauge independent doubly resonant contributions to $\sigma(e^+ e^- \rightarrow \bar{f}_1 \, f_2 \, \bar{f}_3 \, f_4 )$ is afflicted with a series of subtleties. We first briefly review the well known issues in the SM, discussed in part in Ref.\cite{Veltman:1963th,Stuart:1991xk,Grunewald:2000ju,Beenakker:1994vn,Beneke:2003xh}, based on the excellent and extensive discussion in Ref.\cite{Beenakker:1994vn}. These subtleties are also relevant when considering the SMEFT expression for the corrections to this process in a consistent approach.

First consider on-shell $\sigma(e^+ \, e^- \rightarrow  W^+ \, W^-)$. In this case, the three CC03 diagrams\footnote{CC03 diagram contributions to two to four fermion scattering were calculated in Refs.\cite{Tsai:1965hq,Flambaum:1974wp,Alles:1976qv,Bletzacker:1977gi,Gaemers:1978hg,Brown:1978mq,Bilchak:1984ur,Gunion:1985mc,Hagiwara:1986vm,Beenakker:1994vn}.} are manifestly gauge invariant in two sub-expressions for the amplitudes sensitive to a particular coupling in the SM: $\{e,g_2 \}$. So long as the $W^\pm$ are considered to be experimentally reconstructed states, that are effectively treated as asymptotic states of the $S$ matrix, further subtleties can be avoided when considering the tree level expressions for this process. If the precision of SM predictions is desired to reach a level that is sensitive to perturbative corrections, or $\Gamma_W/m_W$ corrections -- which is essentially the percent level and potentially the size of SMEFT corrections -- then this approximation fails.
To incorporate LEPII data that is dominantly off-shell with $s > 4 \, \bar{m}_W^2$, the theoretical expression for off-shell production must be used.

For off-shell production, the situation is more subtle, even when considering a Born approximation to the process. In this case, the CC03 diagrams are not trivially gauge invariant as a subset of the full amplitude. The reason is that the  $W^\pm$ is not being treated as an asymptotic state, so a cut in the Feynman diagram imposing
two simultaneous $W^\pm$ states is no longer well defined. It can be shown that the difference in the axial and t'Hooft-Feynman gauge expressions for the CC03 diagrams when considering off-shell production generates a single-resonant diagram contributing to the $\sigma(e^+ e^- \rightarrow \bar{f}_1 \, f_2 \, \bar{f}_3 \, f_4 )$ process  \cite{Beenakker:1994vn}. Including such singly resonant diagrams in four fermion production are sometimes referred to as the CC11 diagrams in the literature, and some results are reported in Ref.\cite{Bardin:1995uc}. Thus the set of doubly resonant CC03 diagrams is not individually gauge invariant for off-shell $W^\pm$. The sum of the single resonant and double resonant diagrams are in general a gauge invariant subset of diagrams, up to considerations of defining the $W^\pm$ propagator with a finite width. This is the case, since for diagrams where final state fermions are distinct from the initial state fermions, these are the only sets of diagrams contributing \cite{Beenakker:1994vn}.

Naively, once the full set of doubly resonant, singly resonant, and non resonant diagrams are included, one might consider gauge invariance a non-issue.
However, it is still required to define the propagator of the unstable $W^\pm$ bosons. There is no unique prescription for the definition in the field theory. Various choices can be made, defining the contribution of the width to the propagator as $s^2 \, \Gamma_W/m_W$ or $\Gamma_W \, m_W$, leading to gauge invariant results. However, the
individual double resonant, single resonant and non-resonant diagrams are not individually gauge invariant, and this remains the case when a naive substitution of a finite width
in the $W^\pm$ propagator is included.

The effective resolution of these issues in the SM is the use of the double pole scheme to define the process. In this scheme, the full amplitude is decomposed as \cite{Beenakker:1994vn}
\bea
\mathcal{A}(s_{12},s_{34}) &=& \frac{1}{s_{12} - \bar{m}_W^2} \frac{1}{s_{34} - \bar{m}_W^2} {\rm DR}[s_{12},s_{34}, \Omega] + \frac{1}{s_{12} - \bar{m}_W^2}  {\rm SR_1}[s_{12},s_{34},d \Omega], \nonumber \\
&+&  \frac{1}{s_{34} - \bar{m}_W^2}  {\rm SR_2}[s_{12},s_{34},d \Omega] + {\rm NR}[s_{12},s_{34},d \Omega].
\eea
Here ${\rm DR}$, ${\rm SR_{1,2}}$ and ${\rm NR}$ refer to the doubly resonant, singly resonant and non-resonant contributions to the amplitude respectively, and $\Omega$ refers to all angular dependence refined in an $s_{12},s_{34}$ independent manner. Note that the ${\rm SR_{1,2}}$ results include subtractions of components of the CC03 diagrams, and the ${\rm NR}$ results include subtractions of components of the CC11 set of diagrams.

The residues of the double pole contribution are defined as $ {\rm DR}[ \bar{m}_W^2, \bar{m}_W^2, \Omega]$
in a Laurent expansion around the {\it physical poles} in the process. The residues of the poles are then gauge invariant as they can be experimentally measured (in principle). The width of the unstable $W^\pm$ is then added into these pole expressions after the residues are determined, and the individual pieces of the sub-amplitudes are then gauge invariant.
This approach, with perturbative corrections, underlies the SM prediction of this process in the double pole approximation in Refs. \cite{Beenakker:1994vn,Beenakker:1996kt,Bardin:1997gc,Denner:2000bj,Denner:2002cg}. This approach can also be justified in an EFT approach to unstable particles \cite{Beneke:2003xh}.
When considering the doubly resonant contribution defined in this manner, corrections are $\Gamma_W/\bar{m}_W \sim \mathcal{O}(\%)$.

Note that this procedure effectively defines the SM prediction of this process when considering LEP data. The data reported is corrected back to the CC03 set of diagrams
by performing Monte-Carlo studies on the full set of diagrams contributing to this process and comparing the predictions of the doubly resonant contribution.\footnote{This inferred correction factor is modified by SMEFT corrections, but this neglected theoretical error scales as $\Gamma_W/m_W \, \bar{v}_T^2/\Lambda^2 \sim 10^{-2} \bar{v}_T^2/\Lambda^2$, and is accommodated by SMEFT theory errors included in the fit.}
This is the data we incorporate into the global fit in the remainder of the paper, so an understanding of the double pole definition of the cross section in the SMEFT is required.

\subsection{The double pole approximation in the SMEFT}

When considering the definition of the corrections to this process in the SMEFT, the discussion in the previous section on the difficulties present in defining
the two to four fermion scattering process through charged currents, explains some long standing disagreements in the literature. The most naive approach to take when considering higher dimensional operators contributing to LEPII measurements is as follows. Expand out just the effects of the operators
leading to the TGC parameters, add these contributions to the calculation of a narrow width approximation to $\sigma(e^+ \, e^- \rightarrow  W^+ \, W^-)$,
and compare to the data reported for the CC03 off-shell diagrams, defined in a double pole prescription from LEPII. Directly using the data reported in Refs.\cite{Schael:2013ita,Abdallah:2010zj} and treating the TGC parameter as an observable in this manner, is not a gauge and field redefinition invariant procedure.

A TGC parameter is a constructed observable \cite{Trott:2014dma} inferred from the actual measurement, and care must be taken when  using such a measurement to constrain the parameter space of the SMEFT. The main issue can be traced back to approximating the $W^\pm$ boson as effectively an asymptotic external state in the calculation, and the inconsistency of this treatment with the field redefinitions in the SMEFT to define an $\mathcal{L}_6$ operator basis.  Recall that operator bases are defined by first constructing all gauge invariant operators of a mass dimension and then performing small field redefinitions
of $\mathcal{O}(1/\Lambda^2)$ on the field variables. Using the EOM on the transformations that result allows a minimal non-redundant operator basis to be defined
by essentially aligning the field variables with the external states, consistent with the classical equations of motion
conditions.\footnote{So long as the field redefinitions are defined in a gauge invariant manner.} Treating the $W^\pm$ directly as a classical external state (even when it is off shell) and not an internal off-shell field variable in all calculations\footnote{While not using the background field method.} might be considered equivalent to this procedure but this is actually inconsistent with obtaining basis independent constraints on the field theory, as it is simultaneously required
to perform field redefinitions on the $W^\pm$ boson, in order to even define a non-redundant operator basis.

Aspects of this issue has lead to long standing claims that some $\mathcal{L}_6$ operator bases are "better" to use to incorporate constraints due to LEPI and LEPII data,
despite the fact that constraints on the $S$ matrix are basis independent. These claims use the data in a manner
that treats the $W^\pm$ as directly an external state, and are choosing $\mathcal{L}_6$ parameters aligned with such a (mis)treatment of the data.\footnote{In particular factorized expressions are used for charged current processes that
assume a "SM-like" $W^\pm$ and $Z$ decay to fermions, where possible corrections due to $\mathcal{L}_6$ are set to zero in these decays. However, this assumption corresponds to different parameters in different operator bases.} However, these results are problematic as they are not consistent constraints on the SMEFT parameter space
that are basis independent and such a procedure is ambiguous and inconsistent in practice.

The resolution of this issue for the LHC physics program is important, as operator bases can be chosen so that the number of parameters in $\mathcal{L}_6$ contributing to LEPI data, exceeds the number of LEPI pseudo-observables - resulting in the famous two flat directions in LEPI data \cite{Han:2004az}. As such, model independent and basis independent bounds that incorporate the strong constraints of LEPI data, must incorporate LEPII constructed observable data on CC03 cross sections in some manner. It is important to incorporate
these constraints in a basis independent manner when reporting model independent analysis to use in studying LHC data.  The clear resolution to all of these issues is to calculate directly the doubly resonant contribution to LEPII data in the SMEFT and use this result to consistently fit the data. This is the approach we take in this paper.

The procedure to follow to incorporate this data consistently in this formalism is as follows. We define the SMEFT CC03 cross section in direct analogy to the double pole prescription of the SM.
The amplitude is again defined as the residues of the double pole contribution as $ {\rm DR}[ \bar{m}_W^2, \bar{m}_W^2, \Omega]$
in a Laurent expansion around the {\it physical poles} in the process. The relationship between the physical poles taken to define the residues, and the parameters in the SMEFT Lagrangian differ from the SM Lagrangian at leading order in the power counting
\bea
\frac{\delta m_W^2}{\hat{m}_W^2} =\frac{c_{\hat{\theta}} s_{\hat{\theta}}}{(c^2_{\hat{\theta}}-s^2_{\hat{\theta}}) \, 2 \sqrt{2} \hat{G}_F} \, \left[4  \, C_{HWB} + \frac{c_{\hat{\theta}}}{s_{\hat{\theta}}} C_{HD} + 4 \frac{s_{\hat{\theta}}}{c_{\hat{\theta}}} C_{H l}^{(3)} - 2 \frac{s_{\hat{\theta}}}{c_{\hat{\theta}}} C_{ll}\right].
\eea
We take this correction into account when using LEPII data to constrain the SMEFT parameter space.
We emphasize that: {\it The residues of the poles of the doubly resonant CC03 diagrams are fixed to be equal to $s_{12}=s_{34} = \bar{m}_W^2$,
the \it pole value in the SMEFT including the leading $\mathcal{L}_6$ corrections}. We then define the width in the $W^\pm$ propagator to be independent of $s$ and expand the propagator factors in the SMEFT corrections
\bea
\bar{\chi}\left(s_{ij}\right) &=& \bar{D}^W \left(s_{ij}\right)  \bar{D}^{* W} \left(s_{ij}\right)\nonumber \\
&=& \frac{1}{\left(s_{ij} - \bar{m}_W^2 \right)^2 + \left(\bar{\Gamma}_W \bar{m}_W\right)^2} = \frac{1}{\left(s_{ij} -\hat{m}_W^2 \right)^2 + \left(\hat{\Gamma}_W \hat{m}_W\right)^2} \left[1 + \delta \chi \left(s_{ij}\right)\right], \nonumber
\eea
where the modification is given by
\begin{align}
    \delta \chi \left(s_{ij}\right) &= \frac{\left[- 2\left(s_{ij} - \hat{m}_W^2\right) + \hat{\Gamma}_W^2 \right]\delta m_W^2 - 2 \hat{\Gamma}_W \hat{m}_W^2 \delta \Gamma_W}{\left(s_{ij} - \hat{m}_W^2 \right)^2 + \left(\hat{m}_W \hat{\Gamma}_W\right)^2}, \nonumber
\end{align}
and has the same pole structure as the cross section itself. This second step is required to be able to perform a well defined statistical ($\chi^2$) minimization procedure when the input parameter set
$\{\hat{G}_F,\hat{m}_Z,\hat{\alpha} \}$ is used. The difference in this approach and an alternative approach which expands the propagators, and then fixes to the SM tree level value of the $W^\pm$ mass, $s_{12}=s_{34} = \hat{m}_W$,
is conceptually related to considering the calculation to be in the SM or the SMEFT.

Considering this discussion,
the utility of adopting the input parameter set $\{\hat{G}_F,\hat{m}_Z,\hat{m}_W \}$ in future SMEFT studies  is manifest. When incorporating LEPII data, other off-shell data at LHC, or interfacing with the developing Higgs pseudo-observable program \cite{Passarino:2010qk,Isidori:2013cga,Gonzalez-Alonso:2014eva,Greljo:2015sla} such an input set makes double pole calculations required to define off shell data easier to
carry out in the SMEFT. It would be unfortunate to adopt a SMEFT implementation for LHC data that is "hard wired" to the $\{\hat{G}_F,\hat{m}_Z,\hat{\alpha} \}$ input parameter set for this reason, as has been discussed elsewhere at length \cite{Passarino:2138031}.
\subsection{The CC03 Cross Section in the SMEFT}
\begin{figure}[h!]
	\centering
	\begin{subfigure}{0.49 \textwidth}
	\includegraphics[width=1 \textwidth]{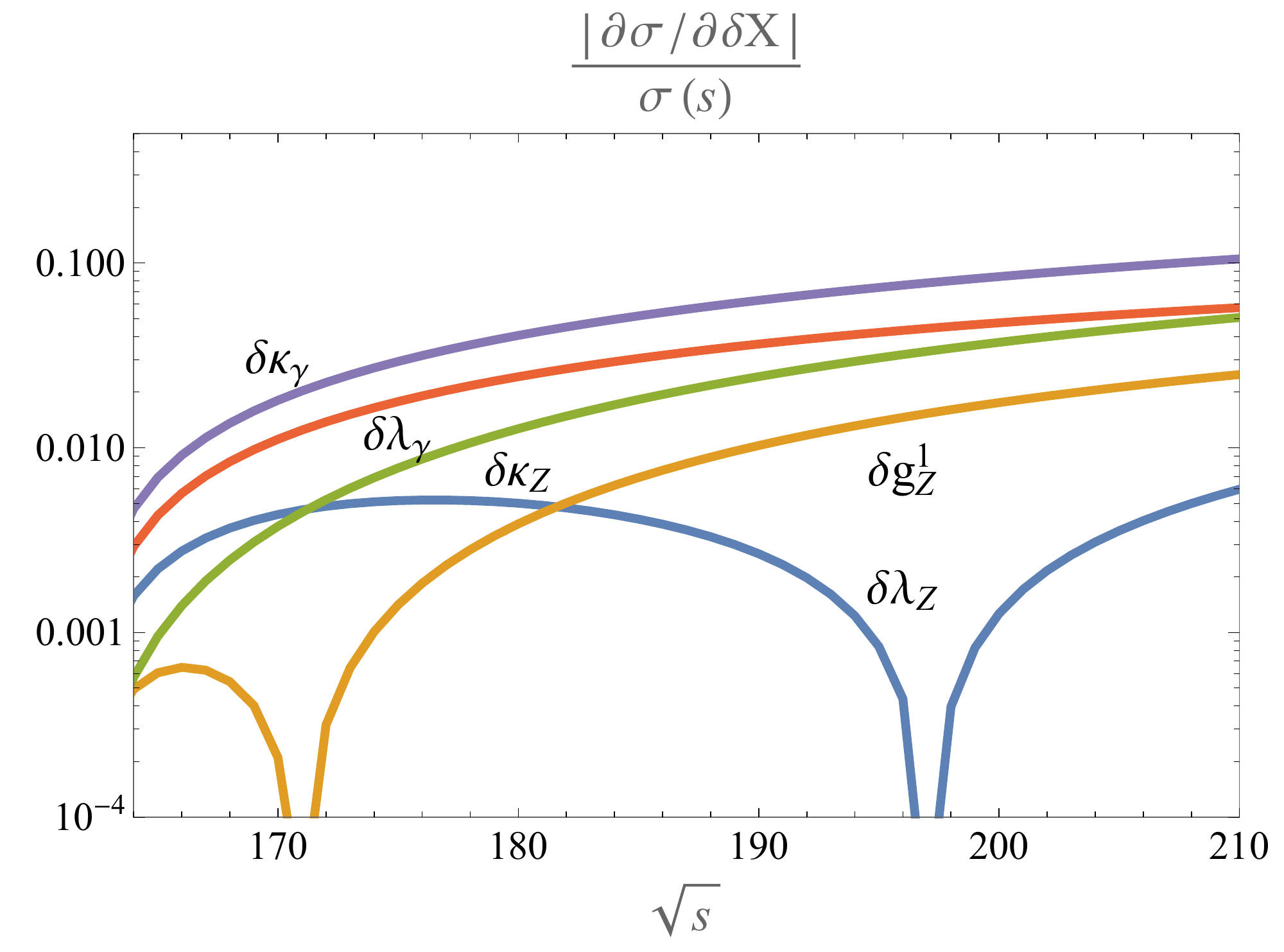}
	\caption{}
	\end{subfigure}
	\begin{subfigure}{0.49 \textwidth}
	\includegraphics[width=1 \textwidth]{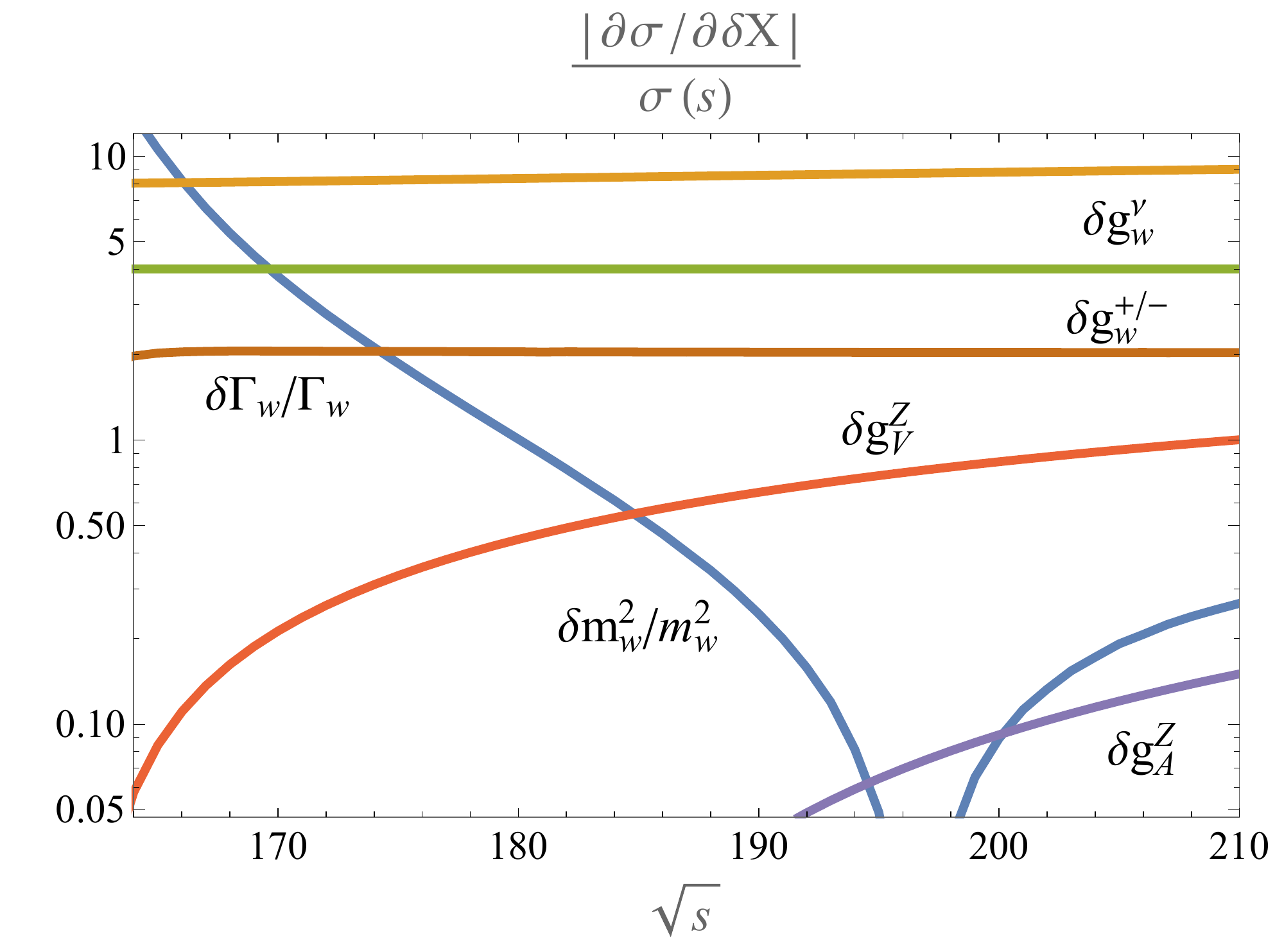}
	\caption{}
	\end{subfigure}
		\begin{subfigure}{0.49 \textwidth}
	\includegraphics[width=1 \textwidth]{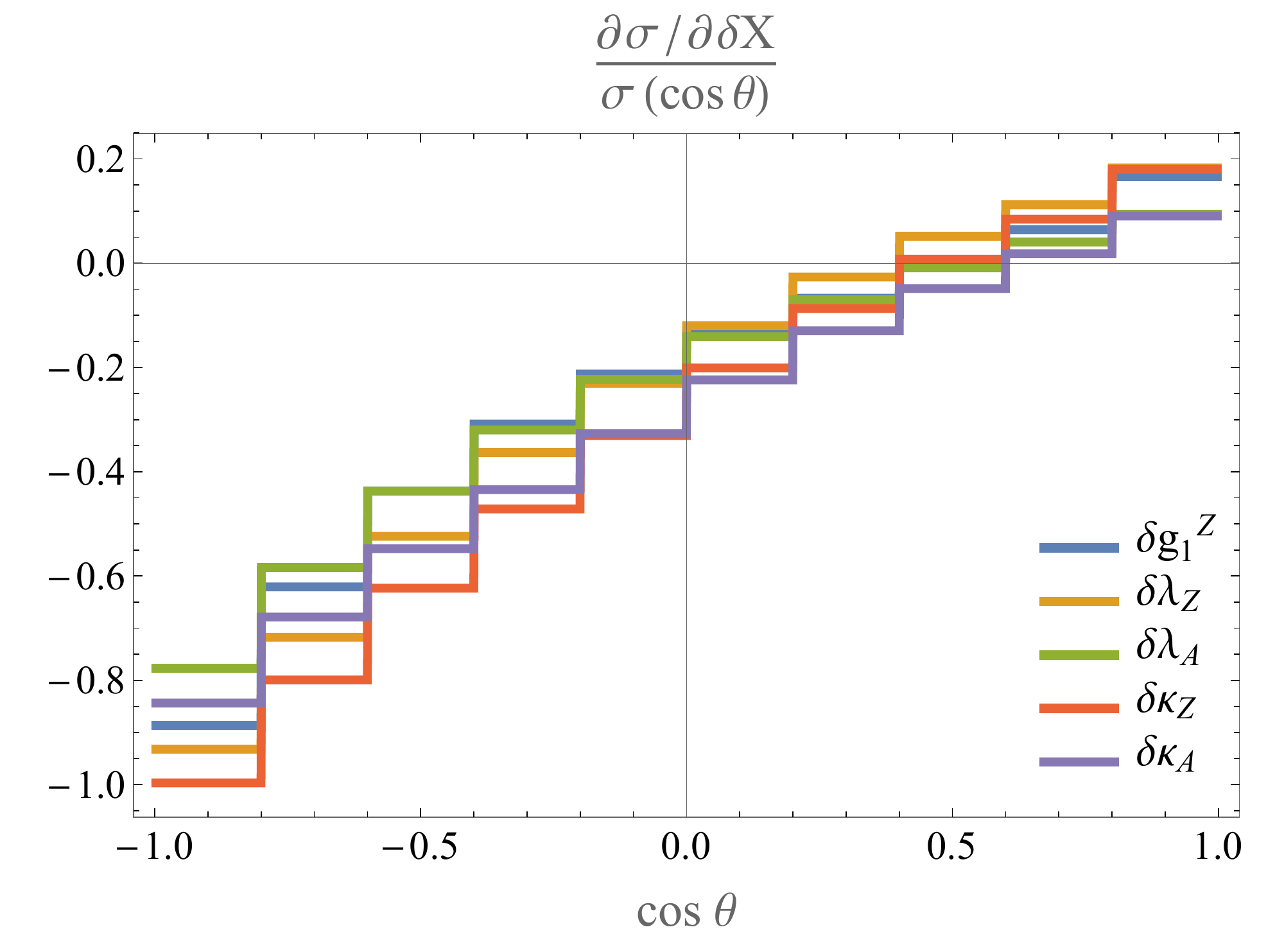}
	\caption{\label{fig:angular1}}
	\end{subfigure}
	\begin{subfigure}{0.49 \textwidth}
	\includegraphics[width=1 \textwidth]{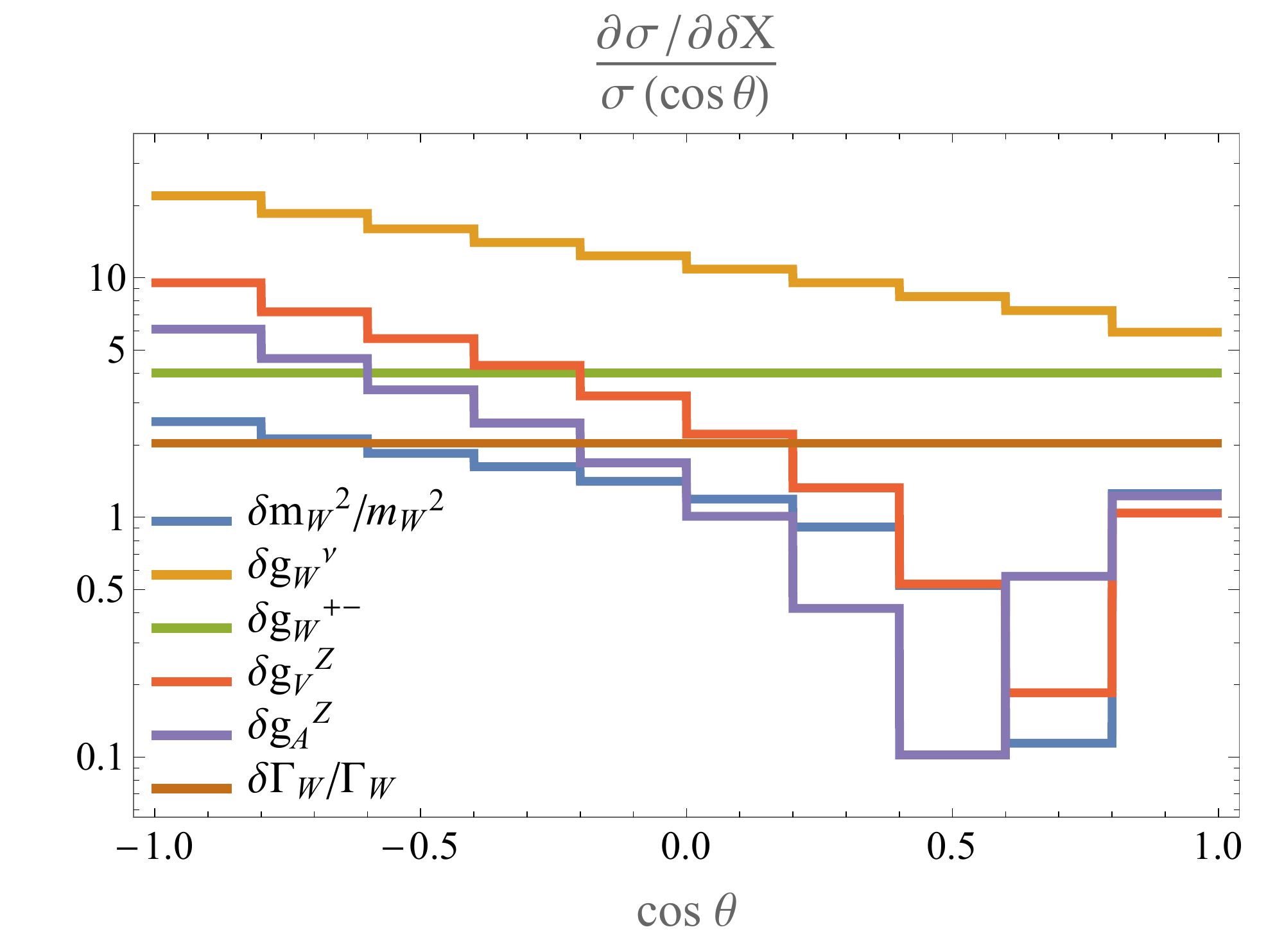}
	\caption{\label{fig:angular2}}
	\end{subfigure}
	\caption{The $s$-dependence of $|\partial \sigma_{CC03}/\partial \delta X|$ for (a) the shifts of the $\delta X$ TGC parameters and (b) the remaining $\delta X$ SMEFT shifts contributing to the $\sigma_{CC03}$
	result in our approach. Fig (c) shows the dependence on the $\delta X$ TGC parameters in $d \sigma_{CC03}/d \cos \theta$ results (note the linear scale), while Fig (d) shows the remaining  $\delta X$ SMEFT shifts impact on $d \sigma_{CC03}/d \cos \theta$. Each shift is normalized by the average value of $d \sigma_{CC03}/d \cos \theta$ for each bin individually.
	 Note we do not plot the $\delta \Gamma_Z/\hat{\Gamma}_Z$ dependence, which is $\mathcal{O}(10^{-4})$, and note the different scales of the left and right plots. The structure in Fig. (a) is due to the effective sign change of the corresponding shift, and the log plot, not resonant behavior.}
	\label{fig:sigma_shift_s_dependence_plot}
\end{figure}

 The total spin averaged cross section, for the process $e^+ e^- \rightarrow WW \rightarrow f_1 \bar{f}_2 f_3 \bar{f}_4$ is
\begin{align}
  \bar{\sigma}_{CC03}(s) &=  \int \frac{\sum |\mathcal{M}|^2}{8 s}  \frac{d s_{12} d s_{34}}{\left(2\pi\right)^2}\left[ \frac{\bar{\beta}_{12}}{8 \pi} \frac{d \cos \tilde{\theta}_{12}}{2}\frac{d \tilde{\phi}_{12}}{2 \pi}\right]\left[ \frac{\bar{\beta}_{34}}{8 \pi} \frac{d \cos \tilde{\theta}_{34}}{2}\frac{d \tilde{\phi}_{34}}{2 \pi}\right] \left[ \frac{\bar{\beta}}{8 \pi} \frac{d \cos {\theta}}{2}\frac{d {\phi}}{2 \pi}\right], \label{eq:cross_section_expression}
\end{align}
where
\begin{align}
 \sum |\mathcal{M}|^2 &= |\bar D^W(s_{12}) \bar D^W(s_{34})|^2 \sum_{\lambda_{12},\lambda_{12}'}  \sum_{\lambda_{34},\lambda_{34}'} \left( \mathcal{M}^{\lambda_{12}}_{W^+} \right)\left( \mathcal{M}^{\lambda_{12}'}_{W^+}\right)^*  \left( \mathcal{M}^{\lambda_{34}}_{W^-}\right)  \left(\mathcal{M}^{\lambda_{34}'}_{W^-}\right)^* \nonumber \\
 &\times  \sum \limits_{\lambda_{+}}  \sum \limits_{\lambda_{-}} \left( \mathcal{M}_{ee \rightarrow WW}^{\lambda_{12}\lambda_{34},\lambda_+,\lambda_-}\right)  \left( \mathcal{M}_{ee \rightarrow WW}^{\lambda_{12}' \lambda_{34}',\lambda_+,\lambda_-}\right)^*,
\end{align}
and we decomposed the the $8$ dimensional four-body phase space as a product of three two-body phase spaces. The angles in the rest frames of the decaying $W^\pm$ bosons are
defined with tilde superscripts. The $\M_{\nu/\gamma/Z}$ are reported in the Appendix. The phase space factors are
\bea
 \tilde{\beta} &=&\sqrt{1- \frac{2\left(s_{12} + s_{34}\right)}{s}+ \frac{\left(s_{12}-s_{34}\right)^2}{s^2} } , \hspace{0.25cm} \tilde{\beta}_{ij} = 1. \nonumber
\eea
and the phase space is given by
\bea
\tilde{\phi}_{12}, \tilde{\phi}_{34},\phi \in [0,2\pi], & \qquad &\cos \tilde{\theta}_{12}, \cos \tilde{\theta}_{34}, \cos\theta \in [-1,1], \notag \\
s_{34} \in [0,(\sqrt{s}-\sqrt{s_{12}})^2], & \qquad & s_{12} \in [0,s]. \notag
\eea
The effects of the SMEFT on the CC03 cross section computation are multi-fold, changing the absolute and relative normalizations of the diagrams, and shifting
$\bar{\Gamma}_W,\bar{m}_W$. When carrying out the integrations in Eqn.\eqref{eq:cross_section_expression}, the angular integrals can be done analytically for the total cross section. We used the Cuba Integration Library \citep{Hahn2005Cuba} for performing the numerical integrals when required to calculate the differential cross sections.
We show $\delta \sigma_{CC03} / \sigma_{CC03}$ as a function of $s$ due to each of these shifts in Figure \ref{fig:sigma_shift_s_dependence_plot}.
Note that some of the $\delta X$ shown on the right hand plots are set to zero in some previous analyses, despite the large numerical enhancement of these shifts.
It is unjustified and unnecessary to set all of the corrections in Figure \ref{fig:sigma_shift_s_dependence_plot} b) -- which contain flat directions in some operator bases -- to vanish when incorporating this data.
We use the predictions for the $\delta X$ parameters in Table \ref{cross} for the global fit. Note that these numerical results
can be mapped to any basis of operators, including the Warsaw basis using the formulii in the Appendix.
Note that these shift variables are correlated theoretically, considering the gauge invariance of the underlying operators.
\begin{table}[t] 
\centering 
\begin{tabular}{ccccccccccccc} 
\toprule
 $\sqrt{s}$ &  $\frac{\delta m_W^2}{m_W^2}$ & $\frac{\delta\Gamma_W}{\Gamma_W}$ & $\delta g_W^\nu$ & $\delta g_W^\pm$ & $\delta g_V^Z$ &$\delta g_A^Z$ & $\delta g_1^Z$ & $ \delta \kappa_\gamma $  & $ \delta \kappa_Z $ & $ \delta \lambda_\gamma $ & $ \delta \lambda_Z $ \\ \midrule
188.6 & $2.6$  & $-17.$  & $72.$  & $34.$  & $5.3$  & $0.3$  & $-0.08$  & $-0.50$  & $-0.19$  & $-0.29$  & $0.026$  \\191.6 & $1.6$  & $-17.$  & $73.$  & $34.$  & $5.8$  & $0.4$  & $-0.10$  & $-0.56$  & $-0.22$  & $-0.32$  & $0.018$   \\195.5 & $0.26$  & $-17.$  & $74.$  & $34.$  & $6.5$  & $0.6$  & $-0.12$  & $-0.64$  & $-0.27$  & $-0.36$  & $0.005$   \\199.5 & $-0.54$  & $-17.$  & $75.$  & $34.$  & $7.1$  & $0.8$  & $-0.15$  & $-0.71$  & $-0.31$  & $-0.40$  & $-0.009$  \\201.6 & $-0.97$  & $-17.$  & $75.$  & $34.$  & $7.4$  & $0.9$  & $-0.16$  & $-0.75$  & $-0.33$  & $-0.42$  & $-0.017$   \\204.8 & $-1.4$  & $-17.$  & $75.$  & $34.$  & $7.8$  & $1.0$  & $-0.18$  & $-0.80$  & $-0.37$  & $-0.44$  & $-0.029$   \\206.5 & $-1.8$  & $-17.$  & $76.$  & $34.$  & $8.0$  & $1.1$  & $-0.19$  & $-0.83$  & $-0.39$  & $-0.46$  & $-0.036$   \\208. & $-2.0$  & $-17.$  & $76.$  & $34.$  & $8.2$  & $1.2$  & $-0.20$  & $-0.85$  & $-0.40$  & $-0.47$  & $-0.042$   \\\bottomrule 
\end{tabular} 
\caption{Total cross section contributions due to $\delta X$ in $pb$. The results are normalized for semileptonic final states.  To normalize to fully leptonic decays the results are divided by $4.04$. For only quark final states, the results are multiplied by $1.01$\label{crosssections}. $\delta g_W^\nu = \delta g_W^\ell$ denotes the $W^\pm$coupling to $e^+e^-$ in the $t$-channel diagrams, whereas $g_W^\pm=g_W^q$ or $g_W^\ell$ denotes $W$-coupling to final state fermions, and depends on which final state is considered. $\delta \Gamma_Z/\hat{\Gamma}_Z$ contributions are $\mathcal{O}(10^{-4})$ pb and not shown in the table, although they are included in the fit for completeness.}\label{cross} 
\end{table}
\subsection{Angular Distributions}

The LEPII collaborations reported combined angular distributions for the CC03 diagrams, as well as  total cross section data.
To incorporate this data, the angular cut for the charged lepton identified in the decay of the $W^\pm$ is restricted
to be $20^{\circ}$ from the beam line \cite{Schael:2013ita}. Explicitly, the angle $\theta_{\ell}$ is the angle between the outgoing charged lepton and the beam line.
We incorporate this cut via the constraint $-0.94 < \cos \theta_\ell < 0.94$ where
\bea \label{cut_eqn}
\cos \theta_\ell = \frac{- \sin \tilde{\theta}_{12} \, \cos \tilde{\phi}_{12} \, \sin \theta+ \gamma_{12}(\beta_{12}+ \cos \tilde{\theta}_{12}) \cos \theta}{\gamma_{12}(\beta_{12} \cos \tilde{\theta}_{12}+ 1)}.
\eea
Here $\gamma_{12} = (s+ s_{12}- s_{34})/2 \sqrt{s \, s_{12}}$.\footnote{We neglect the numerically suppressed correction due to this angular distribution cut in redefining parameters appearing in Eqn.~\ref{cut_eqn} in the SMEFT. Such shifts are lower dimensional in the phase space (proportional to $\delta$ functions) and  our theoretical error is sufficient to account for this neglected correction.}
In order to avoid overfitting when a correlation matrix is unknown, we restrict the angular data that we incorporate in the fit to the bins
${\rm B_1}=\left[-1,-0.8 \right]$, ${\rm B_2}= \left[-0.4,-0.2 \right]$, ${\rm B_3}=\left[0.4,0.6 \right]$, ${\rm B_4}=\left[0.8,1\right]$ for $\sqrt{s} =\{182.66,205.92\}$ ${\rm GeV}$.
This approach is consistent with our treatment of Bhabba scattering angular data in Ref.~\cite{Berthier:2015gja}.
We use the predictions in Table \ref{angular} for the global fit, also shown in Figure \ref{fig:angular1} and \ref{fig:angular2}.
\begin{table}[t] 
\centering 
\begin{tabular}{ccccccccccccc}
 \multicolumn{12}{c}{$\sqrt{s}=182.66$ GeV} \vspace{1mm}\\ \toprule Bin &  $\frac{\delta m_W^2}{m_W^2}$ & $\frac{\delta\Gamma_W}{\Gamma_W}$ & $\delta g_W^\nu$ & $\delta g_W^\pm$ & $\delta g_V^Z$ &$\delta g_A^Z$ & $\delta g_1^Z$ & $ \delta \kappa_\gamma $  & $ \delta \kappa_Z $ & $ \delta \lambda_\gamma $ & $ \delta \lambda_Z $ \\ \midrule
$B_{1}$  & $-1.6$  & $-1.5$  & $12.$  & $2.9$  & $4.1$  & $3.0$  & $-0.44$  & $-0.34$  & $-0.47$  & $-0.32$  & $-0.45$   \\$B_{2}$  & $-1.5$  & $-2.8$  & $16.$  & $5.5$  & $3.5$  & $2.2$  & $-0.30$  & $-0.32$  & $-0.39$  & $-0.26$  & $-0.34$   \\$B_{3}$  & $0.16$  & $-5.3$  & $22.$  & $10.$  & $1.5$  & $0.2$  & $-0.04$  & $-0.14$  & $-0.06$  & $-0.06$  & $0.026$   \\$B_{4}$  & $18.$  & $-14.$  & $39.$  & $27.$  & $-7.7$  & $-8.8$  & $1.2$  & $0.62$  & $1.3$  & $0.63$  & $1.3$   \\\bottomrule 
 \\ 
 \multicolumn{12}{c}{$\sqrt{s}=205.92$ GeV} \vspace{1mm}\\ \toprule Bin  & $\frac{\delta m_W^2}{m_W^2}$ & $\frac{\delta\Gamma_W}{\Gamma_W}$ & $\delta g_W^\nu$ & $\delta g_W^\pm$ & $\delta g_V^Z$ &$\delta g_A^Z$ & $\delta g_1^Z$ & $ \delta \kappa_\gamma $  & $ \delta \kappa_Z $ & $ \delta \lambda_\gamma $ & $ \delta \lambda_Z $ \\ \midrule
$B_{1}$  & $-1.1$  & $-0.9$  & $11.$  & $1.8$  & $4.9$  & $3.0$  & $-0.44$  & $-0.44$  & $-0.50$  & $-0.40$  & $-0.46$   \\$B_{2}$  & $-1.7$  & $-2.1$  & $15.$  & $4.1$  & $5.0$  & $2.8$  & $-0.34$  & $-0.53$  & $-0.55$  & $-0.37$  & $-0.41$   \\$B_{3}$  & $-2.3$  & $-4.6$  & $22.$  & $9.0$  & $3.5$  & $1.2$  & $-0.19$  & $-0.35$  & $-0.25$  & $-0.19$  & $-0.086$  \\$B_{4}$  & $10.$  & $-20.$  & $59.$  & $39.$  & $-9.6$  & $-11.0$  & $1.5$  & $0.86$  & $1.7$  & $0.90$  & $1.7$   \\\bottomrule 
\end{tabular} 
\caption{Angular bin cross section contributions due to $\delta X$ in $pb$. Again, the results are normalized for semi-leptonic final states.\label{angular} $\delta \Gamma_Z/\hat{\Gamma}_Z$ contributions are $\mathcal{O}(10^{-4})$ pb and not shown in the table, although they are included in the fit for completeness.} \label{angular} 
\end{table}

\section{Global fit in the SMEFT} \label{sec:global_fit}

Using the results reported in the previous section, and the data reported by LEPII in Refs.\cite{Achard:2004zw,Abbiendi:2007rs,Heister:2004wr,Collaboration2004Measurementa}
we have extended the global fit developed by two of us to include charged current data. Our fit procedure is to
consider a set of observables  $\Omega_{O} = \left\{O_i \right\}_{i \in \llbracket 1,n\rrbracket}$, and denote by $O_i$, $\bar{O}_i$, $\hat{O}_i$ the SM prediction, SMEFT prediction to first order in the $C^{(6)}$, and experimental value of the observable $O_i$ respectively. Assuming the measured value $\hat{O}_i$ to be a gaussian variable centred about the predicted value $\bar{O}_i$, and introducing the n dimensional vectors $\hat{O} = (\hat{O}_1,...,\hat{O}_n)$ and $\bar{O}=(\bar{O}_1,...,\bar{O}_n)$, we define the likelihood function
\bea
L(C) = \frac{1}{\sqrt{(2 \pi)^n |V|}} \text{exp} \left(-\frac{1}{2} \left( \hat{O} - \bar{O}\right)^T V^{-1} \left( \hat{O} - \bar{O}\right)\right),
\eea
where $V$ is the covariance matrix with determinant $|V|$ and elements
\bea
V_{ij} = \Delta^{exp}_i \rho^{exp}_{ij} \Delta^{exp}_j + \Delta^{th}_i \rho^{th}_{ij} \Delta^{th}_j.
\eea
$\rho^{exp}$/$\rho^{th}$ are the experimental/theoretical correlation matrices and $\Delta^{exp}$/$\Delta^{th}$ the experimental/theoretical error of the observable $O_i$.
The theoretical error $\Delta_i^{th}$ for an observable $O_i$ is defined as
\bea
\Delta^{th}_i = \sqrt{\Delta_{i,SM}^2 + \left(\Delta_{i,SMEFT} \times O_i\right)^2},
\eea
where $\Delta_{i,SM}$, $\Delta_{i,SMEFT}$ correspond to the absolute SM theoretical, and the multiplicative SMEFT theory error for the observable $O_i$.
We use the $\chi^2$ variable defined as $\chi^2 = - 2 \text{Log}[L(C)]$ and the new variable $\Delta \chi^2 \left(C_{true}\right) = \chi^2 \left(C_{true}\right) - \chi^2_{\min} $ to derive bounds on each individual Wilson coefficient.
When profiling parameters we follow the procedure described in Ref.\cite{Berthier:2015gja}.

\subsection{$m_W$ data}
We have also modified our fit procedure to utilise the Tevatron measured central value of $m_W$, replacing the previously used global average value for the following reasons:
\begin{enumerate}
\item{It was found that SMEFT theoretical errors impact Tevatron measurements of transverse variables in a numerically suppressed fashion in Ref.~\cite{todaypaper}. Such measurements are also sensitive to less effective SMEFT parameters
at leading order in the power counting compared to LEPII $m_W$ extractions.}
\item{LEPII measurements of $m_W$ are extracted from data that is sensitive to TGC parameters. We reserve
two to four fermion scattering data through charged currents to lift the flat directions in the global data set in a consistent fashion.
Correlation matrices are unavailable (to our knowledge) to utilise charged current LEPII data to fit for $m_W$, while simultaneously using the same
data set to fit for TGC parameters in the SMEFT.}
\end{enumerate}
Due to these results, the dominant theoretical uncertainty due to the SMEFT is the limited degree of development of the calculation of
shifts in the $W^\pm$ mass parameter (i.e. neglecting dimension eight operators and one loop corrections in the SMEFT).
For this reason we retain our approach developed in  Refs.~\cite{Berthier:2015oma,Berthier:2015gja} for assigning a theoretical error
without any further increase in SMEFT error due to the impact of the EFT on the extracted value of the $W^\pm$ itself \cite{todaypaper}.

\subsection{$\delta X$ Constraints and Correlation matrix}
A set of parameters present in the mass eigenstate SMEFT Lagrangian, labelled $\delta X$, that are algebraically linearly independent
are given by
\bea
\delta X = \{ \delta g_V^{\nu}, \delta g_V^\ell,\delta g_A^\ell,\delta g_V^u,\delta g_A^u,\delta g_V^d,\delta g_A^d, \delta g_1^Z,  \delta \kappa_\gamma \, \delta \lambda_\gamma, \frac{C_{l l}}{\sqrt{2} \hat{G}_F}\},
\eea
where we include the highly correlated $C_{l l}$ in this set of variables. Note that the $\delta X$  variables are defined to be dimensionless. These parameters are added to the relatively uncorrelated dimensionless four fermion operator
Wilson coefficients
\bea
\frac{1}{{\sqrt{2} \hat{G}_F}} \, \{C_{ee}, C_{eu}, C_{ed}, C_{l e}, C_{l u}, C_{l d}, C_{l q}^{(1)}, C_{l q}^{(3)},C_{qe}\}
\eea
when we report fit results for these expressions.
We find the results given in Table \ref{ConstraintsDeltas} that the $\delta X$ parameters are highly constrained as a numerical output of our fit procedure, and
these constraints are only mildly relaxed by a consistent inclusion of $\Delta_{SMEFT}$ theory errors. This is the result of our approach to assign theory errors as percentage corrections
on the most precise prediction of a SM value for an observable.
\begin{table}
\centering
\tabcolsep 4pt
\footnotesize
\begin{tabular}{|c|c|c|c|c|c|}
\hline
$\delta X^i$ & $(1 \sigma,0)$ & $(1 \sigma, 0.1 \%)$ & $(1 \sigma, 0.3 \%)$   & $(1 \sigma, 0.5 \%)$ & $(1 \sigma, 1 \%)$  \\ 
\hline \hline
$\delta g_V^{\nu}$ &$ (-3.2\pm 2.7) 10^{-4}$&	$(-3.8 \pm 3.7) 10^{-4}$&	$(-5.6\pm 6.8)10^{-4}$&$(-6.9 \pm 8.9) 10^{-4}$&$(-7.6 \pm 13) 10^{-4}$\\
$\delta g_V^{\ell}$ & $ (-3.0 \pm 2.8) 10^{-4}$&	$(-2.9 \pm 2.8)10^{-4} $&	$(-2.9 \pm 2.9)10^{-4}$&$ (-2.8 \pm 3.0) 10^{-3}$& $ (-2.6 \pm 3.2) 10^{-4}$\\
$\delta g_A^{\ell}$ &$ (-0.57 \pm 1.2) 10^{-4}$&	$(-0.50 \pm 0.19) 10^{-3}$&$ (0.0 \pm 4.1)10^{-4}$&$ (0.55 \pm 5.9) 10^{-4}$&$( 0.18 \pm 0.90)10^{-3}$\\
$\delta g_V^{u}$ &$(-3.7 \pm 2.8) 10^{-3}$ &	$(1.9 \pm 2.8) 10^{-3} $&		$ (-3.8 \pm 2.8)10^{-3}$&$(-3.9\pm 2.8) 10^{-3}$&$(-4.0 \pm 2.9) 10^{-3}$\\
$\delta g_A^u$ & $ (1.8 \pm 1.2) 10^{-3}$&	$(1.9 \pm 1.2) 10^{-3}$&	 $(1.9 \pm 1.3)10^{-3}$& $(1.9 \pm 1.4)10^{-3}$&$ (1.8 \pm 1.6) 10^{-3}$\\
$\delta g_V^d$ &$(1.0 \pm 0.37) 10^{-2}$&	$ (1.0 \pm 0.37) 10^{-2}$&	$ (1.0 \pm 0.38)10^{-2}$& $(1.0 \pm 0.39) 10^{-2}$&$ (1.0 \pm 0.42)10^{-2}$\\
$\delta g_A^d$ &  $(-7.4 \pm 2.7)10^{-3} $&	$(-7.4 \pm 2.7)10^{-3}$&	$(-7.4 \pm 2.8)10^{-3}$&$(-7.5 \pm 2.9) 10^{-3}$&$(-7.4 \pm 3.2) 10^{-3}$\\
$\delta g_Z^1$ & $-0.98 \pm 0.57$ & 	$-1.0 \pm 0.57$&		$-1.0 \pm 0.58$& $-1.0 \pm 0.58$& $-1.0 \pm 0.59$\\
$\delta \kappa_{\gamma}$ & $0.034\pm 0.12$& $(3.4 \pm 12)10^{-2}$& 	$(2.4 \pm 13)10^{-2}$&	$(1.4 \pm 14)10^{-2}$&$(0.53 \pm 15)10^{-2}$\\
$\delta \lambda_{\gamma}$ &$1.1 \pm 0.67$&	 $1.1 \pm 0.67$& 		$1.2 \pm 0.67$& $1.2 \pm 0.68$&$1.2 \pm 0.69$\\
$\frac{C_{l \, l}}{\sqrt{2} \hat{G}_F}$ &$(-1.1 \pm 1.2) 10^{-3}$&	 $(-0.75 \pm 1.5) 10^{-3}$& $(-0.53 \pm 1.7) 10^{-3}$& $(-0.48 \pm 1.8) 10^{-3}$&$(-0.41 \pm 1.9) 10^{-3}$\\
\hline\end{tabular}
\caption{$1 \sigma$ bounds on the common shift parameters ($\delta X$) appearing in the mass eigenstate effective Lagrangian. These results neglect the effect of the
theoretical correlation matrix discussed in the text. The columns are labeled with the
$\Delta_{SMEFT}$ theory error. These bounds should be interpreted with caution, see the text for further discussion.\label{ConstraintsDeltas}}
\end{table}

\subsubsection{Understanding $\delta X$ Constraints}\label{understanding}
The constraints in Table \ref{ConstraintsDeltas} on the $\delta X$ parameters are unusually strong. This point has been noted in the literature
previously \cite{Falkowski:2014tna} and forms the basis of the assertion in Ref.~\cite{Falkowski:2001958} that possible shifts in leptonic couplings
can be set to zero for LHC analyses. We agree that this numerical behavior exists when fits are done with tree level interference with the SM predictions, and we also agree that setting
the leptonic couplings of the $Z$ to vanish does not dramatically change the numerical values found for the TGC shift parameters to the Z in procedures such as this\footnote{Despite this being a formally ill defined step in a consistent treatment of the SMEFT.} similar to the behavior reported in Ref.~\cite{Falkowski:2014tna}. We
interpret this numerical behavior very differently than in Ref.~\cite{Falkowski:2014tna,Falkowski:2001958}. The reasons that we reach different conclusions are as follows.

\begin{figure}[h]
		\includegraphics[width=0.5\textwidth]{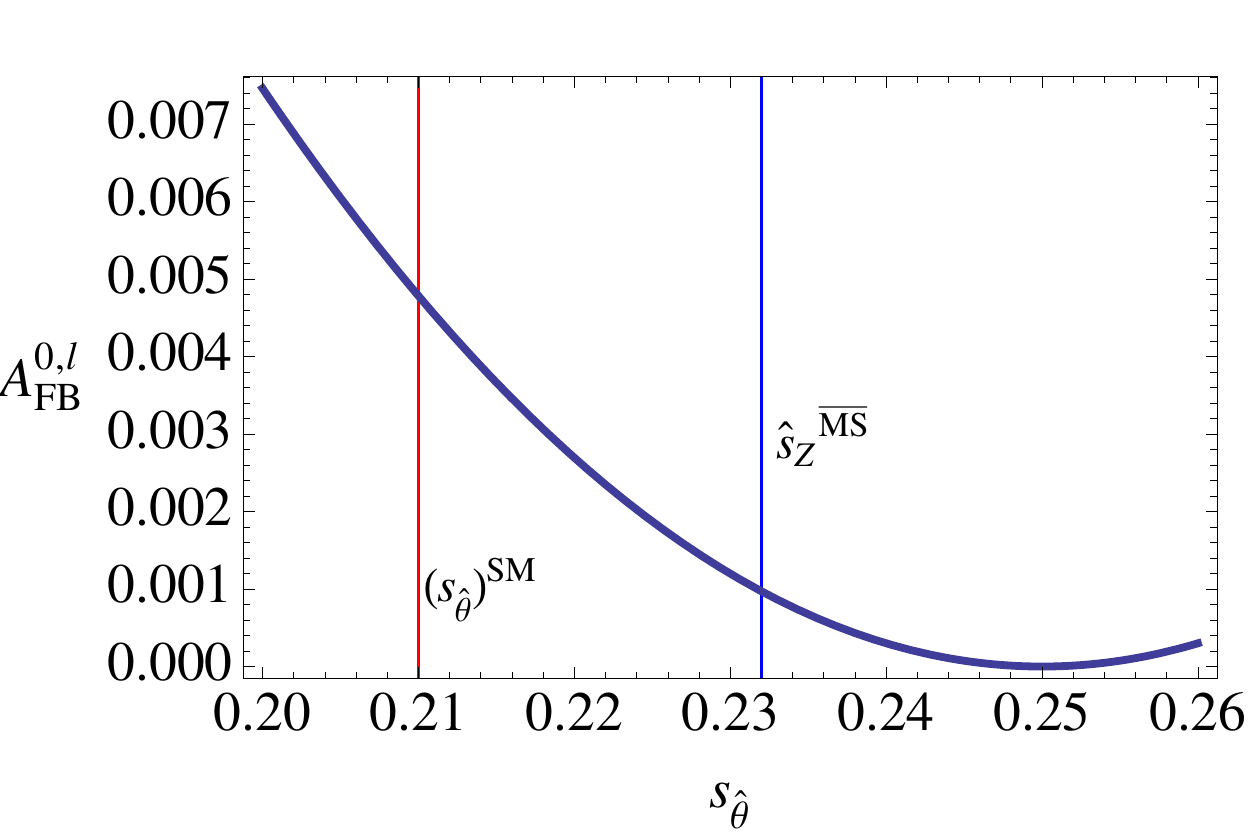}
		\caption{Dependence of the predicted value for $A_{FB}^{0,\ell}$ on $s_{\hat{\theta}}$ used in this fit, and $(\hat{s}_0)^{MS}$ which includes perturbative corrections further suppressing the SM prediction.
		Naively one does not expect a one loop perturbative correction to change the central value for a tree level observable by a factor of five. This accidental numerical
		suppression is due to $g_V^\ell$.}\label{fig:tuning}
\end{figure}

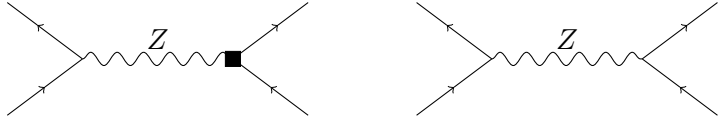
\begin{figure}[t]
\hspace{0.6cm}
\begin{tikzpicture}
\draw[decorate,decoration=snake] (-0.5,0) -- (1.5,0) ;
\filldraw (1.4,-0.1) rectangle (1.6,0.1);
\draw [->](1.5,0) -- (2.1,0.45);
\draw (2.1,0.45) -- (2.5,0.75);
\draw (1.5,0) -- (2,-0.375);
\draw [<-](2,-0.375) -- (2.5,-0.75);

\draw [->](-0.5,0) -- (-1.1,0.45);
\draw (-1.1,0.45) -- (-1.5,0.75);
\draw (-0.5,0) -- (-1,-0.375);
\draw [<-](-1,-0.375) -- (-1.5,-0.75);

\node [above] at (0.5,0) {$Z$};

\end{tikzpicture}\hspace{1.3cm}
\begin{tikzpicture}
\draw[decorate,decoration=snake] (-0.5,0) -- (1.5,0) ;
\draw [->](1.5,0) -- (2.1,0.45);
\draw (2.1,0.45) -- (2.5,0.75);
\draw (1.5,0) -- (2,-0.375);
\draw [<-](2,-0.375) -- (2.5,-0.75);

\draw [->](-0.5,0) -- (-1.1,0.45);
\draw (-1.1,0.45) -- (-1.5,0.75);
\draw (-0.5,0) -- (-1,-0.375);
\draw [<-](-1,-0.375) -- (-1.5,-0.75);

\node [above] at (0.5,0) {$Z$};
\end{tikzpicture}
\caption{\label{fig:1}
Diagrams contributing to near $Z$ pole $2 \rightarrow 2$ scattering in the SMEFT. The black box indicates the insertion of $\lsix$
leading to the effective parameter $\delta g_V^\ell$.}
\end{figure}

The origin of the numerically enhanced bounds on the possible deviations in the leptonic couplings, is in part due to
the accidental numerical suppression of $g_V^\ell$ in the SM. Recall that $(g_V^f)^{SM} = T^f_3/2 - Q_f \, \sin^2 \theta_W$. For the leptons, the numerical accident that at tree level $(g_V^\ell)^{SM} = -0.038$
suppresses the predictions in the SM for forward and backward asymmetries produced from $e^+ e^-$ collisions, and particularly the leptonic forward backward asymmetry $A_{FB}^{0,\ell}$, as is well known \cite{Altarelli:116932}.
This suppression is more pronounced when including radiative corrections in the SM predictions. For example, the PDG value for the radiatively corrected
Weinberg angle in the $\rm \overline{MS}$ scheme\footnote{Using PDG notation.}: $(\hat{s}_Z)^2 = 0.232$,  leads to $(g_V^\ell)^{\overline{MS}} = -0.018$. The effect of this suppression on observables is illustrated in Fig.~\ref{fig:tuning} for the leptonic forward backward asymmetry  $A_{FB}^{0,\ell}$.
When calculating the interference of corrections due to the SMEFT and the SM,
we have used the tree level value of $s_{\hat{\theta}}$. This is due to the fact that no complete set of one loop results is known for the SMEFT for this observable.
When including such corrections, the numerical enhancement of the Weinberg angle present in the SM, and the universal corrections absorbed into the definition of $\hat{s}_Z$,
would lead to a numerical suppression for the interference of the SMEFT corrections with SM.
The bounds on mass eigenstate parameters will then be correspondingly relaxed when fits are performed including loop corrections.
The choice of redefining the Weinberg angle to absorb universal radiative corrections related to the input observables, as is done in the SM predictions,
introduces further numerical sensitivity. To make this scaling argument clearer, consider the shift of the matrix element derived from the interference of the diagrams shown in Fig. \ref{fig:1},
which scales as $| \mathcal{M}|^2 \propto 4 \, [(g_V^\ell)^{SM}]^3 \, \delta g_V^\ell$.
Define the $\rm \overline{MS}$ version of the on-shell scheme for $\hat{s}_Z$ to be $(g_V^\ell)^{\overline{MS}}$. The ratio of this parameter to the tree level value of the vectorial coupling
scales as $(g_V^\ell)^{\overline{MS}}/(g_V^\ell)^{SM} \sim 1/2$. This modifies the dependence of observables on $\delta g_V^\ell$ by an order of magnitude $\sim 2^{-3}$.
As the radiative corrections are absorbed into a redefined parameter $\hat{s}_Z$, three powers of this numerical enhancement are present.
This is a much larger effect than expected to occur naively due to perturbative corrections.

We assign theoretical errors for neglected corrections in the SMEFT in our global analysis \cite{Berthier:2015oma,Berthier:2015gja} to avoid
misleading numerical conclusions in tree level analyses. However, we choose to assign this theoretical error as a percentage of the loop corrected SM value of the observable.

In the case of SM predictions receiving such accidental numerical suppressions, this means the effect of the theory error is artificially suppressed, as is seen in Table \ref{ConstraintsDeltas}.
This explains the weak scaling behavior of the results with an increase in $\Delta_{SMEFT}$.
We have checked that when scaling the theoretical error to compensate for this numerical suppression, the constraints on the parameters weaken to
the percent level, as expected.

Naively interpreting the bounds of the $\delta X$ parameters reported in Table \ref{ConstraintsDeltas} is also challenged by theoretical correlations
of the mass eigenstate parameters. When fitting $\delta X$ to experimental data,
the effect of $n$ copies of the symmetry $(SU(3) \times SU(2) \times U(1))^n$ present due to the
gauge invariant form of each ($n$) operators correlates the $\delta X$ \cite{Passarino:2138031}.\footnote{Linear independence of parameters is not equivalent to a lack of correlation of parameters.
It is also true that the corresponding covariance
matrix of this form requires a variance to be assumed on the Wilson coefficients, and there is no well defined metric on theory space.
We assume that the variance is fixed by the power counting size of the operator corrections, as this dictates the size of the
corrections expected due to the parameters in the SMEFT.} A theoretical correlation matrix of the $\delta X$ parameters, defined through the relation of the  $\delta X$ to the Warsaw basis, is given by
\bea\label{DeltaXcorr}
\rho^{th}_{\delta X} \simeq \left(\begin{array}{cccccccccccc}
1 & 0.053 & 0.39  &0.16 & 0.13  & -0.16 & -0.13  &0.22 & 0 & 0 & 0.43 \\
- & 1 & 0.14  & -0.76  & -0.29  & 0.62  & 0.29  & -0.79 & 0.76 & 0 & -0.33\\
- &- &1 & -0.12 & -0.099 & 0.12  & 0.099  & -0.16  & 0  & 0  & -0.33 \\
- &- &- &1 & 0.51  & -0.63  & -0.37  & 0.83 & -0.55  & 0 & 0.29\\
- &- &- &- &1 & -0.36  & -0.30  & -0.50  & 0  & 0 & 0.24 \\
- &- &- &- &- &1 & 0.55 & -0.75  & 0.36 & 0 & -0.29\\
- &- &- &- &- &- &1 & - 0.50  & 0 & 0  & -0.24  \\
- &- &- &- &- &- &- &1 & -0.41 & 0 & 0.40  \\
- &- &- &- &- &- &- &- & 1 & 0 & 0 \\
- &- &- &- &- &- &- &- & - &1 & 0 \\
- &- &- &- &- &- &- &- & - &0 & 1 \\
\end{array}\right). \nonumber
\eea
This theoretical correlation matrix is obtained from the covariance matrix which is determined using the bilinear property of covariance, multiplied by the variance of the Wilson coefficients.
We choose to take the variance to be dictated by a common power counting size $\sim  C_i^2 /\Lambda^4$. This leads to the dimensionless correction of the
size $\sim  C_i^2 \bar{v}_T^4/\Lambda^4 \lesssim \mathcal{O}(10^{-4})$.

Using this correlation matrix one can directly fit the data in terms of the parameters $\delta X$. This requires
constructing a correlation matrix for the theoretical predictions, using the bilinear nature of covariance as a function of the
$\delta X$ dependence. A more straightforward procedure is to first fit to
the linearly independent Wilson coefficients\footnote{Treated as uncorrelated as the fit spans all possible UV completions consistent with our assumptions.} and then to translate the fit results to the $\delta X$ fit space, by adding
the theoretical correlation matrix into the translation, as a last step. Using this approach and including a correlation matrix of this form or not,
is essentially related to an assumption on the space of possible UV models being aligned with the
$\mathcal{L}_6$ basis used to determine the $\delta X$. By construction, the SMEFT is designed to capture UV theories that generate combinations of gauge invariant $\rm SU(3) \times SU(2) \times U(1)$ operators,
not the mass eigenstate parameters $\delta X$, which justifies this approach. We have performed this analysis. Comparing results when the theory correlation matrix is included relaxes the $1 \sigma$ error for the $\delta X$ parameters
to approximately $\sqrt{C_i^2 \bar{v}_T^4/\Lambda^4}$, due to the assumed variance.

This approach to defining the theoretical covariance matrix is not unique, and does introduce dependence on the $\rm SU(3) \times SU(2) \times U(1)$ operator basis used to fit the
data.
However, again these theoretical errors are larger than quoted in Table \ref{ConstraintsDeltas}, as the numerical suppression of the SM prediction of leptonic observables is avoided in this procedure.
Theoretical errors for the SMEFT are defined by the envelope of the errors found performing different well motivated estimates of neglected theoretical effects \cite{Passarino:2138031}.
This result supports the view that the errors on the bounds on the $\delta X$ are underestimated in Table \ref{ConstraintsDeltas}.
Also, other parameters in the SMEFT will be introduced into the global analysis  at one loop that do not contribute at  tree level,
arguing against bounds that naively rise above the power counting size of the operators by orders of magnitude \cite{Berthier:2015oma,Berthier:2015gja,Passarino:2138031}.
For all of these reasons, the strong constraints on the leptonic mass eigenstate parameters in Table \ref{ConstraintsDeltas} should be interpreted with caution.
We stress that we (approximately) agree with the numerical behavior reported in Ref.~\cite{Falkowski:2014tna} for a similar set of core shift parameters, despite the very significant differences in the analyses.
For the reasons detailed above, we consider the bounds in Table \ref{ConstraintsDeltas} to be overestimating the
degree of constraint on these parameters.\footnote{As the parameters are so highly correlated in LEP data theoretically and experimentally (through the total $Z$ width), the issue of the leptonic observables
numerical sensitivity also feeds into the $\delta X$ quark parameters.}
\subsection{Global Analysis Results on Wilson coefficients} \label{sub:results}
Our previous fit \cite{Berthier:2015gja} contained 19 different Wilson coefficients, contributing to the shifts of the 103 observables.
Only $17$ of the $19$ Wilson coefficients could be constrained due to a $2$ fold degeneracy in the fit with the data considered in this case. The two fold degeneracy is lifted when including the charged current production data from
Tables \ref{L3Data}, \ref{OPALData}, \ref{ALEPHData} and \ref{DiffCrossSectionData}, as has been mentioned in the literature \cite{Han:2004az}. Bounds on each of the 20 Wilson coefficients can now be derived after profiling over the others in a totally data driven fashion.
We define a dimensionless vector $C^G$, now pulling out the cut off scale from the Wilson coefficients explicitly, as
\begin{align}
(C^G)^T= \frac{\bar{v}_T^2}{\Lambda^2} \{C_{He}, C_{Hu}, C_{Hd}, C_{Hl}^{(1)}, C_{Hl}^{(3)}, C_{Hq}^{(1)} , C_{Hq}^{(3)}, C_{HWB}, C_{HD}, C_{ll}, \nn
 C_{ee}, C_{eu}, C_{ed}, C_{l e}, C_{l u}, C_{l d}, C_{l q}^{(1)}, C_{l q}^{(3)},C_{qe},C_W \}.\label{Cvector}
\end{align}
The global fit reported here now contains 177 observables with the inclusion of LEPII data. The $\Delta \chi^2$ obtained in the SM (considering $C^G_{true} = 0$) gives $\Delta \chi^2_{0\%} = 28$ for a chi square distribution with 20 degrees of freedom. This corresponds to a p-value of $0.12$, which indicates the expected very weak evidence against the SM.
\begin{table}
\centering
\tabcolsep 4pt
\begin{tabular}{|c|c|c|c|c|c|}
\hline
$C_i^G$ & $(1 \sigma,0)$ & $(1 \sigma, 0.1 \%)$ & $(1 \sigma, 0.3 \%)$   & $(1 \sigma, 0.5 \%)$ & $(1 \sigma, 1 \%)$  \\ 
\hline \hline
$\chi^2_{min}$ & 153 & 152 & 151 & 149 & 142\\
\hline \hline
$\tilde{C}_{He} $ & $44 \pm 24$& $44 \pm 24$&$44 \pm 24$& $44 \pm 24$& $44 \pm 25$\\
$\tilde{C}_{Hu}$ & $-28 \pm 16$& $-28 \pm 16$&$-28 \pm 16$& $-28 \pm 16$& $-28 \pm 17$\\
$\tilde{C}_{Hd}$ & $11 \pm 8.1$& $11 \pm 8.1$& $11 \pm 8.2$& $11 \pm 8.2$&$11 \pm 8.3$\\
$\tilde{C}_{Hl}^{(1)}$ &$22 \pm 12$& $22 \pm 12$& $22 \pm 12$& $22 \pm 12$&$22 \pm 12$\\
$\tilde{C}_{Hl}^{(3)}$ & $77 \pm 45$& $78 \pm 45$& $80 \pm 45$& $80 \pm 46$& $80 \pm 46$\\
$\tilde{C}_{Hq}^{(1)}$ &$-7.3 \pm 4.0$& $-7.4 \pm 4.0$&$-7.4 \pm 4.0$& $-7.4 \pm 4.1$& $-7.3 \pm 4.1$\\
$\tilde{C}_{Hq}^{(3)}$ &  $77 \pm 45$& $78 \pm 45$& $79 \pm 45$&$80 \pm 46$& $80 \pm 46$\\
$\tilde{C}_{HWB}$ & $1.8 \pm 6.2$& $1.8 \pm 6.3$& $1.2 \pm 6.7$& $0.73 \pm 7.1$& $0.27 \pm 8.0$\\
$\tilde{C}_{HD}$ & $-87 \pm 48$& $-88 \pm 48$& $-88 \pm 49$& $-88 \pm 49$&$-87 \pm 49$\\
$\tilde{C}_{ll}$ & $-0.11 \pm 0.12$& $-0.075 \pm 0.15$& $-0.053 \pm 0.17$& $-0.048 \pm 0.17$& $-0.041 \pm 0.19$\\
$\tilde{C}_{ee}$ & $-0.036 \pm 0.2$& $-0.036 \pm 0.20$&$-0.032 \pm 0.2$&$-0.024 \pm 0.21$& $-0.0066 \pm 0.24$\\
$\tilde{C}_{eu}$ & $-27 \pm 24$& $-26 \pm 24$& $-24 \pm 24$&$-22 \pm 25$& $-20 \pm 25$\\
$\tilde{C}_{ed}$ &$-26 \pm 30$& $-25 \pm 30$& $-24 \pm 31$& $-22 \pm 31$& $-21 \pm 31$\\
$\tilde{C}_{l e}$ &$-0.011 \pm 0.3$& $-0.014 \pm 0.3$& $-0.014 \pm 0.31$& $-0.0096 \pm 0.31$& $0.0036 \pm 0.32$\\
$\tilde{C}_{l u}$ &$-17 \pm 8.4$& $-17 \pm 8.4$& $-17 \pm 8.5$&$-17 \pm 8.5$& $-17 \pm 8.8$\\
$\tilde{C}_{l d}$ & $-33 \pm 16$& $-33 \pm 16$&$-32 \pm 16$&$-32 \pm 16$& $-32 \pm 17$\\
$\tilde{C}_{l q}^{(1)}$ & $-4.1 \pm 1.9$&$-3.5 \pm 2.4$& $-2.4 \pm 3.7$& $-1.7 \pm 4.8$& $-0.94 \pm 6.8$\\
$\tilde{C}_{l q}^{(3)}$ &$-0.52 \pm 0.21$& $-0.47 \pm 0.25$& $-0.39 \pm 0.31$&$-0.35 \pm 0.38$&$-0.25 \pm 0.57$\\
$\tilde{C}_{qe}$ & $-2 \pm 26$&$-2.4 \pm 26$& $-3.0 \pm 26$&$-3.5 \pm 26$& $-4.6 \pm 27$\\
$\tilde{C}_{W}$ &$114 \pm 68$&$115 \pm 68$&$117 \pm 68$& $118 \pm 68$& $118 \pm 70$\\
\hline\end{tabular}
\caption{MLE and their $1\sigma$ confidence region $\tilde{C} \pm \sigma$ for a SMEFT error of $\{ 0\%, 0.1\%, 0.3\%, 0.5 \%, 1\% \}$ where $\tilde{C} = 100  C_{MLE}$
and the error is also scaled by 100. \label{Constraints}}
\end{table}
We give the maximum likelihood estimators (MLE) for the entries in $C^G$ and the $1\sigma$ confidence region $C^G \pm \sigma$ for a SMEFT error of $\{ 0\%, 0.1\%, 0.3\%, 0.5 \%, 1\% \}$ in Table \ref{Constraints}.
The full $\chi^2$ can be reconstructed from these results and the Fisher information matrices, which the authors will supply upon request.

This result can be compared to the one given in Ref.~\cite{Berthier:2015gja}, where two auxiliary conditions were introduced  to break the two dimensional degeneracy of the fit. These auxiliary conditions were taken to be the two null space directions of the fit, and a constraint of $\sim \bar{v}_T^2/\Lambda^2$ was set on them by using a naive dimensional power counting.
The constraints on the Wilson coefficients of the four fermion operators barely change compared to Ref.~\cite{Berthier:2015gja}, as expected.  The one sigma region of Wilson coefficients involved in couplings and $W^\pm$ mass shifts were relaxed by a little over a factor $\sim 10$. This is understandable as the data we have added is roughly $10\%$ precise, which is less constrained by roughly a factor of ten less than the two auxiliary conditions added in Ref.~\cite{Berthier:2015gja}. The way the degeneracy is broken also differs as the charged current data weakly lifts the flat directions in the SMEFFT, and does not correspond exactly to the two null space vectors of the fit.
The issues discussed in Section \ref{understanding} are still present when interpreting bounds on the Wilson coefficients derived from LEP leptonic data. However,
as the constraints are relatively weaker, this issue is not dominant in interpreting the results.
\begin{figure}[t]
	\centering
	%\begin{subfigure}{0.45 \textwidth}
		\includegraphics[width=2.8in,height=2.8in]{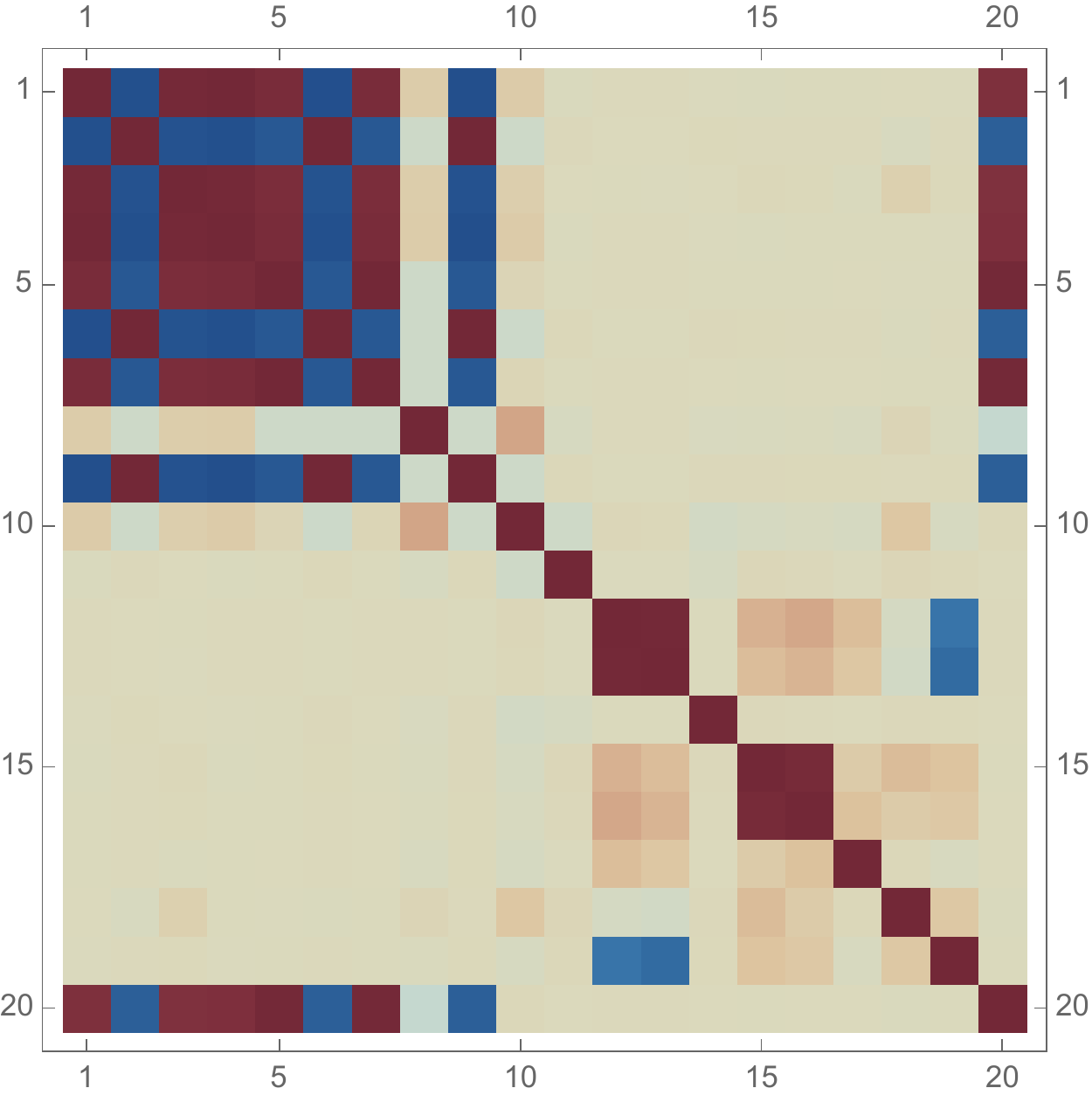}
		\includegraphics[width=0.5in,height=2.8in]{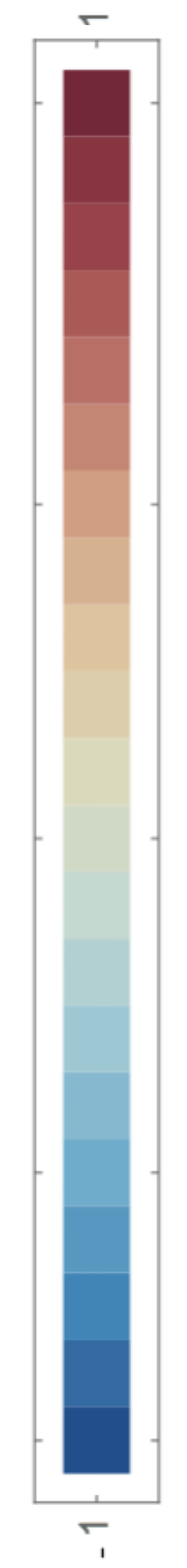}
	%\end{subfigure}
	%\begin{subfigure}{0.45 \textwidth}
		%\includegraphics[width=2.3in,height=2.3in]{PlotConstrainedWC2.pdf}
		%\caption{}
		%\label{fig:CoefficientConstrains}
	%\end{subfigure}
	\caption{Color map of the correlation matrix between the Wilson coefficients when there is no SMEFT error.The Wilson coefficients are ordered as in Eqn.\ref{Cvector}.}
	\label{fig:CorrMatrix}
\end{figure}
The highly correlated fit space of the Wilson coefficients dominates the interpretation of the results.
We illustrate this with a colour map of the correlation matrix between the bounds obtained on the Wilson coefficients in Fig.\ref{fig:CorrMatrix}, which
shows a clear block structure. There are almost no correlations between the Wilson coefficients of the 4 fermions operators (excepting $C_{ll}$), and Wilson coefficients involved in vector boson couplings and mass redefinitions: $C_{He}$, $C_{Hu}$, $C_{Hd}$, $C_{Hl}^{(1)}$, $C_{Hl}^{(3)}$, $C_{Hq}^{(1)}$, $C_{Hq}^{(3)}$, $C_{HWB}$ and $C_{HD}$. The latter are very correlated to each other, and are strongly correlated to $C_W$. This makes clear that a precise and consistent treatment of the charged current data is critical in developing model independent constraints.
Assumptions about UV physics that break the correlations shown in the Wilson coefficient constraint space significantly impact the degree of constraint. The different effects of marginalizing or profiling away parameters
also follow from the highly correlated fit space.
\clearpage
\begin{figure}
	\centering
	\begin{subfigure}{0.25 \textwidth}
		\includegraphics[width=1.4in,height=1.4in]{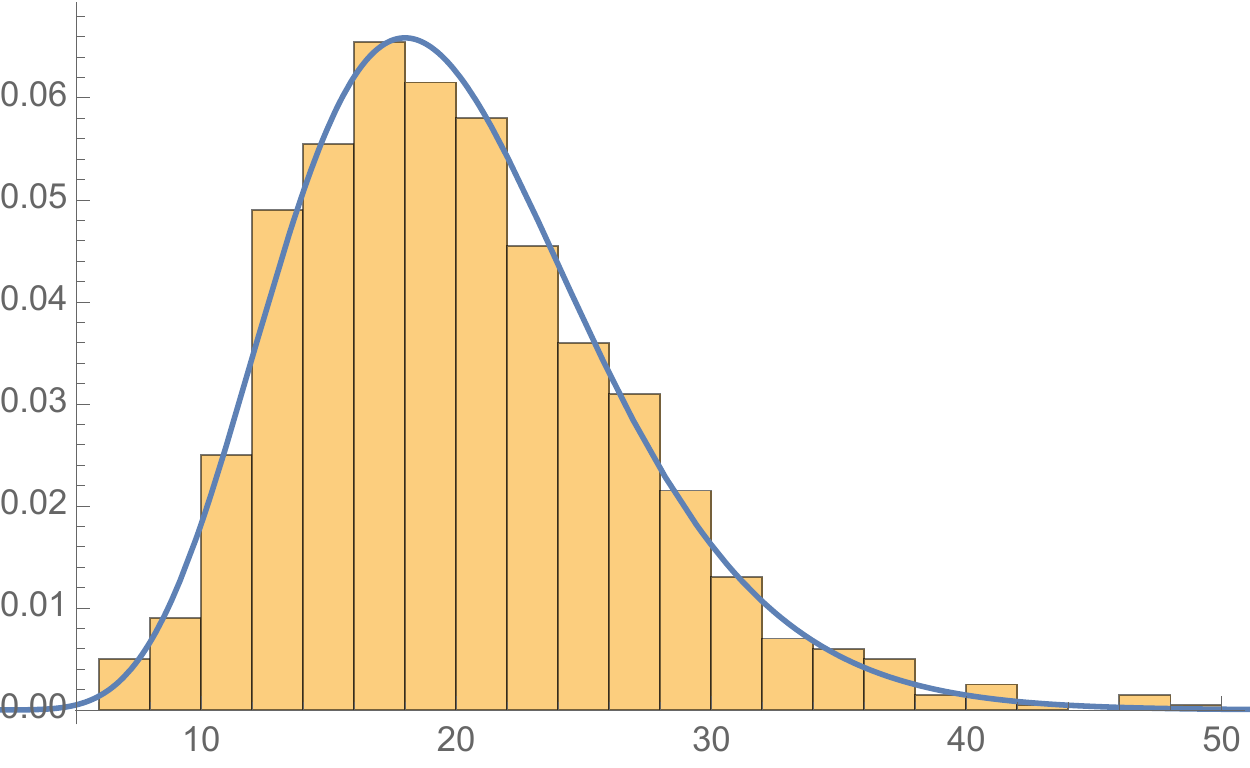}
	\end{subfigure}
	\begin{subfigure}{0.25 \textwidth}
		\includegraphics[width=1.4in,height=1.4in]{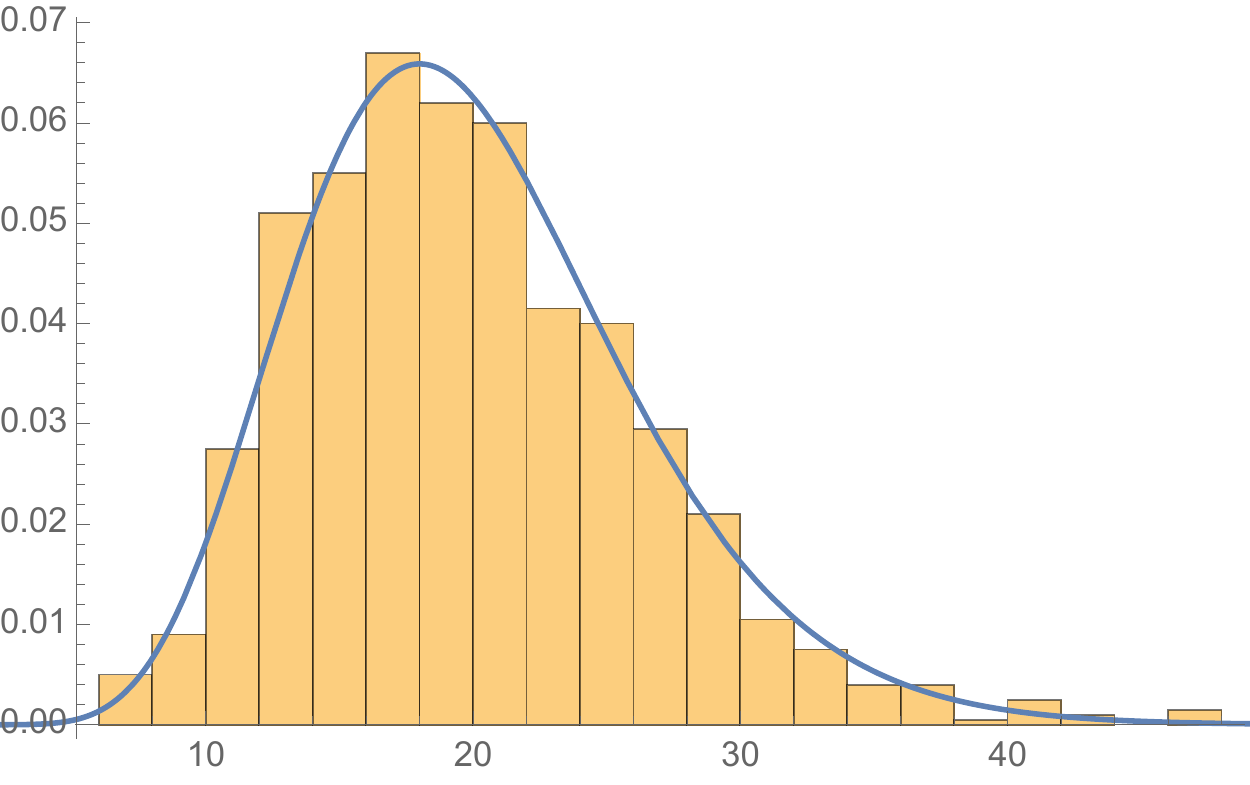}
	\end{subfigure}
	\begin{subfigure}{0.25 \textwidth}
		\includegraphics[width=1.4in,height=1.4in]{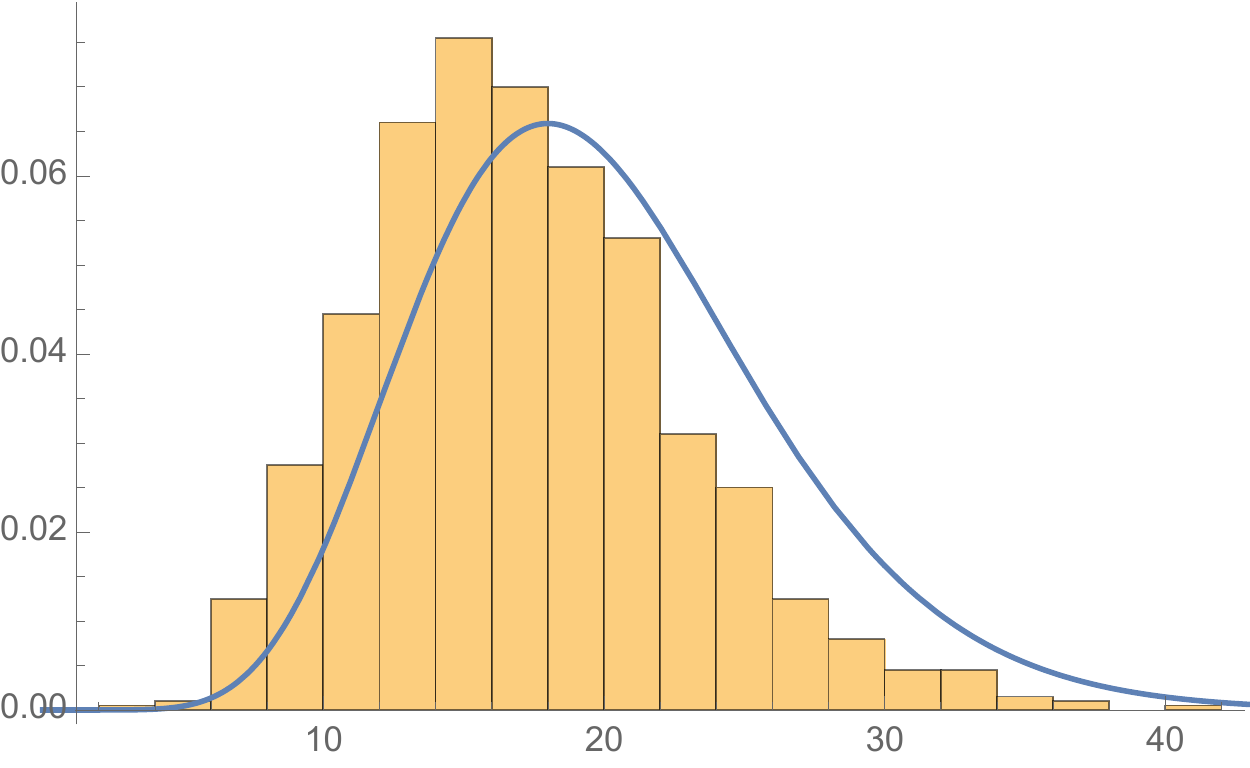}
	\end{subfigure}
\caption{Histograms of the distribution of the $\Delta \chi^2$ variable when (a) no cut has been imposed on the $C_i$, (b) when a cut has been imposed and $C_{cut} = 1$ and in (c) when $C_{cut} = 0.1$. We represent in blue a chi square distribution with 20 degrees of freedom.}
\label{Distribution_Delta_Chi_square}
\end{figure}
\begin{table}
\centering
\tabcolsep 4pt
\begin{tabular}{|c|c|c|c|c|c|}
\hline
$C_i^G$ & $(1 \sigma,0)$ & $(1 \sigma, 0.1 \%)$ & $(1 \sigma, 0.3 \%)$   & $(1 \sigma, 0.5 \%)$ & $(1 \sigma, 1 \%)$  \\ 
\hline \hline
$\tilde{C}_{He} $ &$8.9 \pm 24$&	$8.9 \pm 24$&	$8.6 \pm 24$&$8.4 \pm 24$&$8.4 \pm 25$\\
$\tilde{C}_{Hu}$ & $-4.9 \pm 16$&	$-4.8 \pm 16$&	$-4.6 \pm 16$&$-4.4 \pm 16$& $-4.4 \pm 17$\\
$\tilde{C}_{Hd}$ &$-0.48 \pm 8.1$&	$-0.48 \pm 8.1$&	 $-0.59 \pm 8.2$&$-0.70 \pm 8.2$&$-0.68 \pm 8.3$\\
$\tilde{C}_{Hl}^{(1)}$ &$4.5 \pm 12$&	$4.5 \pm 12$&		$4.4 \pm 12$&$4.3 \pm 12$&$4.3 \pm 12$\\
$\tilde{C}_{Hl}^{(3)}$ & $8.7 \pm 30$&	 $8.8 \pm 30$&	 $8.8 \pm 30$& $8.7 \pm 30$&$8.6 \pm 30$\\
$\tilde{C}_{Hq}^{(1)}$ &$-1.6 \pm 4.0$&	$-1.6 \pm 4.0$&	$-1.5 \pm 4.0$& $-1.4 \pm 4.1$& $-1.5 \pm 4.1$\\
$\tilde{C}_{Hq}^{(3)}$ &  $8.3 \pm 31$&	$8.4 \pm 31$&	$8.4 \pm 31$&$8.4 \pm 31$&$8.2 \pm 31$\\
$\tilde{C}_{HWB}$ & $4.0 \pm 6.2$& 	$4.0 \pm 6.3$&		$3.7 \pm 6.7$& 	$3.4 \pm 7.1$& $3.5 \pm 8.0$\\
$\tilde{C}_{HD}$ & $-18 \pm 27$& 	$-18 \pm 27$& 		$-17 \pm 28$&	 $-16 \pm 28$&$-17 \pm 28$\\
$\tilde{C}_{ll}$ &$-0.11 \pm 0.12$&	 $-0.084 \pm 0.15$& 		$-0.067 \pm 0.17$& $-0.066 \pm 0.17$& $-0.067 \pm 0.19$\\
$\tilde{C}_{ee}$ &$-0.035 \pm 0.20$&	$-0.035 \pm 0.20$&	$-0.035 \pm 0.20$&$-0.033 \pm 0.21$& $-0.029 \pm 0.24$\\
$\tilde{C}_{eu}$ & $-27 \pm 24$&	 $-26 \pm 24$&		$-24 \pm 24$& $-23 \pm 25$& $-22 \pm 25$\\
$\tilde{C}_{ed}$ &$-27 \pm 30$&	 $-26 \pm 30$&		 $-24 \pm 31$& $-23 \pm 31$& $-22 \pm 31$\\
$\tilde{C}_{l e}$ &$-0.01 \pm 0.30$&	$-0.013 \pm 0.30$&		 $-0.015 \pm 0.31$& $-0.013 \pm 0.31$& $-0.0064 \pm 0.32$\\
$\tilde{C}_{l u}$ &$-17 \pm 8.4$&	$-17 \pm 8.4$&		$-17 \pm 8.5$&$-17 \pm 8.5$&$-17 \pm 8.8$\\
$\tilde{C}_{l d}$ & $-33 \pm 16$& 	$-32 \pm 16$&		$-32 \pm 16$&$-32 \pm 16$& $-31 \pm 17$\\
$\tilde{C}_{l q}^{(1)}$ & $-4.1 \pm 1.9$&	$-3.6 \pm 2.4$&	 $-2.9 \pm 3.7$&$-2.6 \pm 4.8$& $-2.7 \pm 6.8$\\
$\tilde{C}_{l q}^{(3)}$ &$-0.51 \pm 0.21$&	$-0.47 \pm 0.25$&	$-0.41 \pm 0.31$& $-0.37 \pm 0.38$&$-0.28 \pm 0.57$\\
$\tilde{C}_{qe}$ &$-1.4 \pm 26$&	$-2.1 \pm 26$&		$-2.7 \pm 26$&$-3.1 \pm 26$& $-4.0 \pm 27$\\
$\tilde{C}_{W}$ &$10 \pm 30$&	$10 \pm 30$&		$10 \pm 30$& $10 \pm 30$& $10 \pm 30$\\
\hline\end{tabular}
\caption{Maximum likelihood estimators and their $ \sigma$ confidence region $\tilde{C} \pm  \sigma$ for a SMEFT error of $\{ 0\%, 0.1\%, 0.3\%, 0.5 \%, 1\% \}$.
We have scaled the results and error by $100$ so that $\tilde{C} = 100  C_{MLE}$. 
A cut $C_{cut} =0.1$ has been imposed on the $C_i$ to obtain their central values, and we impose that $|C_i \pm 3 \sigma | < 1$ . \label{ConstraintsCut}}
\end{table}
If the UV model(s) assumed in profiling or marginalizing breaks the correlations of the parameter space,
stronger bounds are obtained.
 If the likelihood is factorized by hand, and a subset of the parameters are profiled or marginalized away, this can also factor up the fit space in a manner that will significantly effect the bounds obtained.
As the profiling procedure allows the profiled parameters to take on any value when obtaining constraints on the individual Wilson coefficients, it leads to weaker bounds.
As particular correlations are always present in a UV model matched onto the Wilson coefficients, the bounds quoted on the parameters in the SMEFT when profiling or marginalizing has been done
must be interpreted with care. The effect of the correlations relaxing bounds on the individual Wilson coefficients is so strong, the bounds obtained on the Wilson coefficients ($C_{W}$,$C_{Hq}^{(3)}$,$C_{Hl}^{(3)}$,$C_{HD}$) seem to violate the power counting, when the other Wilson coefficients are profiled away.  To ensure that the bounds quoted do not depend on any correlated violation of the power counting in profiling, we impose a cut  $|C_i \bar{v}_T^2/\Lambda^2| < C_{cut}$ so that the $C_i \bar{v}_T^2/\Lambda^2$ are not longer allowed to take value outside this 19-sphere of radius $C_{cut}$ when profiling. This changes the distribution of $\Delta \chi^2 \left(C_{true}\right)$ shifting it to the left compared the chi-squared distribution with 20 degrees of freedom, see Figure \ref{Distribution_Delta_Chi_square}. Being aware of this change in the distribution of the $\Delta \chi^2 \left(C_{true}\right)$, we give in Table \ref{ConstraintsCut} the $1\sigma$ bounds on the $C_i \bar{v}_T^2/\Lambda^2$ when a cut $C_{cut}=0.1$ has been imposed on to get their central values in the profiling procedure, and the $3\sigma$ confidence region has been limited when necessary so that $|\tilde{C}_{i,min} \pm 3 \sigma| < 1$. When imposing this cut, the distribution of $\Delta \chi^2$ is shifted to the left so that the bounds derived are too conservative, in the sense that the $1\sigma$ regions we are reporting in Table \ref{ConstraintsCut} correspond to a (slightly) smaller confidence region.

To further develop an intuition for the degree of constraint present on the Wilson coefficients, and the strong UV dependence on the conclusions drawn, we
consider the case where only one Wilson coefficient is present at a time in constraining the SMEFT parameters. The  results are shown in Table \ref{Constraints1WCon}, which
demonstrate a much stronger degree of constraint. These results are likely too strong in any realistic UV model.
\begin{table}
\centering
\tabcolsep 4pt
\footnotesize
\begin{tabular}{|c|c|c|c|c|c|}
\hline
$C_i^G$ & $(1 \sigma,0)$ & $(1 \sigma, 0.1 \%)$ & $(1 \sigma, 0.3 \%)$   & $(1 \sigma, 0.5 \%)$ & $(1 \sigma, 1 \%)$  \\ 
\hline \hline
$\tilde{C}_{He} $ &$-0.052 \pm 0.036$&	$-0.056 \pm 0.047$&	$-0.078 \pm 0.064$&$-0.083 \pm 0.069$&$-0.072 \pm 0.075$\\
$\tilde{C}_{Hu}$ & $0.021 \pm 0.041$&	$0.021 \pm 0.043$&	$0.018 \pm 0.052$&$0.016 \pm 0.062$& $0.015 \pm 0.08$\\
$\tilde{C}_{Hd}$ &$-0.0096 \pm 0.099$&	$-0.014 \pm 0.1$&	 $-0.042 \pm 0.13$&$-0.076 \pm 0.15$&$-0.13 \pm 0.2$\\
$\tilde{C}_{Hl}^{(1)}$ &$0.025 \pm 0.025$&	$0.019 \pm 0.035$&		$0.013 \pm 0.059$&$0.011 \pm 0.07$&$0.013 \pm 0.082$\\
$\tilde{C}_{Hl}^{(3)}$ & $-0.0064 \pm 0.019$&	$0.0046 \pm 0.027$&	 $0.018 \pm 0.04$& $0.023 \pm 0.044$&$0.018 \pm 0.05$\\
$\tilde{C}_{Hq}^{(1)}$ &$-0.0039 \pm 0.0085$&	$-0.0038 \pm 0.0088$&	$-0.0033 \pm 0.011$& $-0.0026 \pm 0.013$&$-0.0019 \pm 0.017$\\
$\tilde{C}_{Hq}^{(3)}$ &  $0.0027 \pm 0.023$&	$0.011 \pm 0.037$&	$0.032 \pm 0.076$&$0.052 \pm 0.1$&$0.080 \pm 0.14$\\
$\tilde{C}_{HWB}$ & $-0.0092 \pm 0.019$& 	$0.018 \pm 0.026$&		$0.024 \pm 0.027$& $0.025 \pm 0.028$&$0.02 \pm 0.03$\\
$\tilde{C}_{HD}$ & $-0.052 \pm 0.048$& $0.036 \pm 0.092$& 		$0.082 \pm 0.11$&	 $0.085 \pm 0.11$&$0.060 \pm 0.13$\\
$\tilde{C}_{ll}$ &$0.0038 \pm 0.024$&	 $-0.014 \pm 0.037$& 		$-0.036 \pm 0.051$& $-0.041 \pm 0.055$&$-0.036 \pm 0.06$\\
$\tilde{C}_{ee}$ &$-0.00092 \pm 0.19$&	$-0.00092 \pm 0.19$&	$-0.00088 \pm 0.19$&$-0.00055 \pm 0.19$& $0.0027 \pm 0.2$\\
$\tilde{C}_{eu}$ &$-0.54 \pm 0.31$&	 $-0.54 \pm 0.31$&		$-0.55 \pm 0.32$& $-0.55 \pm 0.32$& $-0.59 \pm 0.35$\\
$\tilde{C}_{ed}$ &$0.28 \pm 0.39$&	 $0.28 \pm 0.39$&		 $0.28 \pm 0.39$& $0.28 \pm 0.4$& $0.28 \pm 0.43$\\
$\tilde{C}_{l e}$ &$0.0051 \pm 0.3$&	$0.0051 \pm 0.3$&		$0.0052 \pm 0.3$& $0.0058 \pm 0.3$& $0.011 \pm 0.31$\\
$\tilde{C}_{l u}$ &$0.013 \pm 0.53$&	$0.014 \pm 0.53$&		$0.024 \pm 0.54$&$0.04 \pm 0.54$&$0.09 \pm 0.58$\\
$\tilde{C}_{l d}$ & $0.84 \pm 0.61$& 	$0.84 \pm 0.61$&		$0.84 \pm 0.61$
&$0.83 \pm 0.62$& $0.82 \pm 0.66$\\
$\tilde{C}_{l q}^{(1)}$ & $0.45 \pm 0.34$&$0.45 \pm 0.34$&	$0.46 \pm 0.34$&$0.48 \pm 0.35$& $0.52 \pm 0.37$\\
$\tilde{C}_{l q}^{(3)}$ &$0.019 \pm 0.028$&	$0.047 \pm 0.049$&	$0.11 \pm 0.078$& $0.13 \pm 0.087$&$0.15 \pm 0.1$\\
$\tilde{C}_{qe}$ &$-0.42 \pm 0.41$&	$-0.42 \pm 0.41$&		$-0.42 \pm 0.41$&$-0.42 \pm 0.41$& $-0.42 \pm 0.43$\\
$\tilde{C}_{W}$ &$1.7 \pm 4.4$&	$1.7 \pm 4.4$&		$1.8 \pm 4.4$& $1.8 \pm 4.4$& $1.9 \pm 4.5$\\
\hline\end{tabular}
\caption{The $ 1 \sigma$ confidence region $\tilde{C} \pm  \sigma$ for a SMEFT error of $\{ 0\%, 0.1\%, 0.3\%, 0.5 \%, 1\% \}$.
Here we have multiplied the presented MLE and error by 100 ($\tilde{C} = 100  C_{MLE}$).
The results shown are for when one Wilson coefficient at a time is turned "on". \label{Constraints1WCon}}
\end{table}
Another case of interest is the subset of UV completions to the SM that are weakly coupled and renormalizable where
the Artz-Einhorn-Wudka operator classification scheme \cite{Arzt:1994gp} applies. As we neglect one loop corrections in the results presented, we then neglect the parameters $C_{HWB}$ and $C_W$ in the global fit.
The results are shown in Table \ref{ConstraintsUVren}, which again demonstrate a stronger degree of constraint, by roughly an order of magnitude.
This is another illustration of the important effect of the correlation between near $Z$ pole and charged current LEPII data in these results. Assumptions made
on the parameter $C_W$, contributing to TGC parameters, has a critical impact on analyses of this form.
\begin{table}
\centering
\tabcolsep 4pt
\begin{tabular}{|c|c|c|c|c|c|}
\hline
$C_i^G$ & $(1 \sigma,0)$ & $(1 \sigma, 0.1 \%)$ & $(1 \sigma, 0.3 \%)$   & $(1 \sigma, 0.5 \%)$ & $(1 \sigma, 1 \%)$  \\ 
\hline \hline
$\tilde{C}_{He} $ &$-0.11 \pm 1.6$&	$0.065 \pm 1.7$&	$0.28 \pm 1.8$&$0.38 \pm 1.8$&$0.39 \pm 2$\\
$\tilde{C}_{Hu}$ & $1.2 \pm 1.3$&	$1.1 \pm 1.3$&	$1.0 \pm 1.4$&$0.93 \pm 1.4$& $0.93 \pm 1.5$\\
$\tilde{C}_{Hd}$ &$-3.4 \pm 1.4$&	$-3.4 \pm 1.4$&	$-3.4 \pm 1.4$&$-3.4 \pm 1.4$&$-3.4 \pm 1.6$\\
$\tilde{C}_{Hl}^{(1)}$ &$0.032 \pm 0.82$&	$0.14 \pm 0.85$&		$0.29 \pm 0.93$&$0.35 \pm 0.98$&$0.34 \pm 1.1$\\
$\tilde{C}_{Hl}^{(3)}$ & $-0.33 \pm 3.0$&	$-0.052 \pm 3.1$&	 $0.29 \pm 3.2$& $0.44 \pm 3.3$&$0.42 \pm 3.5$\\
$\tilde{C}_{Hq}^{(1)}$ &$-0.048 \pm 0.32$&	$-0.070 \pm 0.32$&	$-0.10 \pm 0.33$& $-0.11 \pm 0.35$&$-0.11 \pm 0.37$\\
$\tilde{C}_{Hq}^{(3)}$ & $-0.71 \pm 3.0$&	$-0.39 \pm 3.1$&	$-0.02 \pm 3.2$&$0.15 \pm 3.3$&$0.11 \pm 3.5$\\
$\tilde{C}_{HD}$ & $0.21 \pm 3.2$& $0.11 \pm 3.3$& 		$-0.12 \pm 3.3$&	 $-0.21 \pm 3.4$&$-0.16 \pm 3.5$\\
$\tilde{C}_{ll}$ &$-0.14 \pm 0.11$&	 $-0.093 \pm 0.15$& 		$-0.088 \pm 0.16$& $-0.086 \pm 0.17$&$-0.085 \pm 0.19$\\
$\tilde{C}_{ee}$ &$-0.029 \pm 0.19$&	$-0.027 \pm 0.2$&	$-0.023 \pm 0.2$&$-0.019 \pm 0.21$& $-0.012 \pm 0.23$\\
$\tilde{C}_{eu}$ &$-27 \pm 24$&	 $-25 \pm 24$&		$-23 \pm 24$& $-22 \pm 25$& $-20 \pm 25$\\
$\tilde{C}_{ed}$ &$-27 \pm 30$&	 $-25 \pm 31$&		 $-23 \pm 31$& $-22 \pm 31$&$-21 \pm 31$\\
$\tilde{C}_{l e}$ &$-0.003 \pm 0.3$&	$-0.0068 \pm 0.3$&		$-0.006 \pm 0.31$& $-0.0031 \pm 0.31$&$0.0048 \pm 0.32$\\
$\tilde{C}_{l u}$ &$-17 \pm 8.4$&	$-17 \pm 8.4$&		$-17 \pm 8.5$&$-17 \pm 8.5$&$-17 \pm 8.8$\\
$\tilde{C}_{l d}$ & $-32 \pm 16$& 	$-32 \pm 16$&		$-32 \pm 16$
&$-32 \pm 16$& $-32 \pm 17$\\
$\tilde{C}_{l q}^{(1)}$ & $-4 \pm 1.9$&$-2.9 \pm 3$&	$-1.8 \pm 4$&$-1.3 \pm 4.7$& $-0.97 \pm 5.9$\\
$\tilde{C}_{l q}^{(3)}$ &$-0.51 \pm 0.21$&	$-0.43 \pm 0.27$&	$-0.38 \pm 0.32$& $-0.34 \pm 0.38$&$-0.26 \pm 0.57$\\
$\tilde{C}_{qe}$ &$-1.5 \pm 26$&	$-2.1 \pm 26$&		$-2.8 \pm 26$&$-3.3 \pm 26$& $-4.1 \pm 27$\\
\hline\end{tabular}
\caption{$1 \sigma$ confidence region $\tilde{C} \pm  \sigma$ for a SMEFT error of $\{ 0\%, 0.1\%, 0.3\%, 0.5 \%, 1\% \}$ where $\tilde{C} = 100  C_{MLE}$ and $C_{MLE}$ are the MLE when a renormalizable theory is assumed to be the UV completion of the SM. \label{ConstraintsUVren}}
\end{table}
\subsection{The Eigensystem and constraints on the leptonic couplings of the $Z$}
\begin{figure}
  \centering
\centerline{\includegraphics[width=5in,height=1.9in]{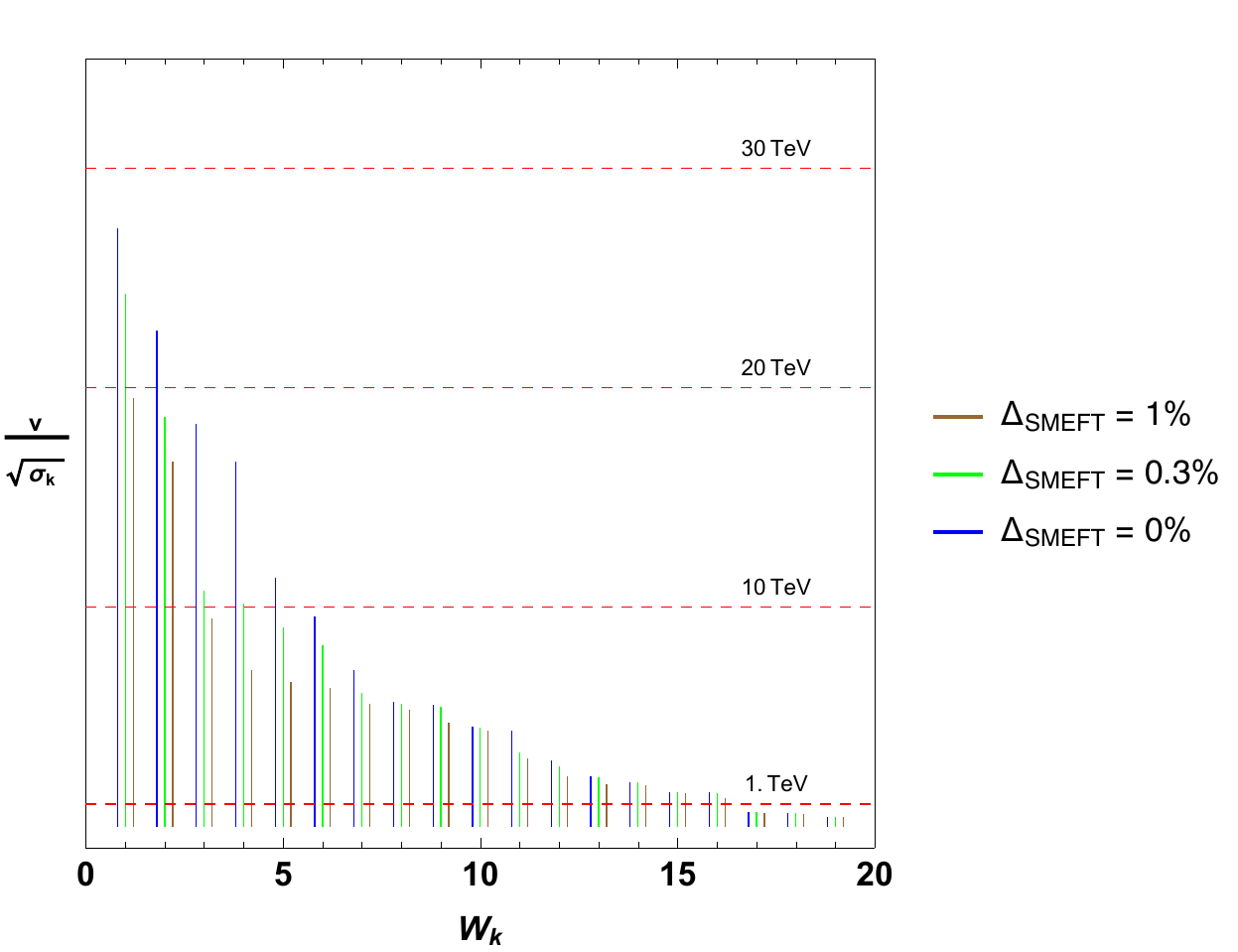}}
  \caption{\label{Fig:eigen} The values $v/ \sqrt{\sigma_k}$ (which corresponds to the effective scale of suppression) for each Eigenvector $W_k$ of the Fisher matrix. We show results for $\Delta_{SMEFT} = \{0 \%, 0.3 \, \%, 1 \, \%\}$. }
\end{figure}
The degree of constraint on orthogonal linear independent combinations of the Wilson coefficients varies for the global fit.
This is related to the different degree of constraint reported for the individual Wilson coefficients and the $\delta X$ parameters.
The normalized Eigenvectors and Eigenvalues of the system are directly obtained from the Fisher matrices. The definition of the Eigensystem is given in
Ref. \cite{Berthier:2015gja}, and the updated Eigensystem is given in Fig \ref{Fig:eigen}.
\begin{figure}[h!]
	\centering
	\begin{subfigure}{0.49 \textwidth}
	\includegraphics[width=1 \textwidth]{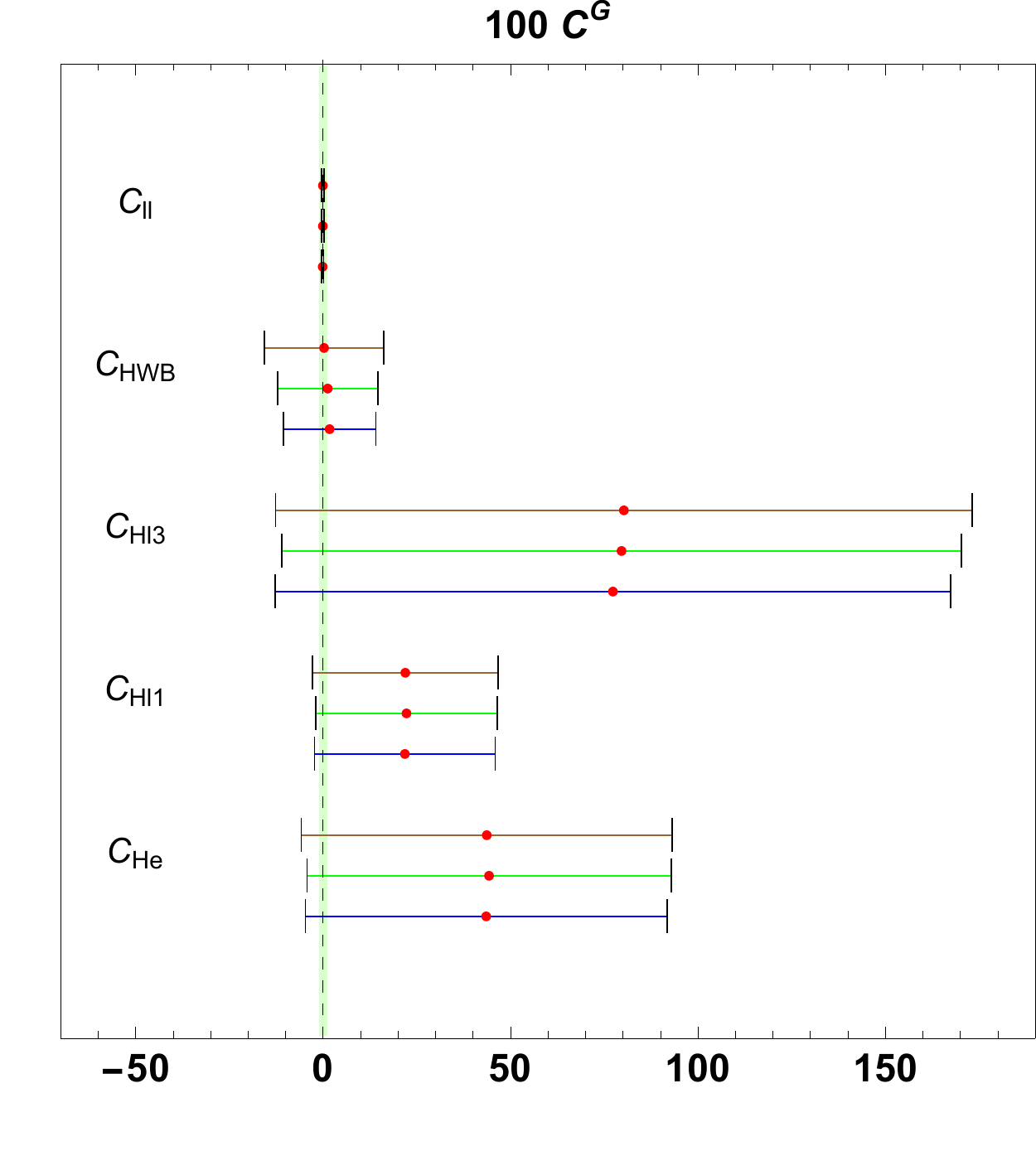}
	\caption{}
	\end{subfigure}
	\begin{subfigure}{0.49 \textwidth}
	\includegraphics[width=1 \textwidth]{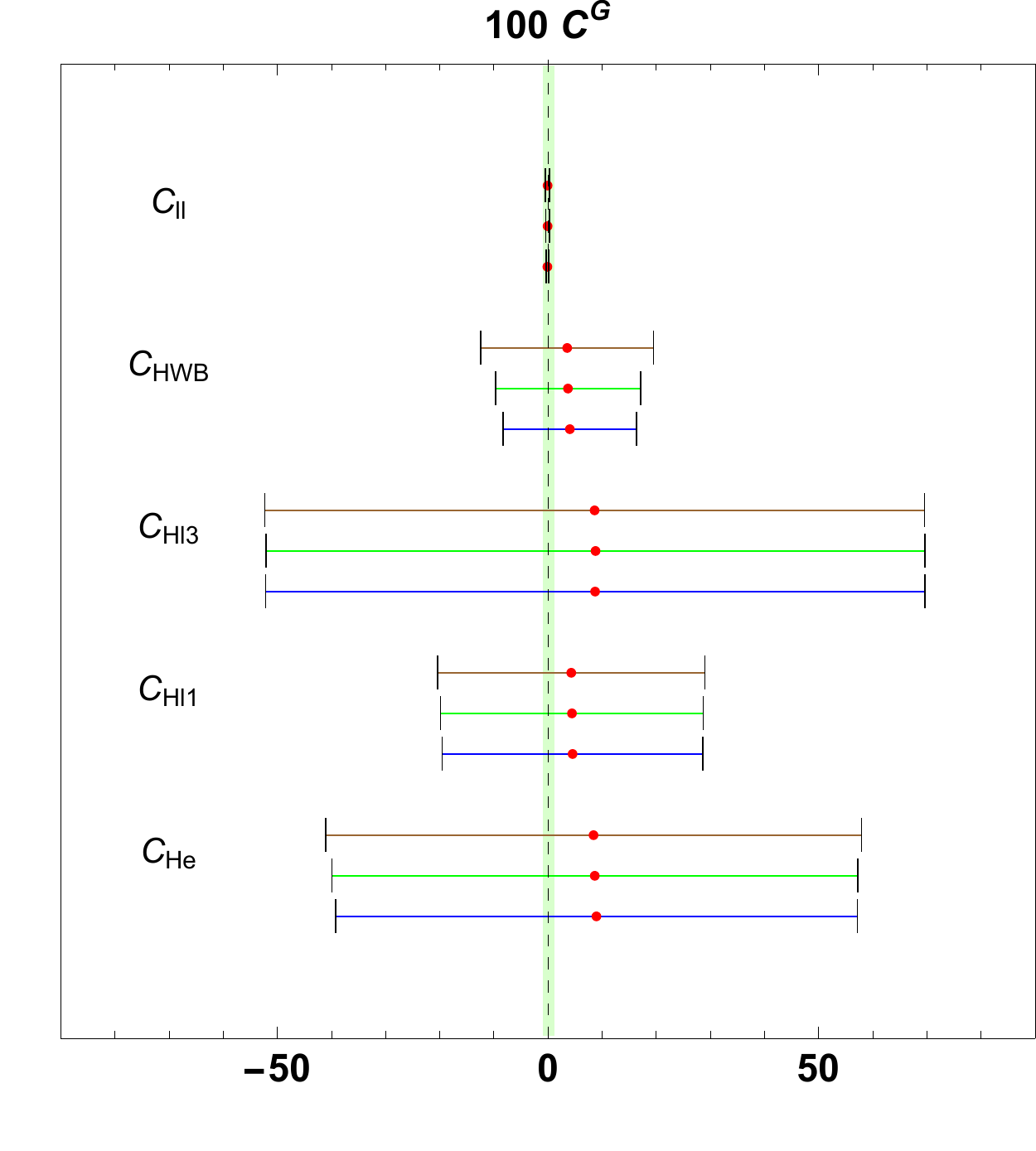}
	\caption{}
	\end{subfigure}
		\begin{subfigure}{0.49 \textwidth}
	\includegraphics[width=1 \textwidth]{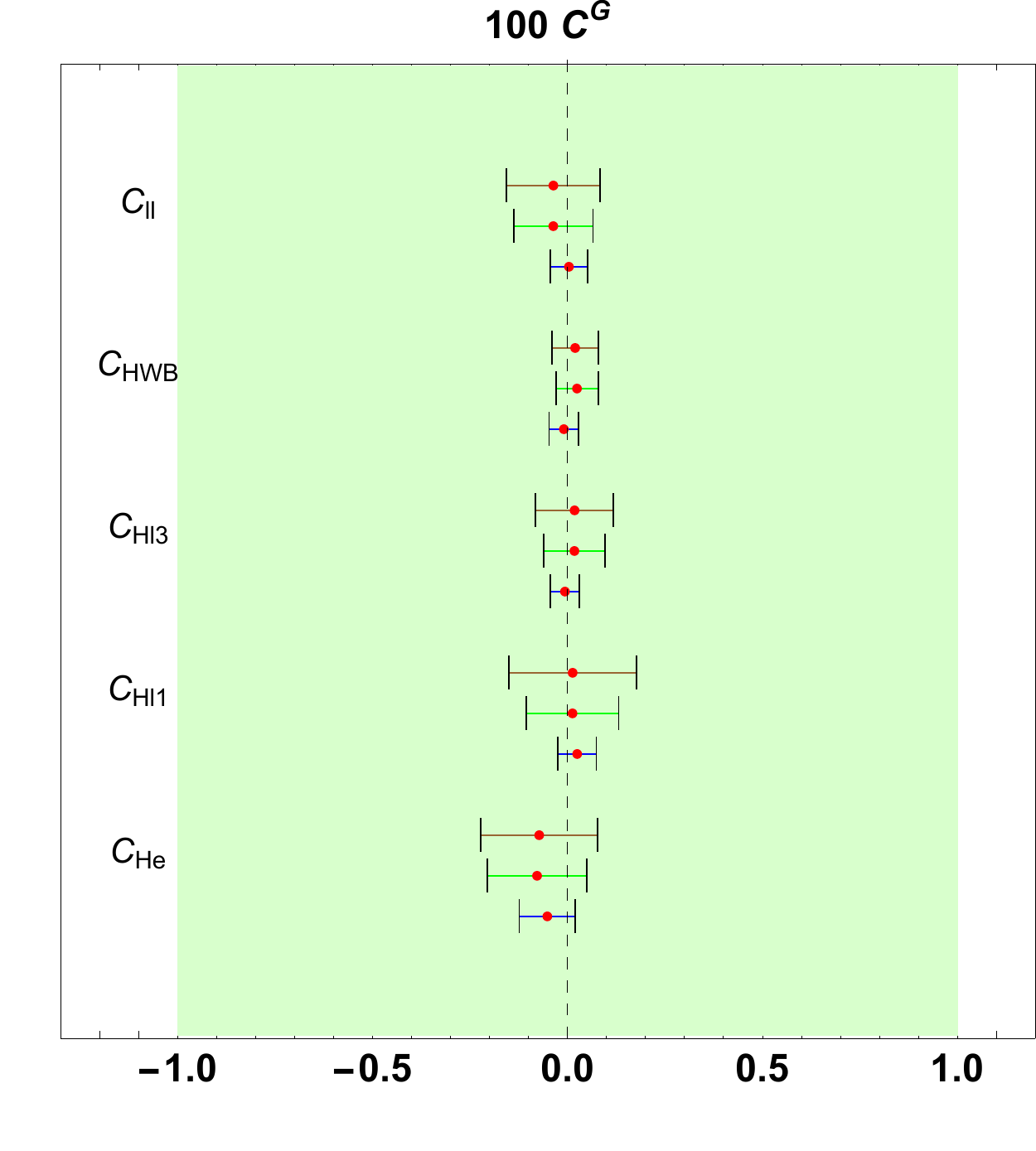}
	\caption{}
	\end{subfigure}
	\begin{subfigure}{0.49 \textwidth}
	\includegraphics[width=1 \textwidth]{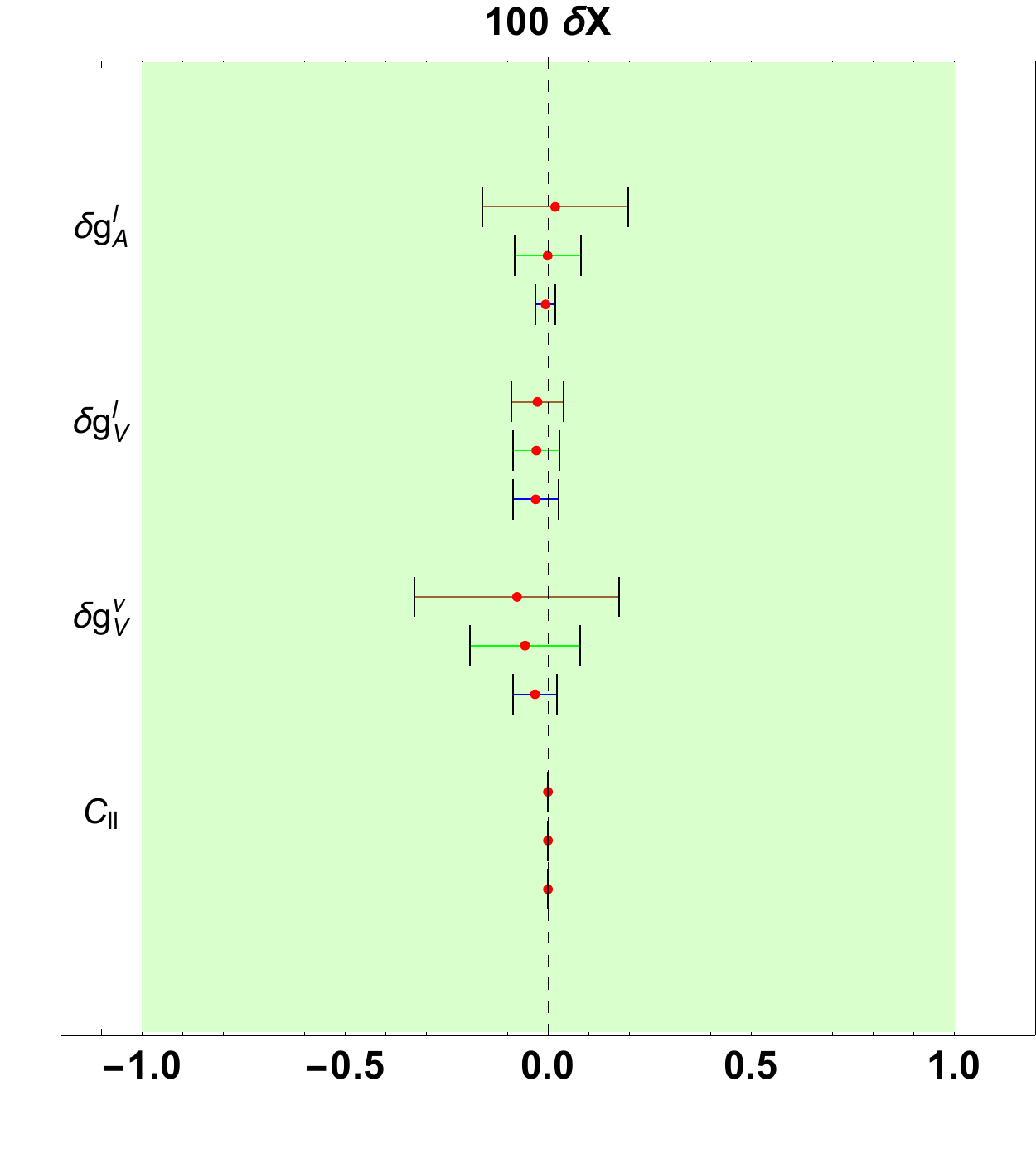}
	\caption{}
	\end{subfigure}
	\caption{Constraints on parameters that contribute to anomalous $Z$ couplings scaled by $10^2$, and their scaled $\pm 2 \sigma$ confidence regions. Fig. (a): Individual bounds on the Wilson coefficients (scaled by $\bar{v}_T^2$), when the other parameters are profiled away in the Warsaw basis. Fig (b): The same constraints when all other parameters are profiled away subject to the cut constraints discussed in the text. Fig. (c): the case where only one parameter is assumed to be present in the global fit at a time. Fig. (d) The constraints found on the $\delta X$ parameters. All results are shown for $\Delta_{SMEFT} = \{0, 0.3 \%, 1\% \}$ for the blue, green and brown lines respectively.
	Note that the shaded green region is the same size for all plots, corresponding to $\%$ level constraints, to make the comparison between cases clear.}
	\label{Fig:span}
\end{figure}
It is the existence of the significant hierarchy in constraints present in the data as illustrated in the Eigensystem, that leads to the span on conclusions drawn on how strongly constrained the parameters are in the SMEFT.
Eigenvectors are not perturbatively stable. Loop corrections in the SMEFT mix the Eigenvectors and modify the interpretation of the bounds obtained away from the $Z$ pole. As the scales of suppression for the orthogonal Eigenvectors differ by over an order of magnitude, this modifies the interpretation of the constraints when applied to LHC data, or higher energy data.
This is the case for parameters that contribute to anomalous leptonic couplings of the $Z$ in particular.
To make the remarkable span in constraints related to anomalous leptonic couplings of the $Z$ clearer, we show in Fig.~\ref{Fig:span} the constraints on such parameters in four cases.
Fig.~\ref{Fig:span} shows that the degree of constraint present on these parameters spans two orders of magnitude in interpreting the same global data set in these cases.

\section{Conclusions} \label{sec:conclusion}
In this paper we have studied the shape of possible physics beyond the Standard Model, building upon the results in Refs.~\cite{Alonso:2013hga,Berthier:2015oma,Berthier:2015gja}
and a companion paper focused on $W^\pm$ mass extractions \cite{todaypaper}. It is required to incorporate LEPII data on four fermion production, reported in terms of CC03  differential and total cross section bounds, to robustly incorporate the impact of the LEP near $Z$ pole pseudo-observable data in the SMEFT. We have developed and reported the results for the double pole prediction of the CC03 four fermion production results in the SMEFT. Using these results, we have simultaneously bounded and studied the constraints on 20 parameters in the SMEFT in this work. The strongest constraints are dictated by the measurements with the highest precision. As we have restricted our attention to flavour conserving observables, the measurements of the $W^\pm$ mass, and measurements of the leptonic observables at LEPI are the strongest bounds. These experimental constraints robustly rise above the percent level in experimental precision, which can exceed the natural power counting size of SMEFT corrections when $1 \,{\rm TeV} \lesssim \Lambda/\sqrt{C_i} \lesssim 3 \, {\rm TeV}$. For this reason,
an interpretation of the corresponding bounds in the SMEFT formalism is important to inform an expectation of possible deviations that can be found at LHC.

The strength of the model independent analysis we have developed
is that it can accommodate a very wide range of UV scenarios. However, at the same time this generality limits the strength of the conclusions that can be drawn in a truly model independent fashion. This is particularly the case when considering how these bounds project onto what deviations are allowed
in LHC measurements off the $Z$ pole. The different conclusions drawn on the degree of constraint of parameters contributing to the leptonic couplings of the $Z$, is summarized in Fig.\ref{Fig:span}. It is reasonable to consider that the constraints shown in the upper/lower panels of Fig.\ref{Fig:span} to be an underestimate/overestimate of the degree of constraint in a realistic UV model. For the bottom left hand plot this is due to the requirement that only one Wilson coefficient is generated at tree level in a UV matching, for the right hand plot, this is due to theoretical errors being underestimated as argued in the text.\footnote{While all of these results are formally correct and follow from their assumptions.}

The global fit shows that the degree of constraint on the SMEFT parameters found, is strongly dependent on the assumptions made about possible UV physics matched onto the SMEFT,
dictating correlations present (or not) among the Wilson coefficients. The theoretical error present in the fit, due to neglected perturbative corrections and $\mathcal{L}_8$, which is also UV dependent in its numerical impact dictates the interpretation. This is consistent with our previous results in Refs.~\cite{Berthier:2015oma,Berthier:2015gja} and basically unsurprising, although the range of conclusions drawn is very significant -- differing by orders of magnitude.

It is reasonable to interpret
the lack of deviations from the Standard Model expectation in measured observables, to mean that the cut off scale and Wilson coefficients are such that the most precise observables
are accommodated without any cancelations between SMEFT parameters. It is also reasonable to consider that physics beyond the SM is present at lower cut off scales with larger Wilson coefficients,
falling in the interesting range $1 \,{\rm TeV} \lesssim \Lambda/\sqrt{C_i} \lesssim 3 \, {\rm TeV}$ motivated by the hierarchy problem, and some partial suppression of UV physics effects is present in near $Z$ pole measurements. This later case is of most interest in understanding LHC data. Our global fit results do not rule out this possibility.
Considering our results, a reasonable approach to LHC analyses is to report data in a manner that does not limit
its interpretation to a subset of UV scenarios where $\Lambda/\sqrt{C_i} \gtrsim 3 \, {\rm TeV}$. This can be done by accommodating $\sim \%$ level constraints on the parameters that are present in this global analysis, when reporting LHC data.
\clearpage
\appendix
\section{Core shifts of parameters due to the SMEFT} \label{CoreShifts}
Our results are expressed in terms the core shifts of parameters present in the SMEFT, given in \citep{Berthier:2015oma} and included below for completeness. Our notational conventions are that a total shifts of a parameter $X$ due to all the operators in $\mathcal{L}_6$ is denoted as $\delta X$. The measured input observables
$\{\hat{G}_F,\hat{m}_Z,\hat{\alpha} \}$ are denoted with hat superscripts. Expressions derived at tree level in the SM from these input parameters are also denoted with hat superscripts.
Here these parameters are used to define $\{c_{\hat{\theta}},s_{\hat{\theta}},\hat{m}_W \}$ at tree level using SM tree level relations.  For more details, see Refs.\cite{Berthier:2015oma,Berthier:2015gja}.
The shifts we use are
\begin{align*}
\frac{\delta m_Z^2}{\hat{m}_Z^2} &\equiv  \frac{1}{\sqrt{2}
\, \hat{G}_F} \, \left(\frac{C_{HD}}{2} + 2 \frac{\hat{m}_W}{\hat{m}_Z} \, s_{\that} \, C_{HWB} \right), \\
\frac{\delta m_W^2}{\hat{m}_W^2} &\equiv - \frac{\delta s_{\that}^2}{s_{\that}^2}  - \frac{\hat{m}_W}{\hat{m}_Z \, s_{\that}} \, \frac{C_{HWB}}{\hat{G}_F} -  \sqrt{2} \, \delta G_F, \\
\frac{\delta s_\theta^2}{s^2_{\hat{\theta}}} &\equiv - \frac{\hat{m}^2_W/\hat{m}^2_Z}{2 \, \sqrt{2} \, \hat{G}_F (1 - 2 s^2_\that)} \left[C_{HD} + 2 \frac{\hat{m}_Z}{s_\that \, \hat{m}_W} \, C_{HWB} + 4 \, \hat{G}_F \, \delta G_F \right],
\end{align*}
further
\begin{align*}
\delta g_Z &=- \frac{\delta G_F}{\sqrt{2}} - \frac{\delta m_Z^2}{2\hat{m}_Z^2} + \frac{s_{\hat{\theta}} \, c_{\hat{\theta}}}{\sqrt{2} \hat{G}_F} \, C_{HWB},
& \delta G_F &= \frac{1}{\sqrt{2} \,  \hat{G}_F} \left(\sqrt{2} \, C^{(3)}_{\substack{Hl}} - \frac{C_{\substack{ll}}}{\sqrt{2}}\right), \\
%%%%%%%%%%%%%%%%%%%%%%%%%%%%%%%%%%%%%%%%%%%%%%%%%%%%%%%%%%%%%%%%
\end{align*}
so that
\bea \label{higherdgvga}
\delta (g^{\ell}_V)_{pr}&=&\delta g_Z \, (g^{\ell}_{V})^{SM}_{pr} - \frac{1}{4 \sqrt{2} \hat{G}_F} \left(C_{\substack{H e \\pr}} + C_{\substack{H l \\ pr}}^{(1)} + C_{\substack{H l \\ pr}}^{(3)} \right) - \delta s_\theta^2, \\
%%%%%%%%%%%%%%%%%%%%%%%%%%%%%%%%%%%%%%%%%%%%%%%%%%%%%%%%%%%%%%%%%
\delta(g^{\ell}_A)_{pr}&=&\delta g_Z \, (g^{\ell}_{A})^{SM}_{pr} + \frac{1}{4 \, \sqrt{2} \, \hat{G}_F}
\left( C_{\substack{H e \\pr}} - C_{\substack{H l \\ pr}}^{(1)} - C_{\substack{H l \\ pr}}^{(3)} \right),
%%%%%%%%%%%%%%%%%%%%%%%%%%%%%%%%%%%%%%%%%%%%%%%%%%%%%%%%%%%%%%%%%
\eea
with $p,r$ flavour index dependence that is trivialized to $\delta_{pr}$ due to our $\rm{U}(3)^5$ assumption, and
\bea
\delta(g^{W_{\pm},\ell/q}_V)_{pr} = \delta(g^{W_{\pm},\ell/q}_A)_{pr}  &=&  \frac{1}{2\sqrt{2} \hat{G}_F} \left(C^{(3)}_{\substack{H l/q \\ pr}} + \frac{1}{2} \frac{c_{\hat{\theta}}}{ s_{\hat{\theta}}} \, C_{HWB} \right)
+ \frac{1}{4} \frac{\delta s_\theta^2}{s^2_{\hat{\theta}}}.
\eea
Here our chosen normalization is $(g^{x}_{V})^{SM} = T_3/2 - Q^x \, \bar{s}_\theta^2, (g^{x}_{A})^{SM} = T_3/2$ where $T_3 = 1/2$ for $u_i,\nu_i$ and $T_3 = -1/2$ for $d_i,\ell_i$
and $Q^x = \{-1,2/3,-1/3 \}$ for $x = \{\ell,u,d\}$. Note that the $g^W_{V,A}$ couplings are normalized to $V_{pr}/2$ and flavour change due to the $W$ couplings shifts above is
also $\propto V_{pr}$. The core shift parameters are useful, but they should not be confused with an operator basis for the SMEFT.

\subsection{Redefinition of $\Gamma_W$} % (fold)
\label{sub:the_redefinition_of_the_w_width}

The partial $W^\pm$ widths are redefined by dimension 6 operators in the following way
\begin{align}
	\bar{\Gamma}_{W \rightarrow f_i f_j} &= \frac{N_C \, |V^f_{ij}|^2 \sqrt{2} \hat{G}_F \, \hat{m}_W^3}{12 \pi} \left(1 + 4 \delta g_{V/A}^{W, f} - \frac{1}{2} \frac{\delta m_W^2}{\hat{m}_W^2}\right).    % \bar \Gamma_W = \frac{3\sqrt{2}}{4\pi}\hat G_F m_W^3\left(1+\frac{4}{3}\delta g^\ell_W + \frac{8}{3}\delta g^q_W -\frac{\delta m_W^2}{2 m_W^2}\right)
\end{align}
Here, $N_C$ depends on the colour representation of final state fermions. $V^f_{ij}$ corresponds to the CKM or PMNS matrix with transitions between the mass eigenstate flavours $i,j$. As the neutrino flavour of the decay of a $W^\pm$ boson is not identified, the sum over the neutrino species gives $ \sum_j |V^\ell_{ij}|^2 = 1$. We have used the short hand notation
$f=\{\ell, q\}$ as we consider the $\rm U(3)^5$ flavour symmetric limit of the SMEFT. This leads to the redefinition of the total width $\Gamma_W$
\begin{align}
\bar{\Gamma}_W = \frac{3\sqrt{2}\hat G_F \hat{m}_W^3}{4\pi}\left(1+\frac{4}{3}\delta g^\ell_W + \frac{8}{3}\delta g^q_W -\frac{\delta m_W^2}{2 \hat{m}_W^2}\right).
\end{align}
At leading order in the SM, $\Gamma_W = 3\sqrt{2}\hat G_F \hat{m}_W^3/(4\pi)$ and $\delta \Gamma_W$ is defined by $\bar{\Gamma}_W = \Gamma_W + \delta \Gamma_W$. Here $\hat{m}_W$ is the standard model value of the W-mass at tree level in terms of the input parameters, $\hat{m}_W = c_{\that} \, \hat {m}_Z$.

% subsection the_redefinition_of_the_w_width (end)

\subsection{Redefinitions of Triple-Gauge-Coupling Parameters}
\label{sub:TGCRedefinitions}
The most general $C$ and $P$ even TGCs between two charged vector bosons and a neutral vector boson are described by the Effective Lagrangian \cite{Hagiwara:1986vm}
\bea
\frac{- \mathcal{L}_{TGC}}{g_{VWW}}=i g_{1}^{V} \left( W_{\mu \nu}^{+} W^{- \mu}- W_{\mu \nu}^{-} W^{+ \mu}\right)V^{\nu} + i \kappa_{V}W^{+}_{\mu}W^-_{\nu}V^{\mu \nu} + i \frac{\lambda_{V}}{M^2_W}V^{\mu \nu} W^{+ \rho}_{\nu}W^{-}_{\rho \mu},
\eea
 where V stands for either the photon field A or the Z field, $W^{\pm}_{\mu \nu}=\partial_{\mu} W^{\pm}_{\nu} - \partial_{\nu}W^{\pm}_{\mu}$ and similarly $V_{\mu \nu} = \partial_{\mu} V_{\nu} - \partial_{\nu} V_{\mu}$.\footnote{The explicit minus sign on the left hand side of $\mathcal{L}_{TGC}$ is due to an opposite $\epsilon$ tensor convention in Ref.~\cite{Hagiwara:1986vm}.} In the SM, the overall coupling constants are $g_{AWW}= e$ and $g_{ZWW}= e \cot \theta$ and the TGC are given by $g_1^V=\kappa_V=1$ and $\lambda_V=0$ at tree level.
Going from the SM to the SMEFT, these couplings get redefined by a subset of dimension 6 operators. The complete Lagrangian for the TGC in the SMEFT is then expressed as
\begin{align}
  \frac{-\mathcal{L}_{TGC}^{SMEFT}}{g_{VWW}} & =i \bar{g}_{1}^{V} \left( \mathcal{W}_{\mu \nu}^{+} \mathcal{W}^{- \mu} - \mathcal{W}_{\mu \nu}^{-} \mathcal{W}^{+ \mu}\right)\mathcal{V}^{\nu} +i \bar{\kappa}_{V} \mathcal{W}^{+}_{\mu}\mathcal{W}^-_{\nu}\mathcal{V}^{\mu \nu} + i \frac{\bar{\lambda}_{V}}{\bar{M}^2_W}\mathcal{V}^{\mu \nu} \mathcal{W}^{+ \rho}_{\nu}\mathcal{W}^{-}_{\rho \mu},
\end{align}
where $\mathcal{W}_{\mu}$, $\mathcal{V}_{\mu}$ are the redefined gauge fields. Once again $\mathcal{W}_{\mu \nu}=\partial_\mu \mathcal{W}_{\nu}-\partial_\nu \mathcal{W}_{\mu}$, $\mathcal{V}_{\mu \nu} =\partial_\mu \mathcal{V}_{\nu}-\partial_\nu \mathcal{V}_{\mu}$, the coupling constants are $g_{AWW} =\hat{e} = \hat{g}_2 s_{\that}=\sqrt{4 \pi \hat{\alpha}}$, $g_{ZWW}=\hat{e} \cot \that=\sqrt{4 \pi \hat{\alpha}} c_{\that}/s_{\that}$ and $\bar{g}_1^V = g_1^V+ \delta g_1^V$, $\bar{\kappa}_V= \kappa_V + \delta \kappa_V$, $\bar{\lambda}_V = \lambda_V+\delta \lambda_V$ are the redefined TGC's, given by
\begin{subequations}
\begin{align*}
\delta g_1^{A}  &= 0,  & \delta g_1^Z &=  \frac{1}{ 2 \sqrt{2}\hat{G}_F}\left(\frac{s_\that}{c_{\that}}+\frac{c_{\that}}{s_{\that}}  \right) C_{HWB} +
\frac{1}{2}\delta s_{\theta}^2\left(\frac{1}{s_{\that}^2}+\frac{1}{c_{\that}^2}\right), \\
\delta \kappa_A &=   \frac{1}{ \sqrt{2}\hat{G}_F}\frac{c_{\that}}{s_{\that}} C_{HWB},
& \delta \kappa_Z &=\frac{1}{ 2 \sqrt{2}\hat{G}_F}\left(- \frac{s_\that}{c_{\that}}+\frac{c_{\that}}{s_{\that}}  \right)C_{HWB} +  \frac{1}{2}\delta s_{\theta}^2\left(\frac{1}{s_{\that}^2}+\frac{1}{c_{\that}^2}\right),  \\
\delta \lambda_{A} &=  6 s_{\that}  \frac{\hat{m}^2_W}{g_{AWW}}C_W,  & \delta \lambda_{Z} &=  6 c_{\that}  \frac{\hat{m}^2_W}{g_{ZWW}}C_W.
\end{align*}
\end{subequations}
Notice that three gauge-invariance conditions (at the level of $\mathcal{L}_6$ \cite{Hagiwara:1993ck}) hold in the SMEFT: $\delta \kappa_Z = \delta g_1^Z - t_{\that}^2 \delta \kappa_A$, $\delta \lambda_A = \delta \lambda_Z$ and $\delta g_1^A = 0$.

\section{Parametrisation of Phase Space} \label{Parametrisation}
\begin{figure}[t]
\begin{subfigure}{0.45 \textwidth}
\begin{tikzpicture}

\draw [dashed] [->] (-2.5,0) -- (2.5,0);
\draw [dashed] [->] (0,-2) -- (0,2);
\draw [->] (0,0) -- (1.5,0);
\draw [->] (0,0) -- (-1.5,0);
\draw [->] (0,0) -- (-1,1.5);
\draw [->] (0,0) -- (1,-1.5);

\small \node [above] at (0.9,0) {$W^+ (k_{12})$};
\small \node [above] at (-1.1,0) {$W^-(k_{34})$};
\node [below] at (1,-1.5) {$e^- (k_{-})$};
\node [above] at (-1,1.5) {$e^+ (k_{+})$};
\node [below] at (2.5,0) {$z_{COM}$};
\node [above] at (0,2) {$x_{COM}$};
\node [right] at (0.5,-0.4) {${\theta}$};

\draw (0.5,0) arc [radius = 0.5, start angle =30, end angle=-48];

\end{tikzpicture}
\caption{}

\end{subfigure}
\begin{subfigure} {0.45 \textwidth}
\begin{tikzpicture}

\draw [dashed] [->] (-1.25,0.5) -- (2.5,-1);
\draw [dashed] [->] (-1.25,-0.25) -- (2.5,0.5);
\draw [dashed] [->] (0,-1) -- (0,2.5);

\draw [dotted] [-] (0,0) -- (2,0);
\draw [dotted] [-] (2,0) -- (2,2);
\draw [->] (0,0) -- (2,2);
\draw [->] (0,0) -- (-2,-2);

% \draw [->] (0,0) -- (1.5,0);
% \draw [->] (0,0) -- (-1.5,0);
% \draw [->] (1.5,0) -- (3.0,0.5);
% \draw [->] (1.5,0) -- (0,-0.5);
% \draw [->] (-1.5,0) -- (-3,1);
% \draw [->] (-1.5,0) -- (0,-1);

% \tiny \node [above] at (0.75,0) {$W^+ (k_{12})$};
% \tiny \node [above] at (-0.75,0) {$W^-(k_{34})$};
\node [above] at (0,2.5) {$z_{W^+\text{\emph{Rest}}}$};
\node [right] at (2.5,-1) {$x_{W^+\text{\emph{Rest}}}$};
\node [right] at (2.5,0.5) {$y_{W^+\text{\emph{Rest}}}$};
\node [below] at (1.85,0) {$\tilde{\phi}_{12}$};
 \node [right] at (0,1.4) {$\tilde{\theta}_{12}$};
% \tiny \node [above] at (-2.3,0) {$\tilde{\theta}_{34}$};
\node [right] at (2,2) {$f_1 (p_1)$};
\node [right] at (-1.75,-2) {$\bar{f}_2 (p_2)$};
% \node [right] at (0,-1) {$\bar{f}_4(p_4)$};
% \node [left] at (-3,1) {$f_3(p_3)$};

\draw (1,0) arc [radius = 2.23, start angle =0, end angle=-10];
\draw (0,1) arc [radius = 1.2, start angle =90, end angle=51];
% \draw (-1.9,0) arc [radius = 0.9, start angle =-161, end angle=-180];

\end{tikzpicture}
\caption{}
\end{subfigure}

\caption{\label{fig:coordinate_system} Parametrisation of the phase space. Fig. (a) shows the definition of $\theta$as the angle between the $W^+$ and the $e^-$ momenta in the Center of Mass (COM) frame. Fig. (b) defines $\tilde \phi_{12}$ and $\tilde \theta_{12}$ in the $W^+$ rest frame. $\tilde \phi_{34}$ and $\tilde \theta_{34}$ are defined in the $W^-$-rest frame
in a similar manner. We take the $z$-axes to be aligned in all three coordinate systems. }
\end{figure}
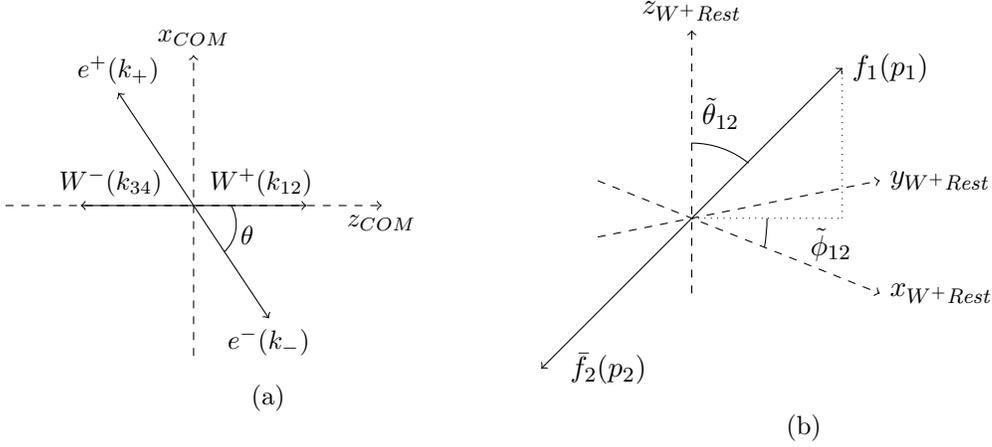

The parametrisation of the phase space shown in Figure \ref{fig:coordinate_system}. Recall that we calculate in the massless fermion limit. The parameterization is given by
$p_i^{\mu} =\left(E_i, \vec{p}_i \right)$ with $E_i = |\vec{p}_i|\label{pi}$ for $i=1 \cdots 4$ and
\begin{align*}
k_{-}^{\mu} &= |\vec{k}_{-}| \left(1,-\sin \theta,0,\cos \theta \right),  &
k_{+}^{\mu}  &= |\vec{k}_{+}| \left(1,\sin \theta,0,- \cos \theta \right),   \\
k_{12}^{\mu}  &= p_1 + p_2 = \left(E_{12}, 0, 0, p\right),  &
k_{34}^{\mu}  &= p_3 + p_4 = \left(E_{34}, 0, 0, - p\right),
\end{align*}
while $E_{-} = E_{+}= |\vec{k}_{-}| = |\vec{k}_+|= \sqrt{s}/2 =m/2, \label{E+-}$ and $k_{+-}^{\mu} = k_{+}^{\mu}  + k_{-}^{\mu}  = \left(m,0,0,0\right)\label{k+-}$.
The $W^+(k_{12})$ and  $W^-(k_{34})$ energy and momentum are
\begin{align*}
E_{12}&= E_1+E_2 = \frac{1}{2\sqrt{s}}\left(s + s_{12} - s_{34}\right), \label{E12} &
E_{34}&= E_3+E_4 = \frac{1}{2\sqrt{s}}\left(s + s_{34} - s_{12}\right),
\end{align*}
while $ p = |\vec{p}_1+\vec{p}_2| = -  |\vec{p}_3+\vec{p}_4| = \frac{1}{2 \sqrt{s}} \lambda^{1/2} \left(\sqrt{s},\sqrt{s_{12}},\sqrt{s_{34}}\right)$.
Here $\lambda$ is the usual K�ll�n function, given by
\begin{align*}
    \lambda \left(\sqrt{s},\sqrt{s_{12}},\sqrt{s_{34}}\right) &= \left[s - \left(\sqrt{s_{12}} + \sqrt{s_{34}}\right)^2\right] \left[s - \left(\sqrt{s_{12}} - \sqrt{s_{34}}\right)^2\right] \nonumber \\
    &= s^2 + s_{12}^2 + s_{34}^2 - 2s_{12}s -2s_{34}s - 2s_{12}s_{34}.
\end{align*}
In the $W^+$ and $W^-$ rest frames respectively the fermion momenta are defined as
\begin{align*}
\tilde{p}_1^{\mu}  &= |\vec{\tilde{p}}_1|\left(1,\sin \tilde{\theta}_{12} \cos \tilde{\phi}_{12}, \sin \tilde{\theta}_{12} \sin \tilde{\phi}_{12}, \cos \tilde{\theta}_{12}\right),  \\
\tilde{p}_3^{\mu}  &= |\vec{\tilde{p}}_3| \left(1,-\sin \tilde{\theta}_{34} \cos \tilde{\phi}_{34},- \sin \tilde{\theta}_{34} \sin \tilde{\phi}_{34},- \cos \tilde{\theta}_{34}\right),
\end{align*}
while $\vec{\tilde{p}}_2  = - \vec{\tilde{p}}_1, \hspace{0.5cm} \vec{\tilde{p}}_4 = - \vec{\tilde{p}}_3, \label{inRF}$ and
$|\vec{\tilde{p}}_1| =  |\vec{\tilde{p}}_2| = \sqrt{s_{12}}/2$ and $|\vec{\tilde{p}}_3| =  |\vec{\tilde{p}}_4| = \sqrt{s_{34}}/2$.
%or equivalently
%\bea
%\tilde{k}_{12}&=& \left(\sqrt{s_{12}},0,0,0\right) = \left(m_{12},0,0,0\right) \\
%\tilde{k}_{34}&=& \left(\sqrt{s_{34}},0,0,0\right)= \left(m_{34},0,0,0\right).
%\eea
The Boson invariants are given by
\begin{align*}
s_{12} &= \left(p_1 + p_2\right)^2 = 2 p_1 \cdot p_2 = E_{12}^2 - p^2= 4 |\vec{\tilde{p}}_1| |\vec{\tilde{p}}_2|,  \\
s_{34} &= \left(p_3 + p_4\right)^2 = 2 p_3 \cdot p_4 = E_{34}^2 - p^2 = 4 |\vec{\tilde{p}}_3| |\vec{\tilde{p}}_4|,
\end{align*}
where $s=\left(k_{+} + k_{-}\right)^2 = 2 k_+ \cdot k_-$ in the massless fermion limit.
\section{Spinor helicity conventions} \label{SpinorHelicity}
Here we define our spinor decomposition and polarization vector conventions using the formalism of Ref.\cite{Hagiwara:1986vm}.
We use the chiral basis of the Dirac matrices and write $\Psi$ in terms of its Weyl components $\Psi^T = \{\Psi_L, \Psi_R\}$
with
\begin{align*}
u \left(p, \lambda \right)_{R/L} &= \omega_{\pm \lambda}\left(p\right) \chi_{\lambda}\left(p\right), &
v \left(p, \lambda \right)_{R/L} &= \pm \lambda \, \omega_{\mp \lambda} \left(p\right) \chi_{-\lambda}\left(p\right),
\end{align*}
and the decomposition
\bea
\chi_{+}\left(p\right)= \left[2 |\vec{p}| \left(|\vec{p}| + p^3\right)\right]^{-1/2} \begin{pmatrix} |\vec{p}| + p^3 \\ p^1 + i p^2 \end{pmatrix}, \quad
\chi_{-}\left(p\right)= \left[2 |\vec{p}| \left(|\vec{p}| + p^3\right)\right]^{-1/2} \begin{pmatrix} -p^1 + i p^2 \\ |\vec{p}| + p^3   \end{pmatrix}. \nonumber
\eea
in terms of the helicity Eigenvalues $\omega_{\pm}\left(p\right) = \left(E \pm |\vec{p}|\right)^{1/2}$.
We define a polarization basis for the polarization vectors of $W^+ \left(k_{12},\lambda\right)$ with $k_{12} = \left(E_{12}, 0, 0, p\right)$
\begin{align*}
\epsilon^{\mu}\left(k_{12},\lambda_{12}=0\right) &= \frac{1}{\sqrt{s_{12}}} \left(p, 0, 0, E_{12}\right), &
\epsilon^{\mu}\left(k_{12},\lambda_{12}=L\right) &= \frac{1}{\sqrt{s_{12}}} \left(E_{12}, 0, 0, p\right), \\
\epsilon^{\mu}\left(k_{12},\lambda_{12}=+\right) &=  \frac{1}{\sqrt{2}} \left(0,-1,-i,0\right),&
\epsilon^{\mu}\left(k_{12},\lambda_{12}=-\right) &=  \frac{1}{\sqrt{2}} \left(0,1,-i,0\right),
\end{align*}
and for the $W^-\left(k_{34},\lambda\right)$ with $k_{34} = \left(E_{34}, 0, 0, -p\right)$
\begin{align*}
\epsilon^{\mu}\left(k_{34},\lambda_{34}=0\right) &= \frac{1}{\sqrt{s_{34}}} \left(p, 0, 0, -E_{34}\right), &
\epsilon^{\mu}\left(k_{34},\lambda_{34}=L\right) &= \frac{1}{\sqrt{s_{34}}} \left(E_{34}, 0, 0, -p\right),\\
\epsilon^{\mu}\left(k_{34},\lambda_{34}=-\right) &=  \frac{1}{\sqrt{2}} \left(0,-1,-i,0\right), &
\epsilon^{\mu}\left(k_{34},\lambda_{34}=+\right) &=  \frac{1}{\sqrt{2}} \left(0,1,-i,0\right).
\end{align*}
In the chiral representation of the Dirac matrices we note
\begin{align*}
\slashed{a} &= a_{\mu} \gamma^{\mu} = \begin{pmatrix} 0 & \slashed{a}_{+} \\ \slashed{a}_{-} & 0 \end{pmatrix}, &
\slashed{a}_{\pm} &= \begin{pmatrix} a_0 \pm a_3 & \pm \left(a_1 - i a_2\right) \\ \pm \left(a_1 + i a_2\right) & a_0 \mp a_3 \end{pmatrix}. \nonumber
\end{align*}
The amplitude decomposition in terms of the helicity Eigenstates are given by
\bea
\mathcal{M}_{W^{+} \rightarrow f_1 \bar{f}_2}^{\lambda_{12}\lambda_1 \lambda_2} &=& C \frac{2 \sqrt{2\pi\hat{\alpha}}}{s_{\that}}\bar{g}_{V}^{W,f_1} \sqrt{s_{12}} \chi^{\dagger}_{-}\left(p_1\right) \slashed{\epsilon}_{-}\left(k_{12},\lambda_{12}\right)\chi_{-}\left(p_2\right), \nonumber \\
\mathcal{M}_{W^{-} \rightarrow f_3 \bar{f}_4}^{\lambda_{34}\lambda_3 \lambda_4} &=& C' \frac{2 \sqrt{2\pi\hat{\alpha}}}{s_{\that}} \bar{g}_{V}^{W,f_3}\sqrt{s_{34}} \chi^{\dagger}_{-}\left(p_3\right) \slashed{\epsilon}_{-}\left(k_{34},\lambda_{34}\right)\chi_{-}\left(p_4\right), \nonumber
\eea
where $C/C'$ are the colour factors that are equal to $\{1,\sqrt{3}\}$ for quarks and leptons respectively,
again using notation consistent with  Ref.\cite{Hagiwara:1986vm}. The remaining amplitudes are
\bea
\frac{\mathcal{M}_{e^+e^- \rightarrow W^- W^+, \nu}^{\lambda_{12}\lambda_{34}\lambda_+\lambda_-}}{4\left(2 \pi \hat{\alpha}\right) \left(\bar{g}_V^{W,l}\right)^2}&=&\frac{\sqrt{s}}{s_{\that}^2 \left(-k_{34} + k_{-}\right)^2}     \chi^{\dagger}_{-}\left(k_{+}\right) \slashed{\epsilon}^{*}_{-}\left(k_{12},\lambda_{12} \right) \left(\slashed{q}\right)_{+}\slashed{\epsilon}^{*}_{-}\left(k_{34},\lambda_{34}\right) \chi_{-}\left(k_{-}\right), \nonumber \\
\frac{\mathcal{M}_{e^+e^- \rightarrow W^- W^+, Z}^{\lambda_{12}\lambda_{34}\lambda_+\lambda_-}}{-\sqrt{s} \,g_{Z,eff} \, g_{ZWW} }&=& \left(\bar{g}_L^e  \chi^{\dagger}_{-}\left(k_{+}\right) \left(\slashed{V}\right)_{-} \chi_{-}\left(k_{-}\right) - \bar{g}_R^e \chi^{\dagger}_{+}\left(k_{+}\right) \left(\slashed{V}\right)_{+} \chi_{+}\left(k_{-}\right)\right)\bar{D}\left(s,\hat{m}_Z^2\right), \nonumber \\
\frac{\mathcal{M}_{e^+e^- \rightarrow W^- W^+, \gamma}^{\lambda_{12}\lambda_{34}\lambda_+\lambda_-}}{-\sqrt{4\pi \hat{\alpha}} \, Q_e \, g_{AWW}}&=&\left(\chi^{\dagger}_{-}\left(k_{+}\right) \left(\slashed{V}\right)_{-} \chi_{-}\left(k_{-}\right) - \chi^{\dagger}_{+}\left(k_{+}\right) \left(\slashed{V}\right)_{+} \chi_{+}\left(k_{-}\right)\right) \frac{\sqrt{s}}{s+ i \epsilon}, \nonumber
\eea
where $q^\mu=\left(-k_{34}^\mu+k_{-}^\mu+k_{12}^\mu-k_{+}^\mu\right)/2$ and $\bar{D}\left(s,\hat{m}_Z^2\right) =1/({s-\hat{m}_Z^2 + i \bar{\Gamma}_Z \hat{m}_Z + i \epsilon})$.
Using the shorthand notations $\epsilon^{*}\left(k_{34}, \lambda_{34}\right)=\epsilon^*_{34,\lambda_{34}}$ and $\epsilon^{*}\left(k_{12},\lambda_{12}\right)=\epsilon^*_{12,\lambda_{12}}$, and $\slashed{V}$ can be written as
\bea
\slashed{V}&=&- \slashed{\epsilon}^{*}_{34,\lambda_{34}} \left[ \bar{g}_1^V + \bar{\kappa}_V  +\frac{\bar{\lambda}_V}{\bar{m}_W^2} s_{12}\right]\left(k_{34} \cdot \epsilon^*_{12,\lambda_{12}}\right)
+ \slashed{\epsilon}^{*}_{12,\lambda_{12}} \left[\bar{g}_1^V + \bar{\kappa}_V+ \frac{\bar{\lambda}_V}{\bar{m}_W^2}s_{34}\right]\left(k_{12} \cdot \epsilon^*_{34,\lambda_{34}}\right), \nonumber \\
&-& \slashed{k}_{12} \left[\left(\bar{g}_1^V   +\frac{s-s_{12}+s_{34}}{2}\frac{\bar{\lambda}_V}{\bar{m}_W^2} \right)\left(\epsilon^*_{12,\lambda_{12}}\cdot \epsilon^*_{34,\lambda_{34}}\right)
- \frac{\bar{\lambda}_V}{\bar{m}_W^2}\left(k_{34} \cdot \epsilon^*_{12,\lambda_{12}}\right)\left(k_{12}\cdot \epsilon^*_{34,\lambda_{34}}\right)  \right], \nonumber \\
&+& \slashed{k}_{34} \left[\left(\bar{g}_1^V + \frac{s+s_{12}-s_{34}}{2} \frac{\bar{\lambda}_V}{\bar{m}_W^2}\right)\left(\epsilon^*_{12,\lambda_{12}}\cdot \epsilon^*_{34,\lambda_{34}},\right)
- \frac{\bar{\lambda}_V}{\bar{m}_W^2}\left(k_{34} \cdot \epsilon^*_{12,\lambda_{12}}\right)\left(k_{12}\cdot \epsilon^*_{34,\lambda_{34}}\right) \right]. \nonumber
\eea
In the fermions massless limit $\omega_{-}\left(p\right) = 0$ and $\omega_{+}\left(p\right) = \sqrt{2E} = \sqrt{2p^0}$ so that in this limit we note
\begin{align*}
\mathcal{M}_{W^{+} \rightarrow f_1 \bar{f}_2}^{\lambda_{12}} = \mathcal{M}_{W^{+} \rightarrow f_1 \bar{f}_2}^{\lambda_{12}-+}, & \hspace{2cm}
\mathcal{M}_{W^{-} \rightarrow f_3 \bar{f}_4}^{\lambda_{34}} = \mathcal{M}_{W^{-} \rightarrow f_3 \bar{f}_4}^{\lambda_{34}- +}
\end{align*}
for simplification while the dependence on $\lambda_+$, $\lambda_-$ is kept as a superscript for the $ee\rightarrow WW$ sub-amplitudes.
\begin{center}
\begin{table}
\tiny
\centering
\tabcolsep 8pt
\begin{tabular}{|c|c|c|}
\hline
$\lambda_{12}$ &$\lambda_{34}$& $\left(\mathcal{M}_{e^+ e^- \rightarrow W^+ W^-,\nu}^{\lambda_{12}\lambda_{34}+-}\right)/\left(2\pi\hat{\alpha} \left(\bar{g}_V^{l,W}\right)^2\right)$   \\ 
\hline \hline
$0$&$0$&$\frac{2 \sin\theta}{s_{\that}^2  \sqrt{s_{12}}\sqrt{s_{34}} \lambda^{1/2}\left(\sqrt{s},\sqrt{s_{12}},\sqrt{s_{34}}\right) }\left(\left(s^2-\left(s_{12}-s_{34}\right)^2\right) - \dfrac{8 s s_{12} s_{34}}{s - s_{12} - s_{34} + \lambda^{1/2}\left(\sqrt{s},\sqrt{s_{12}},\sqrt{s_{34}}\right) \cos \theta}\right)$  \\
$+$&$+$&$- \frac{4 \sin\theta}{s_{\that}^2  \lambda^{1/2}\left(\sqrt{s},\sqrt{s_{12}},\sqrt{s_{34}}\right)}\left(s  + \dfrac{-s\left(s_{12}+ s_{34}\right)-\left(s_{12}-s_{34}\right)\left(-s_{12}+s_{34} + \lambda^{1/2}\left(\sqrt{s},\sqrt{s_{12}},\sqrt{s_{34}}\right) \right)}{s - s_{12} - s_{34} + \lambda^{1/2}\left(\sqrt{s},\sqrt{s_{12}},\sqrt{s_{34}}\right) \cos \theta}\right)$ \\ %$- \frac{4\left(2\pi\hat{\alpha}\right)\left(\bar{g}_V^{l,W}\right)^2}{s_{\that}^2  \lambda^{1/2}\left(\sqrt{s},\sqrt{s_{12}},\sqrt{s_{34}}\right)}\sin\theta\left(s  + \dfrac{-s\left(s_{12}+ s_{34}\right)+\left(s_{12}-s_{34}\right)\left(s_{12}-s_{34} + \lambda^{1/2}\left(\sqrt{s},\sqrt{s_{12}},\sqrt{s_{34}}\right) \right)}{s - s_{12} - s_{34} + \lambda^{1/2}\left(\sqrt{s},\sqrt{s_{12}},\sqrt{s_{34}}\right) \cos \theta}\right)$\\
$-$&$-$&$- \frac{4 \sin\theta}{s_{\that}^2  \lambda^{1/2}\left(\sqrt{s},\sqrt{s_{12}},\sqrt{s_{34}}\right)}\left(s  + \dfrac{-s\left(s_{12}+ s_{34}\right)+\left(s_{12}-s_{34}\right)\left(s_{12}-s_{34} + \lambda^{1/2}\left(\sqrt{s},\sqrt{s_{12}},\sqrt{s_{34}}\right) \right)}{s - s_{12} - s_{34} + \lambda^{1/2}\left(\sqrt{s},\sqrt{s_{12}},\sqrt{s_{34}}\right) \cos \theta}\right)$ \\ %$- \frac{4\left(2\pi\hat{\alpha}\right)\left(\bar{g}_V^{l,W}\right)^2}{s_{\that}^2  \lambda^{1/2}\left(\sqrt{s},\sqrt{s_{12}},\sqrt{s_{34}}\right)}\sin\theta\left(s  + \dfrac{-s\left(s_{12}+ s_{34}\right)-\left(s_{12}-s_{34}\right)\left(-s_{12}+s_{34} + \lambda^{1/2}\left(\sqrt{s},\sqrt{s_{12}},\sqrt{s_{34}}\right) \right)}{s - s_{12} - s_{34} + \lambda^{1/2}\left(\sqrt{s},\sqrt{s_{12}},\sqrt{s_{34}}\right) \cos \theta}\right)$\\
$0$&$-$&$-\frac{4 \left(1-\cos \theta\right) \sqrt{s}}{s_{\that}^2 \sqrt{2}\sqrt{s_{12}}\lambda^{1/2}\left(\sqrt{s},\sqrt{s_{12}},\sqrt{s_{34}}\right)}\left(\left(s+s_{12}-s_{34}\right) - \dfrac{2s_{12} \left(s- s_{12}+ s_{34} - \lambda^{1/2}\left(\sqrt{s},\sqrt{s_{12}},\sqrt{s_{34}}\right) \right)}{s - s_{12} - s_{34} + \lambda^{1/2}\left(\sqrt{s},\sqrt{s_{12}},\sqrt{s_{34}}\right) \cos \theta}\right)$\\%$\frac{4\left(2\pi\hat{\alpha}\right)\left(\bar{g}_V^{l,W}\right)^2 \sqrt{s}}{s_{\that}^2 \sqrt{2}\sqrt{s_{12}}\lambda^{1/2}\left(\sqrt{s},\sqrt{s_{12}},\sqrt{s_{34}}\right)}\left(1+\cos \theta\right)\left(\left(s+s_{12}-s_{34}\right) - \dfrac{2s_{12} \left(s- s_{12}+ s_{34} + \lambda^{1/2}\left(\sqrt{s},\sqrt{s_{12}},\sqrt{s_{34}}\right) \right)}{s - s_{12} - s_{34} + \lambda^{1/2}\left(\sqrt{s},\sqrt{s_{12}},\sqrt{s_{34}}\right) \cos \theta}\right)$\\
$0$&$+$&$-\frac{4 \left(1+\cos \theta\right)\sqrt{s}}{s_{\that}^2 \sqrt{2}\sqrt{s_{12}}\lambda^{1/2}\left(\sqrt{s},\sqrt{s_{12}},\sqrt{s_{34}}\right)}\left(\left(s+s_{12}-s_{34}\right) - \dfrac{2s_{12} \left(s- s_{12}+ s_{34} + \lambda^{1/2}\left(\sqrt{s},\sqrt{s_{12}},\sqrt{s_{34}}\right) \right)}{s - s_{12} - s_{34} + \lambda^{1/2}\left(\sqrt{s},\sqrt{s_{12}},\sqrt{s_{34}}\right) \cos \theta}\right)$\\%$\frac{4\left(2\pi\hat{\alpha}\right)\left(\bar{g}_V^{l,W}\right)^2 \sqrt{s}}{s_{\that}^2 \sqrt{2}\sqrt{s_{12}}\lambda^{1/2}\left(\sqrt{s},\sqrt{s_{12}},\sqrt{s_{34}}\right)}\left(1-\cos \theta\right)\left(\left(s+s_{12}-s_{34}\right) - \dfrac{2s_{12} \left(s- s_{12}+ s_{34} - \lambda^{1/2}\left(\sqrt{s},\sqrt{s_{12}},\sqrt{s_{34}}\right) \right)}{s - s_{12} - s_{34} + \lambda^{1/2}\left(\sqrt{s},\sqrt{s_{12}},\sqrt{s_{34}}\right) \cos \theta}\right)$\\
$+$&$0$&$\frac{4 \left(1-\cos \theta\right)\sqrt{s}}{s_{\that}^2 \sqrt{2}\sqrt{s_{34}}\lambda^{1/2}\left(\sqrt{s},\sqrt{s_{12}},\sqrt{s_{34}}\right)}\left(\left(s-s_{12}+s_{34}\right) - \dfrac{2s_{34} \left(s+ s_{12}- s_{34} - \lambda^{1/2}\left(\sqrt{s},\sqrt{s_{12}},\sqrt{s_{34}}\right) \right)}{s - s_{12} - s_{34} + \lambda^{1/2}\left(\sqrt{s},\sqrt{s_{12}},\sqrt{s_{34}}\right) \cos \theta}\right)$\\ %$\frac{4\left(2\pi\hat{\alpha}\right)\left(\bar{g}_V^{l,W}\right)^2 \sqrt{s}}{s_{\that}^2 \sqrt{2}\sqrt{s_{34}}\lambda^{1/2}\left(\sqrt{s},\sqrt{s_{12}},\sqrt{s_{34}}\right)}\left(1+\cos \theta\right)\left(\left(s-s_{12}+s_{34}\right) - \dfrac{2s_{34} \left(s+ s_{12}- s_{34} + \lambda^{1/2}\left(\sqrt{s},\sqrt{s_{12}},\sqrt{s_{34}}\right) \right)}{s - s_{12} - s_{34} + \lambda^{1/2}\left(\sqrt{s},\sqrt{s_{12}},\sqrt{s_{34}}\right) \cos \theta}\right)$\\
$-$&$0$&$\frac{4  \left(1+\cos \theta\right)\sqrt{s}}{s_{\that}^2 \sqrt{2}\sqrt{s_{34}}\lambda^{1/2}\left(\sqrt{s},\sqrt{s_{12}},\sqrt{s_{34}}\right)}\left(\left(s-s_{12}+s_{34}\right) - \dfrac{2s_{34} \left(s+ s_{12}- s_{34} + \lambda^{1/2}\left(\sqrt{s},\sqrt{s_{12}},\sqrt{s_{34}}\right) \right)}{s - s_{12} - s_{34} + \lambda^{1/2}\left(\sqrt{s},\sqrt{s_{12}},\sqrt{s_{34}}\right) \cos \theta}\right)$\\%$\frac{4\left(2\pi\hat{\alpha}\right)\left(\bar{g}_V^{l,W}\right)^2 \sqrt{s}}{s_{\that}^2 \sqrt{2}\sqrt{s_{34}}\lambda^{1/2}\left(\sqrt{s},\sqrt{s_{12}},\sqrt{s_{34}}\right)}\left(1-\cos \theta\right)\left(\left(s-s_{12}+s_{34}\right) - \dfrac{2s_{34} \left(s+ s_{12}- s_{34} - \lambda^{1/2}\left(\sqrt{s},\sqrt{s_{12}},\sqrt{s_{34}}\right) \right)}{s - s_{12} - s_{34} + \lambda^{1/2}\left(\sqrt{s},\sqrt{s_{12}},\sqrt{s_{34}}\right) \cos \theta}\right)$\\
$+$&$-$&$- \frac{4}{s_{\that}^2}\dfrac{s\sin\theta\left(1-\cos\theta\right)}{s - s_{12} - s_{34} + \lambda^{1/2}\left(\sqrt{s},\sqrt{s_{12}},\sqrt{s_{34}}\right) \cos \theta}$\\ %$\frac{4\left(2\pi\hat{\alpha}\right)\left(\bar{g}_V^{l,W}\right)^2}{s_{\that}^2}\dfrac{s\sin\theta\left(1+\cos\theta\right)}{s - s_{12} - s_{34} + \lambda^{1/2}\left(\sqrt{s},\sqrt{s_{12}},\sqrt{s_{34}}\right) \cos \theta}$\\
$-$&$+$&$\frac{4}{s_{\that}^2}\dfrac{s\sin\theta\left(1+\cos\theta\right)}{s - s_{12} - s_{34} + \lambda^{1/2}\left(\sqrt{s},\sqrt{s_{12}},\sqrt{s_{34}}\right) \cos \theta}$\\ %$- \frac{4\left(2\pi\hat{\alpha}\right)\left(\bar{g}_V^{l,W}\right)^2}{s_{\that}^2}\dfrac{s\sin\theta\left(1-\cos\theta\right)}{s - s_{12} - s_{34} + \lambda^{1/2}\left(\sqrt{s},\sqrt{s_{12}},\sqrt{s_{34}}\right) \cos \theta}$\\
\hline
\end{tabular}
\caption{ \label{nu-exchange} The $W^\pm$ pair production matrix elements for helicities $\lambda_{12},\lambda_{34}=\{0,+,-\}$.  }
\end{table}
\end{center}
%%%%%%%%%%%%%%%%%%%%%%%%%%%%%%%%%%%%%%%%%%%%%%%
\begin{center}
\begin{table}
\centering
\tabcolsep 8pt
\begin{tabular}{|c|c|c|}
\hline
$\lambda_{12}$ &$\lambda_{34}$& $\mathcal{M}_{e^+ e^- \rightarrow W^+ W^-,V}^{\lambda_{12}\lambda_{34}-+}$   \\ 
\hline \hline
$0$&$0$&$- \frac{ \bar{F}_2^V\left(\bar{g}_1^V\left(s_{12}+s_{34}\right)+\bar{\kappa}_V s\right)\lambda^{1/2}\left(\sqrt{s},\sqrt{s_{12}},\sqrt{s_{34}}\right)\sin \theta \bar{D}^{V}\left(s\right)}{2\sqrt{s_{12}}\sqrt{s_{34}}}$  \\
$+$&$+$&$\frac{\bar{F}_2^V \left(2 \bar{g}_1^V \bar{M}_W^2 +\bar{\lambda}_V s\right)\lambda^{1/2}\left(\sqrt{s},\sqrt{s_{12}},\sqrt{s_{34}}\right)\sin\theta \bar{D}^{V}\left(s\right)}{2\bar{M}_W^2}$\\
$-$&$-$&$\frac{\bar{F}_2^V \left(2 \bar{g}_1^V \bar{M}_W^2 +\bar{\lambda}_V s\right)\lambda^{1/2}\left(\sqrt{s},\sqrt{s_{12}},\sqrt{s_{34}}\right)\sin\theta \bar{D}^{V}\left(s\right)}{2\bar{M}_W^2}$\\
$0$&$-$&$-\frac{\sqrt{s}\lambda^{1/2}\left(\sqrt{s},\sqrt{s_{12}},\sqrt{s_{34}}\right)\left(\bar{F}_2^V\cos \theta  +\bar{F}_2^V \right) \left(\bar{g}_1^V \bar{M}_W^2 +\bar{\kappa}_V \bar{M}_W^2 +\bar{\lambda}_V s_{12}\right)\bar{D}^{V}\left(s\right)}{2\sqrt{2}\sqrt{s_{12}}\bar{M}_W^2}$ \\
$0$&$+$&$\frac{\sqrt{s}\lambda^{1/2}\left(\sqrt{s},\sqrt{s_{12}},\sqrt{s_{34}}\right)\left(\bar{F}_2^V\cos \theta  -\bar{F}_2^V \right) \left(\bar{g}_1^V \bar{M}_W^2 +\bar{\kappa}_V \bar{M}_W^2 +\bar{\lambda}_V s_{12}\right)\bar{D}^{V}\left(s\right)}{2\sqrt{2}\sqrt{s_{12}}\bar{M}_W^2}$  \\
$+$&$0$&$\frac{\sqrt{s}\lambda^{1/2}\left(\sqrt{s},\sqrt{s_{12}},\sqrt{s_{34}}\right)\left(\bar{F}_2^V\cos \theta+\bar{F}_2^V \right) \left(\bar{g}_1^V \bar{M}_W^2 +\bar{\kappa}_V \bar{M}_W^2 +\bar{\lambda}_V s_{34}\right)\bar{D}^{V}\left(s\right)}{2\sqrt{2}\sqrt{s_{34}}\bar{M}_W^2}$ \\
$-$&$0$&$-\frac{\sqrt{s}\lambda^{1/2}\left(\sqrt{s},\sqrt{s_{12}},\sqrt{s_{34}}\right)\left(\bar{F}_2^V\cos \theta  -\bar{F}_2^V \right) \left(\bar{g}_1^V \bar{M}_W^2 +\bar{\kappa}_V \bar{M}_W^2 +\bar{\lambda}_V s_{34}\right)\bar{D}^{V}\left(s\right)}{2\sqrt{2}\sqrt{s_{34}}\bar{M}_W^2}$  \\
$+$&$-$&$0$\\
$-$&$+$&$0$\\
\hline
\end{tabular}
\caption{ \label{V-exchange-+}The W-production matrix elements $\mathcal{M}_{e^+ e^- \rightarrow W^+ W^-,V-exchange}^{\lambda_{12},\lambda_{34},-,+}$  for $\lambda_{12},\lambda_{34}=\{0,+,-\}$ with in our notations: $\bar{F}_1^Z = -g_{Z,eff} . g_{ZWW}. \bar{g}_L^e$, $\bar{F}_2^Z = -g_{Z,eff}. g_{ZWW} .\bar{g}_R^e$, $\bar{F}_1^{A} = \bar{F}_2^{A}=\sqrt{4\pi \hat{\alpha}}g_{AWW}$  and $\bar{D}^Z\left(s\right) = \bar{D}\left(s,\hat{m}_Z^2\right)$ and $\bar{D}^A\left(s\right) =1/s$. }
\end{table}
\end{center}
%%%%%%%%%%%%%%%%%%%%%%%%%%%%%%%%%%%%%%%%%%%%%%%%%
\begin{center}
\begin{table}
\centering
\tabcolsep 8pt
\begin{tabular}{|c|c|c|}
\hline
$\lambda_{12}$ &$\lambda_{34}$& $\mathcal{M}_{e^+ e^- \rightarrow W^+ W^-,V}^{\lambda_{12}\lambda_{34}+-}$   \\ 
\hline \hline
$0$&$0$&$- \frac{\bar{F}_1^V\left(\bar{g}_1^V\left(s_{12}+s_{34}\right)+\bar{\kappa}_V s\right)\lambda^{1/2}\left(\sqrt{s},\sqrt{s_{12}},\sqrt{s_{34}}\right)\sin \theta \bar{D}^{V}\left(s\right)}{2\sqrt{s_{12}}\sqrt{s_{34}}}$  \\
$+$&$+$&$\frac{\bar{F}_1^V \left(2 \bar{g}_1^V \bar{M}_W^2 +\bar{\lambda}_V s\right)\lambda^{1/2}\left(\sqrt{s},\sqrt{s_{12}},\sqrt{s_{34}}\right)\sin\theta \bar{D}^{V}\left(s\right)}{2\bar{M}_W^2}$\\
$-$&$-$&$\frac{\bar{F}_1^V \left(2 \bar{g}_1^V \bar{M}_W^2 +\bar{\lambda}_V s\right)\lambda^{1/2}\left(\sqrt{s},\sqrt{s_{12}},\sqrt{s_{34}}\right)\sin\theta \bar{D}^{V}\left(s\right)}{2\bar{M}_W^2}$\\
$0$&$-$&$-\frac{\sqrt{s}\lambda^{1/2}\left(\sqrt{s},\sqrt{s_{12}},\sqrt{s_{34}}\right)\left(\bar{F}_1^V\cos \theta - \bar{F}_1^V\right) \left(\bar{g}_1^V \bar{M}_W^2 +\bar{\kappa}_V \bar{M}_W^2 +\bar{\lambda}_V s_{12}\right)\bar{D}^{V}\left(s\right)}{2\sqrt{2}\sqrt{s_{12}}\bar{M}_W^2}$ \\
$0$&$+$&$\frac{\sqrt{s}\lambda^{1/2}\left(\sqrt{s},\sqrt{s_{12}},\sqrt{s_{34}}\right)\left(\bar{F}_1^V\cos \theta + \bar{F}_1^V  \right) \left(\bar{g}_1^V \bar{M}_W^2 +\bar{\kappa}_V \bar{M}_W^2 +\bar{\lambda}_V s_{12}\right)\bar{D}^{V}\left(s\right)}{2\sqrt{2}\sqrt{s_{12}}\bar{M}_W^2}$  \\
$+$&$0$&$\frac{\sqrt{s}\lambda^{1/2}\left(\sqrt{s},\sqrt{s_{12}},\sqrt{s_{34}}\right)\left(\bar{F}_1^V\cos \theta - \bar{F}_1^V \right) \left(\bar{g}_1^V \bar{M}_W^2 +\bar{\kappa}_V \bar{M}_W^2 +\bar{\lambda}_V s_{34}\right)\bar{D}^{V}\left(s\right)}{2\sqrt{2}\sqrt{s_{34}}\bar{M}_W^2}$ \\
$-$&$0$&$-\frac{\sqrt{s}\lambda^{1/2}\left(\sqrt{s},\sqrt{s_{12}},\sqrt{s_{34}}\right)\left(\bar{F}_1^V\cos \theta + \bar{F}_1^V  \right) \left(\bar{g}_1^V \bar{M}_W^2 +\bar{\kappa}_V \bar{M}_W^2 +\bar{\lambda}_V s_{34}\right)\bar{D}^{V}\left(s\right)}{2\sqrt{2}\sqrt{s_{34}}\bar{M}_W^2}$  \\
$+$&$-$&$0$\\
$-$&$+$&$0$\\
\hline
\end{tabular}
\caption{ \label{V-exchange+-}The W-production matrix elements $\mathcal{M}_{e^+ e^- \rightarrow W^+ W^-,V-exchange}^{\lambda_{12},\lambda_{34},+,-}$ for $\lambda_{12},\lambda_{34}=\{0,+,-\}$ with in our notations: $\bar{F}_1^Z = -g_{Z,eff} . g_{ZWW}. \bar{g}_L^e$, $\bar{F}_2^Z = -g_{Z,eff}. g_{ZWW} .\bar{g}_R^e$, $\bar{F}_1^{A} = \bar{F}_2^{A}=\sqrt{4\pi \hat{\alpha}}g_{AWW}$  and $\bar{D}^Z\left(s\right) = \bar{D}\left(s,\hat{m}_Z^2\right)$ and $\bar{D}^A\left(s\right) =1/s$. }
\end{table}
\end{center}
%%%%%%%%%%%%%%%%%%%%%%%%%%%%%%%%%%%%%%%%%%%%%%%%%
%%%%%%%%%%%%%%%%%%%%%%%%%%%%%%%%%%%%%%%%%%%%%%%%%%%%%%
\begin{center}
\begin{table}
\centering
\tabcolsep 8pt
\begin{tabular}{|c|c|c|c|c|c|}
\hline
$\sigma$ & $\sqrt{s}$ [GeV] & Experimental value [pb] &Ref.& Theoretical value [pb] &Ref. \\ 
\hline \hline
$\sigma_{\ell \nu \ell \nu}$ &$188.6$&$1.88 \pm 0.16 \pm 0.07$&\cite{Achard:2004zw}&$1.72\left(1\pm 0.5\%\right)$&\cite{Achard:2004zw} \\
&$191.6$&$1.66 \pm 0.39 \pm 0.07$&\cite{Achard:2004zw}&$1.76\left(1\pm 0.5\%\right)$&\cite{Achard:2004zw}\\
&$195.5$&$1.78\pm0.24\pm0.07$&\cite{Achard:2004zw}&$1.79\left(1\pm 0.5\%\right)$&\cite{Achard:2004zw}\\
&$199.6$&$1.75 \pm 0.25 \pm 0.06$&\cite{Achard:2004zw}&$1.80\left(1\pm 0.5\%\right)$&\cite{Achard:2004zw}\\
&$201.8$&$1.51\pm0.34\pm0.07$&\cite{Achard:2004zw}&$1.81\left(1\pm 0.5\%\right)$&\cite{Achard:2004zw}\\
&$204.8$&$1.58\pm 0.24 \pm 0.05$&\cite{Achard:2004zw}&$1.82\left(1\pm 0.5\%\right)$&\cite{Achard:2004zw}\\
&$206.5$&$1.44 \pm 0.18 \pm 0.06$&\cite{Achard:2004zw}&$1.82\left(1\pm 0.5\%\right)$&\cite{Achard:2004zw}\\
&$208.0$&$2.23 \pm 0.86 \pm 0.06$&\cite{Achard:2004zw}&$1.82\left(1\pm 0.5\%\right)$&\cite{Achard:2004zw}\\
\hline
 $\sigma_{\ell \nu q q}$ &$188.6$&$7.19 \pm 0.24 \pm 0.08$&\cite{Achard:2004zw}&$7.14\left(1\pm 0.5\%\right)$& \cite{Achard:2004zw}\\
&$191.6$&$7.69 \pm 0.61 \pm 0.09$&\cite{Achard:2004zw}&$7.26\left(1\pm 0.5\%\right)$&\cite{Achard:2004zw}\\
&$195.5$&$7.58\pm0.36\pm0.08$&\cite{Achard:2004zw}&$7.38\left(1\pm 0.5\%\right)$&\cite{Achard:2004zw}\\
&$199.6$&$6.81 \pm 0.35 \pm 0.08$&\cite{Achard:2004zw}&$7.44\left(1\pm 0.5\%\right)$&\cite{Achard:2004zw}\\
&$201.8$&$7.34\pm0.54\pm0.08$&\cite{Achard:2004zw}&$7.47\left(1\pm 0.5\%\right)$&\cite{Achard:2004zw}\\
&$204.8$&$7.68\pm 0.39 \pm 0.13$&\cite{Achard:2004zw}&$7.50\left(1\pm 0.5\%\right)$&\cite{Achard:2004zw}\\
&$206.5$&$7.60 \pm 0.30 \pm 0.08$&\cite{Achard:2004zw}&$7.50\left(1\pm 0.5\%\right)$&\cite{Achard:2004zw}\\
&$208.0$&$8.18 \pm 1.21 \pm 0.09$&\cite{Achard:2004zw}&$7.50\left(1\pm 0.5\%\right)$&\cite{Achard:2004zw}\\
\hline
 $\sigma_{q q q q}$ &$188.6$&$7.17 \pm 0.24 \pm 0.12$&\cite{Achard:2004zw}&$7.42\left(1\pm 0.5\%\right)$&\cite{Achard:2004zw} \\
&$191.6$&$6.78 \pm 0.56 \pm 0.12$&\cite{Achard:2004zw}&$7.56\left(1\pm 0.5\%\right)$&\cite{Achard:2004zw}\\
&$195.5$&$6.92\pm0.34\pm0.11$&\cite{Achard:2004zw}&$7.68\left(1\pm 0.5\%\right)$&\cite{Achard:2004zw}\\
&$199.6$&$7.91 \pm 0.36 \pm 0.13$&\cite{Achard:2004zw}&$7.76\left(1\pm 0.5\%\right)$&\cite{Achard:2004zw}\\
&$201.8$&$7.09\pm0.52\pm0.12$&\cite{Achard:2004zw}&$7.79\left(1\pm 0.5\%\right)$&\cite{Achard:2004zw}\\
&$204.8$&$7.66\pm 0.37 \pm 0.13$&\cite{Achard:2004zw}&$7.81\left(1\pm 0.5\%\right)$&\cite{Achard:2004zw}\\
&$206.5$&$8.07 \pm 0.29 \pm 0.13$&\cite{Achard:2004zw}&$7.82\left(1\pm 0.5\%\right)$&\cite{Achard:2004zw}\\
&$208.0$&$7.29 \pm 1.16 \pm 0.11$&\cite{Achard:2004zw}&$7.82\left(1\pm 0.5\%\right)$&\cite{Achard:2004zw}\\
\hline
\end{tabular}
\caption{ \label{L3Data} Measured cross section of the process $e^+ e^- \rightarrow \ell \nu \ell \nu$, $e^+ e^- \rightarrow qq \ell \nu$, $e^+ e^- \rightarrow qqqq$ by the L3 collaboration assuming charged-lepton universality. The SM theory error is taken to be 0.5\% of the SM values following \cite{Achard:2004zw}.}
\end{table}
\end{center}
%%%%%%%%%%%%%%%%%%%%%%%%%%%%%%%%%%%%%%%%%%%%%%%%%%%%%%%%
\begin{center}
\begin{table}
\centering
\tabcolsep 8pt
\begin{tabular}{|c|c|c|c|c|c|}
\hline
$\sigma$ & $\sqrt{s}$ [GeV] & Experimental value [pb] &Ref.& Theoretical value [pb] &Ref. \\ 
\hline \hline
$\sigma_{\ell \nu \ell \nu}$ &$188.63$&$1.69 \pm 0.11 \pm 0.02$&\cite{Abbiendi:2007rs}&$1.72\left(1\pm 0.5\%\right)$&\cite{Abbiendi:2007rs} \\
&$191.61$&$2.04 \pm 0.30 \pm 0.02$&\cite{Abbiendi:2007rs}&$1.75\left(1\pm 0.5\%\right)$&\cite{Abbiendi:2007rs}\\
&$195.54$&$2.03\pm0.19\pm0.02$&\cite{Abbiendi:2007rs}&$1.78\left(1\pm 0.5\%\right)$&\cite{Abbiendi:2007rs}\\
&$199.54$&$1.91 \pm 0.18 \pm 0.02$&\cite{Abbiendi:2007rs}&$1.79\left(1\pm 0.5\%\right)$&\cite{Abbiendi:2007rs}\\
&$201.65$&$2.50\pm0.29\pm0.03$&\cite{Abbiendi:2007rs}&$1.80\left(1\pm 0.5\%\right)$&\cite{Abbiendi:2007rs}\\
&$204.88$&$1.82\pm 0.17 \pm 0.02$&\cite{Abbiendi:2007rs}&$1.81\left(1\pm 0.5\%\right)$&\cite{Abbiendi:2007rs}\\
&$206.56$&$1.83 \pm 0.13 \pm 0.02$&\cite{Abbiendi:2007rs}&$1.81\left(1\pm 0.5\%\right)$&\cite{Abbiendi:2007rs}\\
\hline
 $\sigma_{\ell \nu q q}$ &$188.63$&$6.98 \pm 0.22 \pm 0.05$&\cite{Abbiendi:2007rs}&$7.13\left(1\pm 0.5\%\right)$&\cite{Abbiendi:2007rs}\\
&$191.61$&$6.48 \pm 0.54 \pm 0.05$&\cite{Abbiendi:2007rs}&$7.26\left(1\pm 0.5\%\right)$&\cite{Abbiendi:2007rs}\\
&$195.54$&$7.94\pm0.37\pm0.05$&\cite{Abbiendi:2007rs}&$7.38\left(1\pm 0.5\%\right)$&\cite{Abbiendi:2007rs}\\
&$199.54$&$7.01 \pm 0.35 \pm 0.05$&\cite{Abbiendi:2007rs}&$7.46\left(1\pm 0.5\%\right)$&\cite{Abbiendi:2007rs}\\
&$201.65$&$7.39\pm0.51\pm0.05$&\cite{Abbiendi:2007rs}&$7.48\left(1\pm 0.5\%\right)$&\cite{Abbiendi:2007rs}\\
&$204.88$&$6.85\pm 0.33 \pm 0.05$&\cite{Abbiendi:2007rs}&$7.50\left(1\pm 0.5\%\right)$&\cite{Abbiendi:2007rs}\\
&$206.56$&$7.67 \pm 0.27 \pm 0.05$&\cite{Abbiendi:2007rs}&$7.51\left(1\pm 0.5\%\right)$&\cite{Abbiendi:2007rs}\\
\hline
 $\sigma_{q q q q}$ &$188.63$&$7.66 \pm 0.25 \pm 0.12$&\cite{Abbiendi:2007rs}&$7.41\left(1\pm 0.5\%\right)$&\cite{Abbiendi:2007rs}\\
&$191.61$&$7.51 \pm 0.62 \pm 0.12$&\cite{Abbiendi:2007rs}&$7.54\left(1\pm 0.5\%\right)$&\cite{Abbiendi:2007rs}\\
&$195.54$&$8.35\pm0.40\pm0.12$&\cite{Abbiendi:2007rs}&$7.67\left(1\pm 0.5\%\right)$&\cite{Abbiendi:2007rs}\\
&$199.54$&$7.42 \pm 0.38 \pm 0.11$&\cite{Abbiendi:2007rs}&$7.75\left(1\pm 0.5\%\right)$&\cite{Abbiendi:2007rs}\\
&$201.65$&$8.16\pm0.57\pm0.12$&\cite{Abbiendi:2007rs}&$7.77\left(1\pm 0.5\%\right)$&\cite{Abbiendi:2007rs}\\
&$204.88$&$7.40\pm 0.37 \pm 0.11$&\cite{Abbiendi:2007rs}&$7.79\left(1\pm 0.5\%\right)$&\cite{Abbiendi:2007rs}\\
&$206.56$&$8.19 \pm 0.30 \pm 0.12$&\cite{Abbiendi:2007rs}&$7.80\left(1\pm 0.5\%\right)$&\cite{Abbiendi:2007rs}\\
\hline
\end{tabular}
\caption{ \label{OPALData} Measured cross section of the process $e^+ e^- \rightarrow \ell \nu \ell \nu$, $e^+ e^- \rightarrow qq \ell \nu$, $e^+ e^- \rightarrow qqqq$ by the OPAL collaboration assuming charged-lepton universality. The SM theory error is taken to be 0.5\% of the SM values following \cite{Achard:2004zw}.}
\end{table}
\end{center}
%%%%%%%%%%%%%%%%%%%%%%%%%%%%%%%%%%%%%%
\begin{center}
\begin{table}
\centering
\tabcolsep 8pt
\begin{tabular}{|c|c|c|c|c|c|}
\hline
$\sigma$ & $\sqrt{s}$ [GeV] & Experimental value [pb] &Ref.& Theoretical value [pb] &Ref. \\ 
\hline \hline
%$\sigma_{\ell \nu \ell \nu}$ &$182.65$&$1.45 \pm 0.20 \pm 0.02$&\cite{Abbiendi:2007rs}&$1.72\left(1\pm 0.5\%\right)$&\cite{Abbiendi:2007rs} \\
$\sigma_{\ell \nu \ell \nu}$ &$188.63$&$1.78 \pm 0.13 \pm 0.02$&\cite{Heister:2004wr}&$1.75\left(1\pm 0.5\%\right)$&\cite{Abbiendi:2007rs}\\
&$191.58$&$1.45\pm0.29\pm0.02$&\cite{Heister:2004wr}&$1.78\left(1\pm 0.5\%\right)$&\cite{Abbiendi:2007rs}\\
&$195.52$&$1.78 \pm 0.19 \pm 0.02$&\cite{Heister:2004wr}&$1.79\left(1\pm 0.5\%\right)$&\cite{Abbiendi:2007rs}\\
&$199.52$&$1.83\pm0.19\pm0.03$&\cite{Heister:2004wr}&$1.80\left(1\pm 0.5\%\right)$&\cite{Abbiendi:2007rs}\\
&$201.62$&$1.78\pm 0.27 \pm 0.02$&\cite{Heister:2004wr}&$1.81\left(1\pm 0.5\%\right)$&\cite{Abbiendi:2007rs}\\
&$204.86$&$1.51 \pm 0.18 \pm 0.02$&\cite{Heister:2004wr}&$1.81\left(1\pm 0.5\%\right)$&\cite{Abbiendi:2007rs}\\
&$206.53$&$1.69 \pm 0.15 \pm 0.02$&\cite{Heister:2004wr}&$1.81\left(1\pm 0.5\%\right)$&\cite{Abbiendi:2007rs}\\
\hline
 %$\sigma_{\ell \nu q q}$ &$182.65$&$6.82 \pm 0.39 \pm 0.06$&\cite{Abbiendi:2007rs}&$7.13\left(1\pm 0.5\%\right)$&\cite{Abbiendi:2007rs}\\
$\sigma_{\ell \nu q q}$&$188.63$&$7.14 \pm 0.23 \pm 0.06$&\cite{Heister:2004wr}&$7.26\left(1\pm 0.5\%\right)$&\cite{Abbiendi:2007rs}\\
&$191.58$&$7.40\pm0.56\pm0.06$&\cite{Heister:2004wr}&$7.38\left(1\pm 0.5\%\right)$&\cite{Abbiendi:2007rs}\\
&$195.52$&$7.31 \pm 0.34 \pm 0.06$&\cite{Heister:2004wr}&$7.46\left(1\pm 0.5\%\right)$&\cite{Abbiendi:2007rs}\\
&$199.52$&$7.70\pm0.33\pm0.06$&\cite{Heister:2004wr}&$7.48\left(1\pm 0.5\%\right)$&\cite{Abbiendi:2007rs}\\
&$201.62$&$7.92\pm 0.49 \pm 0.06$&\cite{Heister:2004wr}&$7.50\left(1\pm 0.5\%\right)$&\cite{Abbiendi:2007rs}\\
&$204.86$&$7.47 \pm 0.34 \pm 0.06$&\cite{Heister:2004wr}&$7.51\left(1\pm 0.5\%\right)$&\cite{Abbiendi:2007rs}\\
&$206.53$&$7.96 \pm 0.27 \pm 0.06$&\cite{Heister:2004wr}&$7.51\left(1\pm 0.5\%\right)$&\cite{Abbiendi:2007rs}\\
\hline
 %$\sigma_{q q q q}$ &$182.65$&$7.57 \pm 0.42 \pm 0.09$&\cite{Abbiendi:2007rs}&$7.41\left(1\pm 0.5\%\right)$&\cite{Abbiendi:2007rs}\\
$\sigma_{q q q q}$ &$188.63$&$6.88 \pm 0.23 \pm 0.09$&\cite{Heister:2004wr}&$7.54\left(1\pm 0.5\%\right)$&\cite{Abbiendi:2007rs}\\
&$191.58$&$8.21\pm0.61\pm0.09$&\cite{Heister:2004wr}&$7.67\left(1\pm 0.5\%\right)$&\cite{Abbiendi:2007rs}\\
&$195.52$&$7.51 \pm 0.35 \pm 0.09$&\cite{Heister:2004wr}&$7.75\left(1\pm 0.5\%\right)$&\cite{Abbiendi:2007rs}\\
&$199.52$&$7.40\pm0.33\pm0.09$&\cite{Heister:2004wr}&$7.77\left(1\pm 0.5\%\right)$&\cite{Abbiendi:2007rs}\\
&$201.62$&$6.96\pm 0.47 \pm 0.09$&\cite{Heister:2004wr}&$7.79\left(1\pm 0.5\%\right)$&\cite{Abbiendi:2007rs}\\
&$204.86$&$7.79 \pm 0.35 \pm 0.09$&\cite{Heister:2004wr}&$7.80\left(1\pm 0.5\%\right)$&\cite{Abbiendi:2007rs}\\
&$206.53$&$7.73 \pm 0.27 \pm 0.09$&\cite{Heister:2004wr}&$7.80\left(1\pm 0.5\%\right)$&\cite{Abbiendi:2007rs}\\
\hline
\end{tabular}
\caption{ \label{ALEPHData} Measured cross section of the process $e^+ e^- \rightarrow \ell \nu \ell \nu$, $e^+ e^- \rightarrow qq \ell \nu$, $e^+ e^- \rightarrow qqqq$ by the ALEPH collaboration assuming charged-lepton universality. The SM predictions are taken to be the same as the ones for OPAL data. The SM theory error is  0.5\% of the SM values following \cite{Achard:2004zw}.}
\end{table}
\end{center}
%%%%%%%%%%%%%%%%%%%%%%%%%%%%%%%%%%%%%%%%%%%%%%%%%%%%%%
\begin{center}
\begin{table}
\centering
\tabcolsep 8pt
\begin{tabular}{|c|c|c|c|c|c|}
\hline
$\cos \theta \, \text{bin}$ & $\sqrt{s}$ [GeV] & Experimental value [pb] &Ref.& Theoretical value [pb] &Ref. \\ 
\hline \hline
 $\text{bin 1}$ &$182.66$&$0.502 \pm 0.114$&\cite{Schael:2013ita}&$0.74\left(1\pm 0.2\%\right)$& \cite{Achard:2004zw}\\
$\text{bin 4}$&$182.66$&$1.281 \pm 0.203 $&\cite{Schael:2013ita}&$1.20\left(1\pm 0.2\%\right)$&\cite{Achard:2004zw}\\
$\text{bin 7}$&$182.66$&$2.583\pm0.270$&\cite{Schael:2013ita}&$2.16\left(1\pm 0.2\%\right)$&\cite{Achard:2004zw}\\
$\text{bin 10}$&$182.66$&$5.372 \pm 0.419$&\cite{Schael:2013ita}&$5.47\left(1\pm 0.2\%\right)$&\cite{Achard:2004zw}\\
\hline
$\text{bin 1}$&$205.92$&$0.495\pm0.058$&\cite{Schael:2013ita}&$0.52\left(1\pm 0.2\%\right)$&\cite{Achard:2004zw}\\
$\text{bin 4}$&$205.92$&$1.057\pm 0.094$&\cite{Schael:2013ita}&$0.98\left(1\pm 0.2\%\right)$&\cite{Achard:2004zw}\\
$\text{bin 7}$&$205.92$&$2.294 \pm 0.140$&\cite{Schael:2013ita}&$2.06\left(1\pm 0.2\%\right)$&\cite{Achard:2004zw}\\
$\text{bin 10}$&$205.92$&$7.584 \pm 0.262$&\cite{Schael:2013ita}&$7.80\left(1\pm 0.2\%\right)$&\cite{Achard:2004zw}\\
\hline
\end{tabular}
\caption{ \label{DiffCrossSectionData} Combined measured $d\sigma_{\ell \nu q q}/d\text{cos}[\theta]$ reported by the LEPII collaboration \cite{LEP-2}. The SM theory error is taken to be 0.2\% of the SM values following \cite{Achard:2004zw}.}
\end{table}
\end{center}
%%%%%%%%%%%%%%%%%%%%%%%%%%%%%%%%%%%%%%%%%%%%%%%%%%%%%%%%
%%%%%%%%%%%%%%%%%%%%%%%%%%%%%%%%%%%%%%%%%%%%%%%%%%%%
\section*{Acknowledgements}
M.T. acknowledges generous support by the Villum Fonden and partial support by the Danish National Research Foundation (DNRF91).
The project leading to this application has received funding from the European Union's Horizon 2020 research
and innovation programme under the Marie Sklodowska-Curie grant agreement No 660876, HIGGS-BSM-EFT.
We thank Christine Hartmann, Andre Tinoco-Mendes, Giampiero Passarino, Troels Petersen, members of the Tevatron $W$ mass analysis group,
and particularly William Shepherd for helpful discussions.
We thank J. Erler and A. Freitas for helpful correspondence on SM predictions when developing the fit results, and patience.
MT also thanks members of the Higgs Cross Section Working Group 2 for feedback and discussions
related to the material presented here.
%--------------------------------------------------------------------------------------------
%%%%%%%%%%%%%%%%%%%%%%%%%%%%%%%%%%%%%%%%%%%%%%%%%%%%%
%%%%%%%%%%%%%%%%%%%%%%%%%%%%%%%%%%%%%%%%%%%%%%%%%%%%
\bibliographystyle{JHEP}
\bibliography{bibliography2}

\end{document}